\pdfoutput=1
\documentclass{article}
\usepackage[utf8]{inputenc}
\usepackage{graphicx}
\usepackage{float}
\usepackage{physics}
\usepackage{color}
\usepackage{hyperref}
\usepackage{amssymb}
\usepackage{xcolor}
\usepackage{wrapfig}
\usepackage{amsmath}
\usepackage{microtype}
\usepackage{wrapfig}
\DeclareMathOperator\arctanh{arctanh}
\usepackage{blindtext}
\newcommand{\HRule}[1]{\rule{\linewidth}{#1}}
\numberwithin{equation}{section}
\usepackage{geometry}
\title{\vspace*{2\baselineskip}
		\HRule{0.5pt} \\
		\LARGE \textbf{\uppercase{Entropy and replica geometry in generic two-dimensional dilaton gravity theories}\footnote{This is an updated version of writing sample for PhD applications which contains original non-trivial results.}}
		\HRule{2pt} \\ [0.5cm]
		\normalsize  \vspace*{2\baselineskip}}
\date{\vspace{-5ex}}
\author{Yueren Xing\\\ \href{mailto: y.xing1@students.uu.nl}{ y.xing1@students.uu.nl}\thanks{Institute for Theoretical Physics, Utrecht University, Princetonplein 5, 3584 CC Utrecht, The Netherlands}}  

\begin{document}

\maketitle

\begin{abstract}
       We set up a new version of black hole information paradox in an eternal Narayan black hole, a generic two-dimensional dilaton gravity theory with non-trivial on-shell bulk action and a product of dimensional reduction from higher-dimensional AdS black brane, joined to Minkowski bath on both sides. We also report both similarities as well as important differences between our model and the famous model of JT gravity coupled with baths. The contradiction of Hawking's result of entanglement entropy with unitarity is resolved by including a new saddle in the Euclidean gravitational path integral. As part of ongoing and developing research, we attempt, and have had partial success, to explicitly construct the replica wormhole geometry for our model to fully justify the quantum extremal surface calculations with Euclidean gravitational path integral, without using holography. \\

     %  \underline{\textcolor{red}{\textbf{\large [UPDATES TO THE EARLIER VERSION OF WRITING SAMPLE]}}}\\
       
     % \textbf{This latest version of writing sample of my thesis research receives many updates, corrections and is also greatly expanded.} \\
      
     % Subsection \ref{eternal BH + bath} is, to a significant degree, rewritten. The boundary conditions of joining the baths to the black hole are laid down with more rigor. The eternal Narayan black holes we construct are now considerably reinforced with a calculation of their ADM energy, a derivation of EOM for boundary modes plus multiple consistency checks including stress-energy tensor at the boundary explicitly calculated. All of them combined proves the kinetic constraint for the boundary modes work well in eternal black holes.\\

      %In subsection \ref{one interval section}, important corrections are made to the extremity conditions and the existence of QES in one-interval case, substantiated with plots.\\

      %A whole new section \ref{rep} is added. In subsection \ref{general replica formalism}, we provide justifications from the replica trick for calculating von Neumann entropy through evaluating Euclidean path integral, supported by graphics, for results obtained in section \ref{BHIP}, the formalism in which is slightly different from that in \cite{almheiri2020replica}. In subsection \ref{attempt}, an serious attempt is made to construct the geometry of the replica wormhole for one-interval scenario and recover the extremity condition, the bulk of detailed calculation can be found in the newly added Appendix \ref{B}.

\end{abstract}

\newpage

\tableofcontents

\newpage

\section{Introduction}\label{intro}
In recent years, there have been many papers dedicated to the development of self-consistent solutions, built from first principles of theoretical physics, to certain versions of the black hole information paradox(puzzle) in various gravity theories, see \cite{almheiri2021entropy} for a general and heuristic review on recent progress. Among them, a simple model of black holes in the two-dimensional Jackiw-Teitelbolm (JT) gravity \cite{jackiw1985lower}\cite{teitelboim1983gravitation}\cite{almheiri2015models} stands out as the playground of one version of the information paradox first formulated in \cite{almheiri2019entropy} and \cite{penington2020entanglement} where a black hole in anti-de Sitter spacetime radiates into an attached non-gravitational flat region. \hfill\break

An answer to this version of the paradox was proposed using different theoretical tools in a series of paper \cite{almheiri2020page} \cite{rozali2020information} \cite{chen2020information}\cite{almheiri2019islands}\cite{almheiri2020replica}\cite{penington2022replica}\cite{goto2021replica}. Their results converge on the point that a quantum extremal island appears in the interior of the black hole after Page time through an extension of the generalized gravitational entropy formula \cite{almheiri2019entropy} \cite{penington2020entanglement} \cite{ryu2006holographic} \cite{hubeny2007covariant} \cite{lewkowycz2013generalized} \cite{barrella2013holographic} \cite{faulkner2013quantum} \cite{engelhardt2015quantum}, namely the following 'island formula'\cite{almheiri2019islands},

\begin{equation} \label{island formula}
    S(R) = \min \text{ext}_{\textit{I}} \bigg[ \frac{\text{Area} (\partial I)}{4 G_N} + S_{\text{QFT} (I \cup R)} \bigg]
\end{equation}
where $R$ is the radiation, $I$ is the island, and $S_{\text{QFT}}$ is the traditional semi-classical entropy of quantum fields computed by quantum field theory in curved spacetime. In the context of two-dimensional dilaton gravity, 'Area' is represented by the value of the dilaton. \hfill\break

While the island formula (\ref{island formula}) was first put forward based on - and a unitary Page curve \cite{page1993information} \cite{page2013time} is consistently produced derived through- holography, a calculation, without holography while recovering the same result, was done using the gravitational replica method (developed in \cite{lewkowycz2013generalized} and \cite{dong2016deriving}) originally by Almheiri 
et.al. \cite{almheiri2020replica} for eternal JT black holes joined by Minkowski space and later generalized into evaporating black hole scenario in \cite{goto2021replica}. As shown in \cite{almheiri2020replica}, new saddles - replica wormholes - in the gravitational path integral dominate, at late time, over the Hawking saddle which gives the ever-increasing entanglement entropy of the radiation initially derived by Hawking \cite{hawking1976breakdown} and thus inconsistent with unitarity. \hfill\break

Although in \cite{almheiri2020replica} the results were explicitly obtained only in some simple examples with regards to the information paradox for eternal black hole in JT gravity, the same method based on Euclidean path integral description nonetheless can, at least in principle, be applied to eternal black holes in other more general two-dimensional dilaton gravity theories, again producing large corrections to the fine-grained entropy from the effects brought by replica wormholes. \hfill\break

One interesting type of the more general two-dimensional gravity theories arises from the two-dimensional subsector of higher-dimensional gravity on spaces of the form $\mathcal{M}_2 \cross X^d$ on Kaluza-Klein compactification over the compact space $X^d$ \cite{narayan2021aspects}, while $\text{AdS}_2$ arising in the near horizon geometry of extremal black holes and branes is one well-known example of it. \cite{sarosi2017ads} \cite{rosenhaus2019introduction} \cite{trunin2021pedagogical}. Here we study one specific example of this type, two dimensional dilaton gravity theory as the product of dimensional reduction of higher dimensional gravity with a negative cosmological constant, or Narayan theories as we name them. The on-shell metric of Narayan theories is always conformally $AdS_2$, whatever higher dimension the gravity theory lives in before dimensional reduction. But unlike the constant curvature $AdS_2$ arising from $AdS_3$ by dimensional reduction, which is the simplest example of these conformally $AdS_2$ metrics, it generally has an IR singularity - a curvature singularity where the dilaton is also vanishingly small, and a UV divergence in the \textbf{bulk} on-shell action where the dilaton grows large \cite{narayan2021aspects}. Both of these are, however, operationally curable if we introduce a black hole horizon to regulate the curvature singularity and add appropriate counterterms to regulate the on-shell action. The divergences the two-dimensional theories exhibit suggest they are best regarded as UV incomplete thermodynamic low energy effective theories universal to all UV completions $AdS_D \cross X$ and encode higher dimensional gravity intrinsically, quite different from the near extremal near horizon $AdS_2 \cross X$ throats in extremal objects, where the $X$-compactification leads to an intrinsically two-dimensional theory. What we are concerned with the most in this article, is to formulate a version of black hole information problem for eternal black holes in a model of generic two-dimensional dilaton gravity theory featuring both on-shell curvature singularity and a divergent bulk action as opposed to the constant non-singular curvature and a trivial on-shell bulk action in JT gravity, compute the fine-grained entropy of the black hole and Hawking radiation and then set up replica wormhole geometry to verify the calculations by recovering the same quantum extremal surface (QES) result. We also very much intend, in addition, to compare the results obtained in our version of black hole information problem and the underlying reasoning process with that in the context of JT gravity \cite{almheiri2020replica}. \hfill\break

\newpage

This article is organized as follows. \\

Section 2 is dedicated to setting the stage for our version of black hole information paradox. In section 2.1 we review generic two-dimensional dilaton gravity theories arising from dimensional reduction of higher-dimensional gravity theories by writing down the bare action with an on-shell UV divergence and the metric and dilaton solutions bearing IR singularity. We then show the theory can be regulated via a black hole horizon leading to a black hole solution of the same theory and of non-zero Hawking temperature. We also show there is a renormalized action made up of the bulk and boundary terms where all the UV divergence of the action cancel out. We then, in section 2.2, proceed to maximally extend it to a black hole with two asymptotic regions connected by a wormhole and possessing spacelike singularities in the black hole and white hole regions cloaked by the event horizon. In order to let the black hole radiate  away Hawking quanta, in section 2.3, we join the maximal extension of the black hole by two copies of Minkowski half-space, one on each side, and then couple it to a CFT with $c \gg 1$ to build an eternal black hole assuming a global vacuum state, mimicking the setup in \cite{almheiri2019islands} and \cite{almheiri2020replica}. Of course, the boundary conditions at the physical boundary gluing the black hole spacetime and Minkowski space change accordingly, differing from those in JT gravity (see section 2 of \cite{almheiri2019entropy} or \cite{goto2021replica}), and impose new kinematic constraint on the dynamics of 'boundary particle' \cite{maldacena2016conformal} \cite{engelsoy2016investigation} \cite{jensen2016chaos}by identifying the Poincaré time with the boundary proper time to the leading order. This new kinematic constraint proves to be consistent with an eternal black hole solution in Narayan theory.  \\

In section 3 we formulate our version of black hole information paradox and show it can be resolved after a calculation similar to that in \cite{almheiri2019islands} and \cite{almheiri2020replica}. In section 3.1, we begin with QES calculation in the simple case of a single interval tethered to the boundary of eternal black hole on one end, which constitutes a warm-up for the more physically relevant case of two-intervals in the eternal black hole in section 3.2 where our version of black hole information paradox is formally presented. We show, at late times, the island always exists and dominates the entropy in two-interval scenario, just like in JT gracity \cite{almheiri2019islands} \cite{almheiri2020replica}, producing large non-perturbative corrections to the true entropy of the Hawking radiation and the two-side eterbal black hole according to the QES prescription.\\

In section 4, we attempt to do the replica calculations in Euclidean signature. We first review, in section 4.1, an extension of the general replica trick for computing the von Neumann entropy of a system coupled with gravity and then apply it to our theory. We then continue to define the "conformal welding problem" in the most general form, which will characterize joining the replica geometry to the flat space in Euclidean signature. We then apply this formalism to our model to set up the replica geometry. In section 4.2, for the one-interval scenario built in section 3,1,  we propose a central hypothesis about the on-shell replica geometry on the covering manifold, inspired by its parallel in JT gravity. Based on this hypothesis, we construct the on-shell action on the orbifold where conical singularities and twist fields are present. We then try to recover the extremity condition for the one-interval scenario by deriving the equation of motion for the boundary mode. Although this hypothesis proves not to account for the whole story and is thus incomplete, there is still much to discuss and the method employed in this subsection to sidestep conical singularities in the calculation of the bulk action can be easily generalized/modified to be used in other theories.  \\

In section 5, we conclude the results we obtained and then share some thoughts about how we can fully recover the results we obtained in section 3 by accounting for the complete on-shell replica geometry. We also compare our version of the black hole information paradox and the JT version and discuss the underlying reason and list some possibilities of further research.

\newpage

\section{Generic dilaton gravity theories in 2D: Narayan theories}\label{narayan}
 The bulk action of the most general two-dimensional dilaton gravity theory (without kinetic terms of the dilaton) takes the form

\begin{equation} \label{general action}
    I = \frac{1}{16 \pi G_2} \int d^2 x \sqrt{-g} (\phi \mathcal{R} - U (\phi)),
\end{equation}
including the two-dimensional dilaton $\phi$ coupled to gravity and a general dilaton potential $U (\phi)$. (For the definition of more general dilaton gravity theories in 2D, see \cite{grumiller2022generalized}.) The equations of motion of the above general 2D dilaton gravity theory are

\begin{equation} \label{general EOM}
     g_{\mu \nu} \nabla^2 \phi - \nabla_\mu \nabla_\nu \phi + \frac{1}{2} g_{\mu \nu} U = 0, \quad \mathcal{R} - \frac{\partial U}{\partial \phi} = 0.
\end{equation}

In JT gravity, the potential takes the simple form $U (\phi) = -2 \phi$ in (\ref{general action}). The EOMs (\ref{general EOM}) dictate the metric part of the 
 classical solution is a constant negative curvature spacetime where $\mathcal{R} = -2$, identified as $\text{AdS}_2$ spacetime, and the on-shell bulk action vanishes. JT gravity, as the simplest 2D dilaton gravity theory of physical relevance, has been under active investigation following discussions of the nearly $\text{AdS}_2$ holography. \cite{almheiri2015models} \cite{maldacena2016conformal} \cite{engelsoy2016investigation} \cite{jensen2016chaos}. \\
 
 It is widely known that $\text{AdS}_2$ arises in the near horizon geometry of extremal black holes and branes \cite{sarosi2017ads} \cite{rosenhaus2019introduction} \cite{trunin2021pedagogical}. Nonetheless in its own right, upon Kaluza-Klein compactification over the compact space $X^d$, 2D dilaton gravity theories arise generically from the 2D subsector of higher dimensional gravity on spaces of the form $\mathcal{M}_2 \cross X^d$. We now review an interesting example of the dilaton gravity theories produced by dimensional reduction from well-behaved theories in higher dimensions.

 \subsection{Basics of Narayan theory and its regularization} \label{basics}

The dimensional reduction of higher dimensional gravity theories with a negative cosmological constant is a prototypical example of arising generic dilaton gravity. The gravity sector of such higher dimensional gravity theories gives the action

\begin{equation} \label{D action}
    \begin{split}
       & I_D = \frac{1}{16 \pi G_D} \int d^D x \sqrt{-g^{(D)}} (\mathcal{R}^{(D)} - 2 \Lambda),\\
       & \Lambda = - \frac{1}{2} d_i (d_i + 1), \quad D = d_i + 2.
    \end{split}
\end{equation} 
Following \cite{narayan2021aspects}, we adopt the reduction conventions of \cite{kolekar2018ad} \cite{kolekar2019ads2} \cite{bhattacharya2020cosmological}. The higher dimensional space takes the form  $\mathcal{M}_2 \cross X^{d_i}$ by assuming translational and rotational invariance in the $d_i$ spatial dimensions on the boundary. If $X^{d_i}$ is taken to be $T^{d_i}$ - a torus and is compactified with the following reduction ansatz 

\begin{equation} \label{reduction metric}
    ds^2_D = g_{\mu \nu}^{(2)} dx^\mu dx^\nu + \phi^{\frac{2}{d_i}} d\sigma^2_{d_i}; \quad g_{\mu \nu} = \phi^{\frac{d_i - 1}{d_i}} g_{\mu \nu}^{(2)},
\end{equation}
the following 2D dilaton gravity theory or \textit{Narayan theory}, as we call it, with a monomial potential comes out of the reduction,

\begin{equation} \label{bare action 0}
   I_{\text{bulk}} = \frac{1}{16 \pi G_2} \int d^2 x \sqrt{-g} (\phi \mathcal{R} - U (\phi)), \quad U = 2 l^{-2} \Lambda \phi^{\frac{1}{d_i}} = - l^{-2} d_i (d_i + 1) \phi^{\frac{1}{d_i}}, 
\end{equation}
where $l$ is a constant length scale, similar to the $\text{AdS}_2$ radius in JT gravity.
See the appendix of \cite{narayan2021aspects} for how $T^{D - 2}$ reduction of the $D$-dimensional space with $D = d_i + 2$ gives (\ref{bare action 0}). The $d_i = D - 2$ spatial coordinates defines the transverse part of the bulk spacetime, we also identify the dilaton defined in (\ref{reduction metric}) as the transverse area in $D$-dimensional space, $g_{ii}^{(D-2)/2} = \phi$. Note that if we set $d_i = 1$, $\text{AdS}_3$ leads to $U = -2 l^{-2} \phi$ in (\ref{bare action 0}) marking the JT action.  \\

We now present one simple solution to the theory (\ref{bare action 0}). We assume the 2D space here is expressed in "Poincaré" coordinates - a time coordinate, $t$ and a radial coordinate, $r$ - and the vector field parameterized by $t$ is a Killing symmetry, then in conformal gauge, the metric ansatz can be written as 

\begin{equation}
    ds^2 = g_{\mu \nu} dx^\mu dx^\nu = e^{h(r)} (-dt^2 + dr^2)
\end{equation}
and the Einstein equations can be written as 

\begin{equation}
    -\partial^2_r \phi + \partial_r h \partial_r \phi = 0, \; \partial^2_r \phi + e^h U = 0, \; -\partial^2_r h - e^{h} \partial_{\phi} U = 0.
\end{equation}

It can be shown (\cite{narayan2021aspects}) that the following conformally $\text{AdS}_2$ solution is a time-independent solution to the above Einstein equations
\begin{equation} \label{unregulated solution 1}
    ds^2 = \frac{l^{d_i + 1}}{r^{d_i + 1}} (-dt^2 + dr^2) = \frac{l^{d_i - 1}}{r^{d_i - 1}} ds^2_{\text{AdS}_2}, \quad \phi = \frac{l^{d_i}}{r^{d_i}}.
\end{equation}
where $l$ is still a constant length scale.
According to (\ref{general EOM}), the curvature is 

\begin{equation} \label{curvature 0}
    \mathcal{R} = -(d_i + 1) \, l^{-(d_i + 1)} \,  r^{d_i - 1},
\end{equation}
displaying an curvature (IR) singularity in the deep interior as $r \rightarrow \infty$ for $d_i > 1$. Note that if we set $d_i = 1$, the metric is exactly $\text{AdS}_2$ with a constant curvature $\mathcal{R} = -2$. However, if $d_i > 1$, the geometry is only conformally $\text{AdS}_2$. As one moves towards the boundary $r = 0$, the conformal factor and the value of dilaton both blow up while the Ricci curvature scalar $\mathcal{R} \rightarrow 0$, showing the space "flattens" near the boundary. Another thing worth much attention is that the on-shell bulk action of the solution (\ref{unregulated solution 1}) has a UV-divergence if $d_i \neq 1$. The reasoning is follows. Starting from (\ref{bare action 0}) and (\ref{general EOM}), we have 

\begin{equation} \label{on-shell bare action}
    \begin{split}
        I_{\text{bulk, on-shell}} & = \frac{1}{16 \pi G_2} \int d^2 x \sqrt{-g} \bigg( \phi \frac{\partial U}{\partial \phi} - U (\phi) \bigg) \\
        & = \frac{({d_i}^2 - 1) }{16 \pi G_2} \int d^2 x \sqrt{-g} \, \phi^{\frac{1}{d_1}}\\
        & \propto ({d_i}^2 - 1) \int dt \int_0dr  \, \frac{1}{r^{d_i + 1}} \cdot \frac{1}{r} =  ({d_i}^2 - 1) \int dt \int_0dr  \, \frac{1}{r^{d_i + 2}}. 
    \end{split}
\end{equation}
While in JT case, $d_i = 1$, so the on-shell action vanishes due to the same equation (\ref{on-shell bare action}).\\

The IR curvature singularity deep inside the solution (\ref{unregulated solution 1}) can nonetheless be regulated by introducing a black hole horizon as follows,

\begin{equation} \label{metric Hr}
    ds^2 = e^{h(r)} (-dt^2 + dr^2) \longrightarrow e^{h(r)} \bigg(- H(r) dt^2 + \frac{1}{H(r)}dr^2 \bigg);
\end{equation}
with $H \rightarrow 1$ as $r \rightarrow 0$ and $H \rightarrow 0$ as $r \rightarrow r_0$, where $r = r_0$ is the radius of the event horizon. 
The Einstein equations now become

\begin{equation} \label{EOM black factor}
    -\partial^2_r \phi + \partial_r h \partial_r \phi = 0, \quad \partial_r (H \partial_r \phi) + e^h U = 0, \quad   H \partial^2_r h + \partial_r H \partial_r h + \partial^2_r H + e^h \frac{\partial U}{\partial \phi} = 0. 
\end{equation}

Combining (\ref{EOM black factor}) and the boundary condition for $H(r)$ below (\ref{metric Hr}), we get the regulated black hole solution to Narayan theory (\ref{bare action 0})

\begin{equation} \label{BH solution 0}
  ds^2 = \frac{l^{d_i + 1}}{r^{d_i + 1}}  \bigg(- H(r) dt^2 + \frac{1}{H(r)}dr^2 \bigg), \; H(r) = 1 - \frac{r^{d_i + 1}}{{r_0}^{d_i + 1}};\quad \phi = \frac{l^{d_i}}{r^{d_i}}, \end{equation}
and $\mathcal{R} =  -(d_i + 1) \, l^{-(d_i + 1)} \,  r^{d_i - 1}$ still holds.
Essentially, this is  not the dimensional reduction of global $\text{AdS}_D$ spacetime but the $\text{AdS}_D$ black brane \cite{narayan2021aspects}.\\

To obtain the temperature of the black hole solution (\ref{BH solution 0}), we do a Wick rotation  $t \rightarrow i \theta$,

\begin{equation} \label{Euclidean BH}
     ds^2_E = \frac{l^{d_i + 1}}{r^{d_i + 1}}  \bigg( H(r) d\theta^2 + \frac{1}{H(r)}dr^2 \bigg), \; H(r) = 1 - \frac{r^{d_i + 1}}{{r_0}^{d_i + 1}}.
\end{equation}
We then expand the Euclidean black hole metric (\ref{Euclidean BH}) at the horizon $r = r_0$ 
\begin{equation} \label{near-horizon metric 0}
    ds^2_{\text{near horizon}} = -l^{d_i + 1} \biggl[ \frac{(d_i + 1)(r - r_0)}{{r_0}^{d_i + 2}} d\theta^2 + \frac{1}{(d_i + 1) {r_0}^{d_i} (r - r_0)} dr^2
    \biggr],
\end{equation}
and define the near-horizon radial coordinate 

\begin{equation}
    \rho = \frac{2}{{r_0}^{\frac{d_i}{2}}\sqrt{d_i + 1}} \sqrt{r -r_0}.
\end{equation}
In $\rho$ and Euclidean time $\theta$, the near-horizon geometry (\ref{near-horizon metric 0}) takes the form

\begin{equation} \label{near-horizon metric 1}
    ds^2_{\text{near horizon}} = -l^{d_i + 1}  \biggl[  d\rho^2 + \frac{(d_i + 1)^2}{ 4{r_0}^{2}} \rho^2 d\theta^2 \biggr]. 
\end{equation}
That there is no conical singularity at the horizon $r = r_0$ in ( (\ref{near-horizon metric 1})) gives the Hawking temperature of the black hole (\ref{BH solution 0}) in Narayan theory

\begin{equation}
    T = \frac{1}{\beta} = \frac{d_i + 1}{4 \pi r_0}. 
\end{equation}

As shown in (\ref{on-shell bare action}), the on-shell bulk action for the original conformally $\text{AdS}_2$ solution (\ref{unregulated solution 1}) and the black hole solution (\ref{BH solution 0}) in Narayan theory has a UV divergence for $d_i \neq 1$. This can be fixed by adding the Gibbons-Hawking-York boundary term $I_{\text{GHY}}$ and a holographic counterterm $I_{\text{ct}}$ \cite{narayan2021aspects}

\begin{equation} \label{regularized action}
\begin{split}
    I_{\text{ren}} & = I_{\text{bulk}} + I_{\text{GHY}} + I_{\text{ct}} \\
                  & = \frac{1}{16 \pi G_2} \biggl[ \int d^2 x \sqrt{-g} (\phi \mathcal{R} - U (\phi)) - 2 \int dt \sqrt{- \gamma} \phi K - 2 d_i \int dt \sqrt{- \gamma} \phi^{\frac{d_i + 1}{2 d_i}} \biggr], \ U =  - l^{-2} d_i (d_i + 1) \phi^{\frac{1}{d_i}},
\end{split}
\end{equation} 
where $\gamma_{\mu \nu}$ is the induced metric on the timelike boundary $\partial \mathcal{M}$ at $r = 0$ and $K$ is the extrinsic curvature at $\partial \mathcal{M}$ for the outward pointing normal with $n^r > 0$. It can be easily proved that the UV divergence in the action disappears as

\begin{equation} \label{UV cancel 0}
    I_{\text{ren}}^{\text{div}} \propto \frac{1}{16 \pi G_2} \bigg( \frac{d_i - 1}{\epsilon^{d_i + 1}} + \frac{d_i + 1}{\epsilon^{d_i + 1}} - \frac{2 d_i}{\epsilon^{d_i + 1}} \bigg) \longrightarrow 0 \; (\epsilon \rightarrow 0),
\end{equation}
therefore the action is renormalized. The variation of the renormalized action (\ref{UV cancel 0}) with the metric can be written as \cite{grumiller2017menagerie}
\begin{equation} \label{variation}
\begin{aligned}
   \delta I_{\text{ren}} & = \frac{1}{16 \pi G_2} \int d^2 x \sqrt{-g} \bigg(  g_{\mu \nu} \nabla^2 \phi - \nabla_\mu \nabla_\nu \phi + \frac{1}{2} g_{\mu \nu} U \bigg) \delta g^{\mu \nu} \\
    & + \bigg( \frac{\delta  I_{\text{ct}}}{\delta \gamma^{\mu \nu}} -\frac{1}{16 \pi G_2} \int dt \sqrt{- \gamma} \gamma_{\mu \nu} n^\rho  \nabla_\rho \phi \bigg) \delta \gamma^{\mu \nu}, 
\end{aligned}
\end{equation}
which gives the correct Einstein equation (\ref{general EOM}) in the bulk.

 \subsection{The maximally extended black holes in Narayan theories when $d_i = 3$}

 Now that we have the black hole solution (\ref{BH solution 0}) with a regulated metric and the renormalized action, (\ref{regularized action}) of the Narayan theory, we are ready to construct the playground of our version of black hole information paradox. \textbf{For algebraic simplicity and without loss of generality, we set $d_i = 3$ throughout the rest of this article.}\\

 To maximally extend the black hole solution in Narayan theory, we start from the metric in the black hole solution of inverse temperature $\beta = \pi r_0$, expressed in "Poincaré" coordinates $\{t , r\}$ with $d_i = 3$, 
 
 \begin{equation} \label{BH solution 1}
ds^2  =  \frac{l^4}{r^4} \bigg[ -\bigg( 1- \frac{r^4}{r_0^4} \bigg) dt^2 + {\bigg( 1 - {\frac{r^4}{r_0^4}} \bigg)}^{-1} dr^2 \bigg] = \frac{l^4}{r^4} \bigg( 1- \frac{\pi^4 r^4}{\beta^4} \bigg) \bigg[ -dt^2 + {\bigg( 1 - {\frac{\pi^4 r^4}{\beta^4}} \bigg)}^{-2} dr^2 \bigg].
\end{equation}
The above metric has a coordinate singularity at the horizon $r = r_0 = \frac{\beta}{\pi}$ and therefore $r \in (0, r_0) $ in the Poincaré patch.\\

We first define the tortoise coordinate in the Poincaré patch
\begin{equation} \label{tortoiseDef}
\begin{split}
 r_*  & = r_0 \bigg( \frac{1}{2} \arctan \frac{r}{r_0} + \frac{1}{4} \log \abs{\frac{r+r_0}{r-r_0}}\bigg) = \frac{1}{2} r_0 \bigg(\arctan{\frac{r}{r_0}} +\arctanh{\frac{r}{r_0}}  \bigg) \\
 & = \frac{\beta}{2 \pi} \bigg(\arctan{\frac{\pi r}{\beta}} +\arctanh{\frac{\pi r}{\beta}}  \bigg),
\end{split}
\end{equation}
where $r_*$ is a monotonically non-decreasing function of r and $r_* = 0$ when $r = 0$, $r_* \rightarrow +\infty $ when $r \rightarrow r_0$. It is not feasible to analytically invert (\ref{tortoiseDef}). The black hole metric in (\ref{BH solution 1}) can thus be written in tortoise coordinates as 

\begin{equation} \label{BH metric in tortoise}
    ds^2 = l^4 \bigg( \frac{1}{r^4} - \frac{\pi^4 }{\beta^4} \bigg) (-dt^2 + dr_*^2).
\end{equation}

\begin{figure}[H]
    \centering
    \includegraphics[width = 0.7\linewidth]{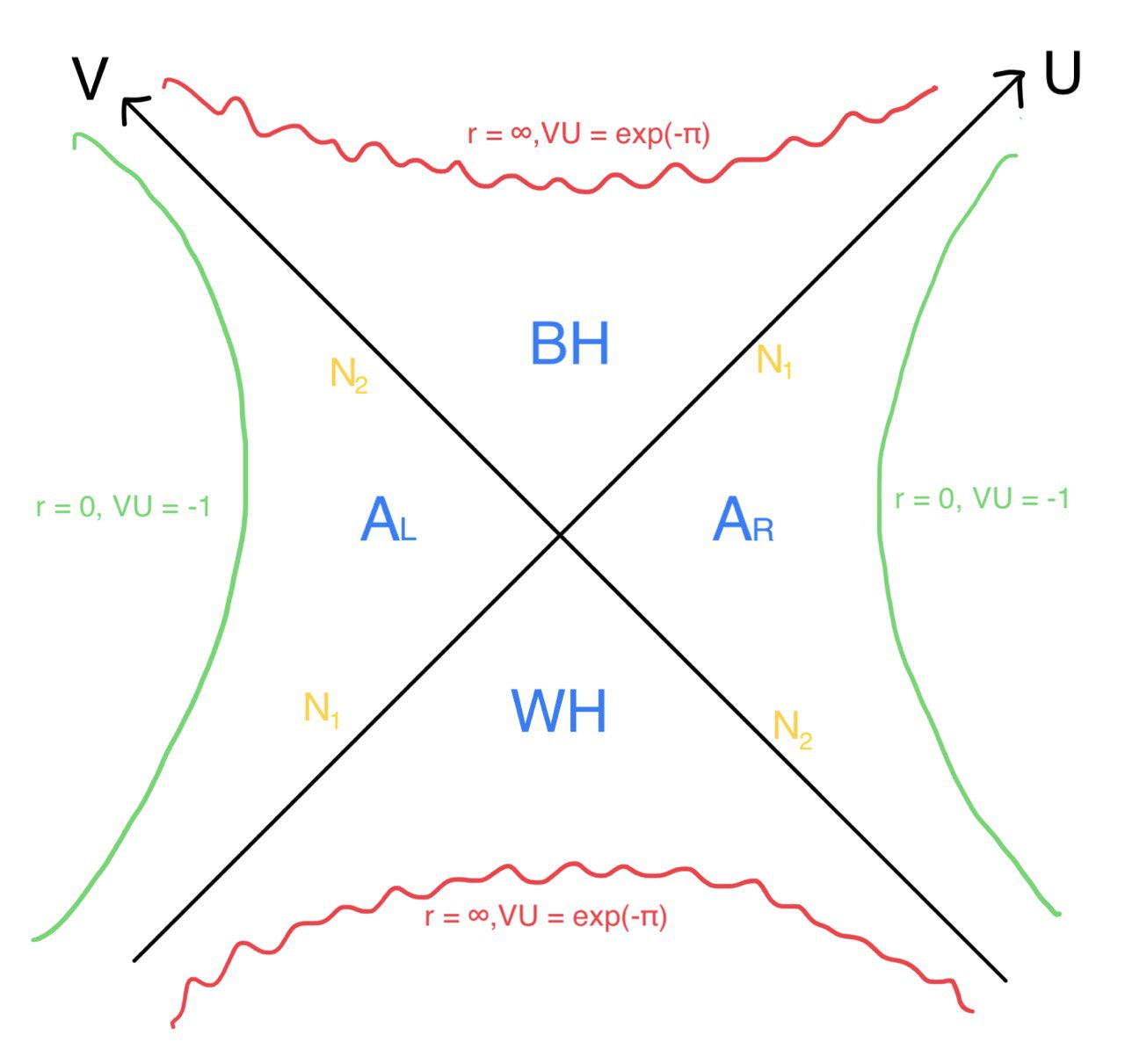}
    \caption{Maximally extended Narayan black hole in Kruskal coordinates. The timelike boundaries are denoted as two green curves and the spacelike singularities are shown as red wavy lines. Meanwhile, the event horizons comprise two diagonal black lines where $r = r_0 = \frac{\beta}{\pi}, \, VU = 0$. }
    \label{Max VU}
\end{figure}

Then, based on (\ref{tortoiseDef}), we define the Kruskal(-like) coordinates

\begin{equation} \label{Kruskaldef}
    V = -e^{-\frac{2 \pi}{\beta} (t + r_*)}, \quad U = e^{\frac{2 \pi}{\beta} (t - r_*)},
\end{equation}
where $V \in ( -\infty, 0 )$  , $ U \in ( 0, \infty) $. The (implicit) inverse coordinate transformations in "Poincaré" patch are

\begin{equation}
    VU = -e^{- \frac{4 \pi}{\beta}r_*} = e^{-2 \arctan{\frac{\pi r}{\beta}}} \frac{r - \frac{\beta}{\pi}}{r + \frac{\beta}{\pi}} 
\end{equation} 
and
\begin{equation}
    t = - \frac{\beta}{4 \pi} \log \left(-\frac{V}{U} \right) .
\end{equation}
The black hole metric can now be cast into the following form in Kruskal coordinates 

\begin{equation} \label{max BH metric}
ds^2 = -\frac{l^4 (r^2 + {r_0}^2) (r + r_0)^2}{4 {r_0}^2 r^4} e^{2 \arctan \frac{r}{r_0}} dVdU = -\frac{\pi^2 l^4 \Big( r^2 + \frac{\pi^2 }{\beta^2} \Big) {\Big(r +  \frac{\pi }{\beta } \Big)}^2}{4 \beta^2 r^4} e^{2 \arctan \frac{\pi r}{\beta}} dVdU,
\end{equation}
which is regular at the event horizon $r= r_0 =\beta/\pi$ and can thus be extended beyond the event horizon to a maximally extended black hole spacetime with two asymptotic regions, $A_R$ and $A_L$, connected by a wormhole and two spacelike singularities, one in the black hole region $BH$ and the other in the white hole region $WH$, cloaked by the event horizons $N_1$ and $N_2$. see Figure \ref{Max VU}.

Next, we list the relation among Poincaré coordinates, tortoise coordinates and Kruskal coordinates in all 4 regions

\begin{equation} \label{coordinateAR}
 A_R:
 \begin{cases}
      V = - e^{ - \frac{2\pi}{\beta} (t + r_*)},\\
      U = e^{ \frac{2\pi}{\beta} (t - r_*)},
 \end{cases}
r_* = \frac{\beta}{2 \pi} \bigg(\arctan{\frac{\pi r}{\beta}} +\arctanh{\frac{\pi r}{\beta}}  \bigg);
\end{equation}

\begin{equation} \label{coordinateAL}
 A_L:
 \begin{cases}
      V =  e^{ - \frac{2\pi}{\beta} (t + r_*)},\\
      U = - e^{ \frac{2\pi}{\beta} (t - r_*)},
 \end{cases}
 r_* = \frac{\beta}{2 \pi} \bigg(\arctan{\frac{\pi r}{\beta}} +\arctanh{\frac{\pi r}{\beta}}  \bigg);
\end{equation}

\begin{equation} \label{coordinateB}
 BH:
 \begin{cases}
      V = e^{ - \frac{2\pi}{\beta} (t + r_*)},\\
      U = e^{ \frac{2\pi}{\beta} (t - r_*)},
 \end{cases}
  r_* = \frac{\beta}{2 \pi} \bigg(\arctan{\frac{\pi r}{\beta}} +\arctanh{\frac{\beta}{\pi r}}  \bigg);
\end{equation}

\begin{equation} \label{coordinateW}
 WH:
 \begin{cases}
      V = - e^{ - \frac{2\pi}{\beta} (t + r_*)},\\
      U = - e^{ \frac{2\pi}{\beta} (t - r_*)},
 \end{cases}
  r_* = \frac{\beta}{2 \pi} \bigg(\arctan{\frac{\pi r}{\beta}} +\arctanh{\frac{\beta}{\pi r}}  \bigg);
\end{equation}
while we stress that following two identities regarding the relations between the Kruskal and the Poincaré coordinates hold throughout the maximally extended spacetime:
\begin{equation}\label{coordinate inverse Vu}
     VU = e^{-2 \arctan{\frac{\pi r}{\beta}}} \cdot \frac{r - \frac{\beta}{\pi}}{r + \frac{\beta}{\pi}},
\end{equation}
and
\begin{equation}\label{t in VU}
    t = - \frac{\beta}{4 \pi} \log \abs{\frac{V}{U}}. 
\end{equation}

Also, from (\ref{coordinate inverse Vu}), we see the boundary is located at $r = 0,r_* = 0, VU = -1$. Note that we cannot extend the metric beyond the boundary even though, in principle, the metric could still be extended if we disregarded the dilaton.  However, the dilaton $\phi = l^3 / r^3$ diverges when approaching the boundary, and if the metric is extended beyond the boundary where $r<0$, the value of the dilaton will suddenly change from plus infinity to minus infinity once we cross the boundary, leading to an incurable singularity. Since the dilaton is an integral part of the Narayan black hole solutions, we cannot just consider the extension of the metric so extension beyond the boundary $r = 0,r_* = 0, VU = -1$ is a no-go. Meanwhile, the event horizons $N_1$ and $N_2$ lie at $r = r_0 = \beta/\pi, r_* = \infty , VU = 0$; the spacelike singularities where $R = - 4 l^{-4} r^2$ blows up in $BH$ and $WH$ are found to be at $r = \infty , VU = e^{-\pi}$. \\

\begin{figure}[H]
    \centering
    \includegraphics[width = 0.6\linewidth]{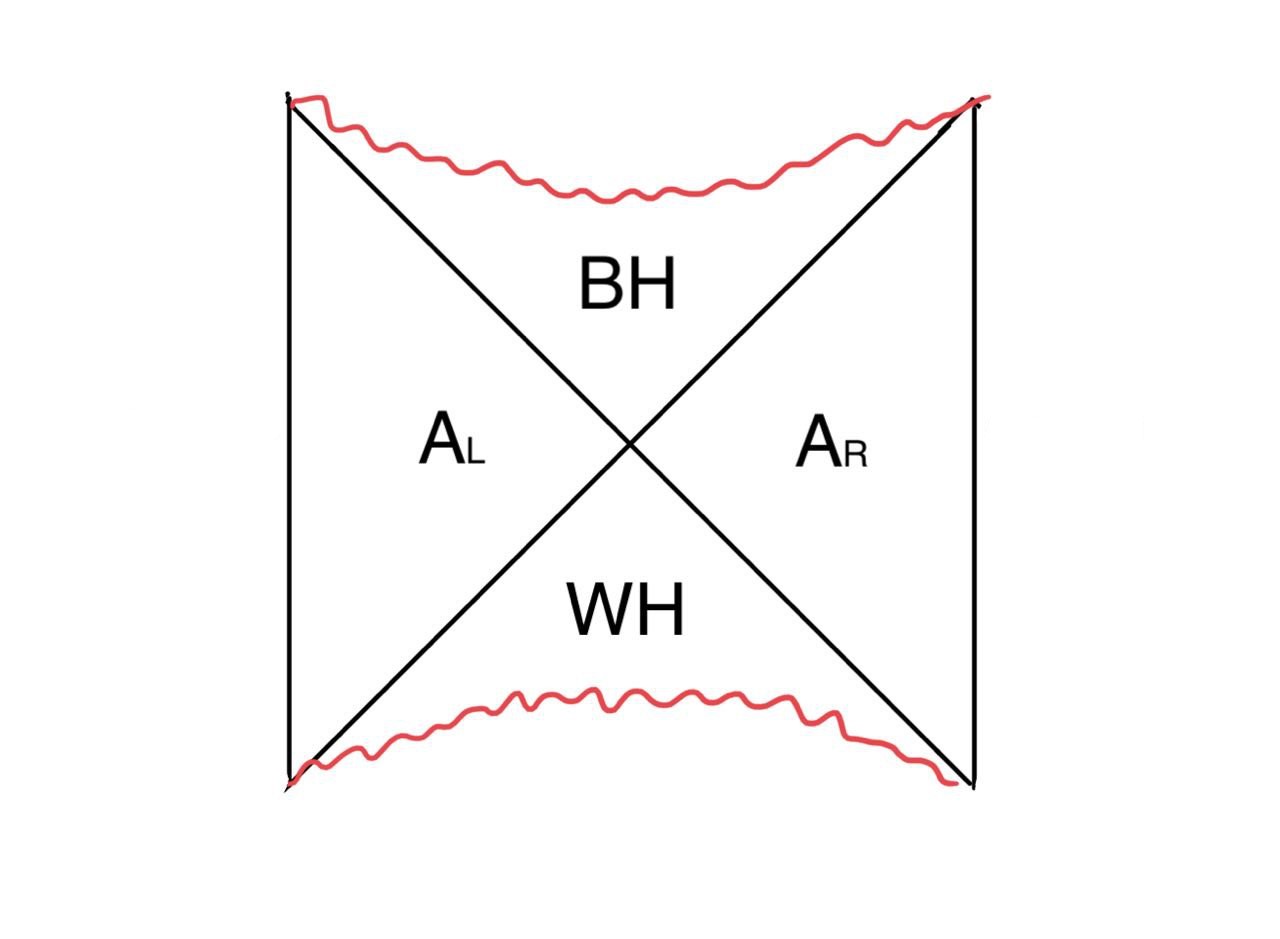}
    \caption{The Penrose diagram of the maximally extended Narayan black hole in conformal coordinates $\xi$ and $\chi$. Again, the spacelike singularities are denoted by red wavy lines.}
    \label{Pensrose 0}
\end{figure}

Based on the Kruskal coordinates $V$ and $U$, we can further define the conformal coordinates 
\begin{equation}
    \xi = \arctan U + \arctan V, \quad \chi = \arctan U - \arctan V,
\end{equation}
such that the maximally extended black hole spacetime can be illustrated as the Penrose diagram Figure \ref{Pensrose 0} . Note that this maximally extended black hole spacetime bears some resemblance to the maximally extended Schwarzschild spacetime though we have two wormhole-connected asymptotically conformally $\text{AdS}_2$ regions each with a timelike conformal boundary rather than two asymptotically flat regions.

\subsection{Narayan theory plus a CFT: eternal black hole coupled to baths} \label{eternal BH + bath}

\begin{figure}[H]
    \centering
    \includegraphics[width = 1\linewidth]{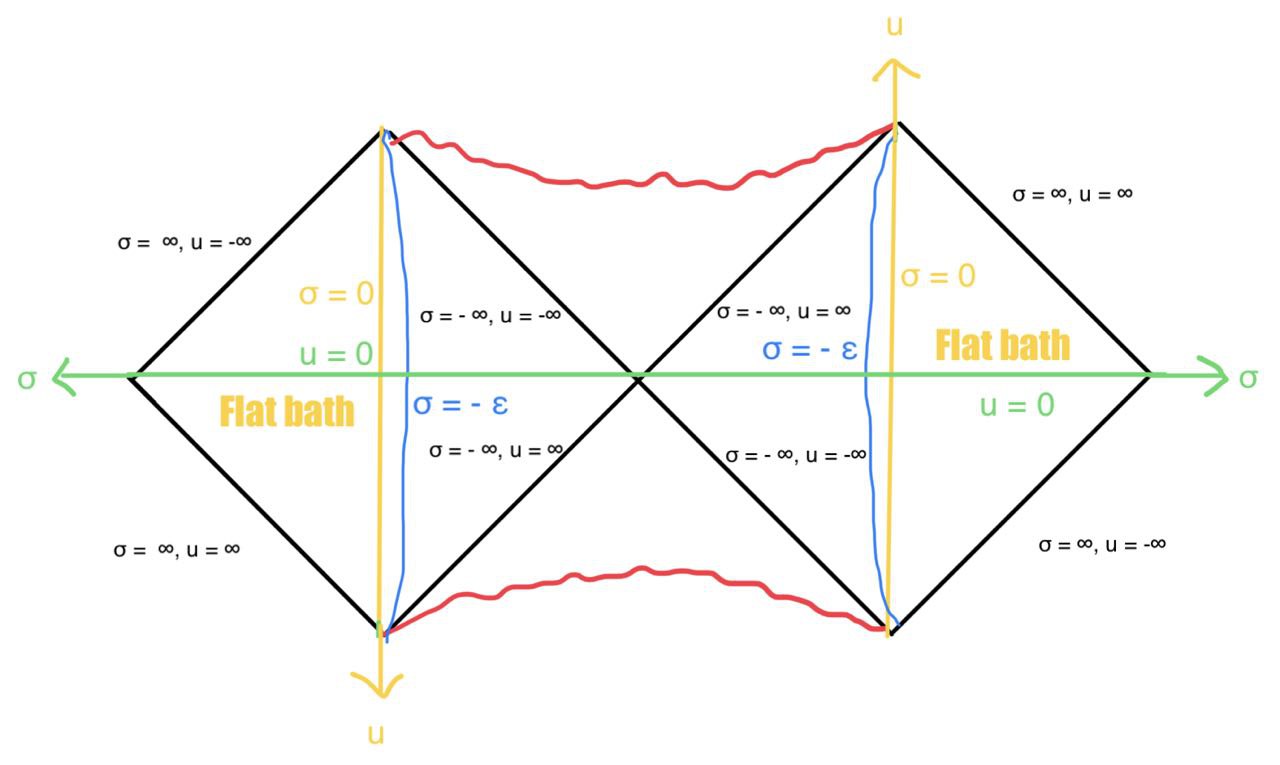}
    \caption{The maximally extended Narayan black hole is joined by a flat baths on both sides. Our choice of Rindler coordinates are also shown in this figure. }
    \label{Pensrose bath figure}
\end{figure}

To set up an example of the black hole information problem, we intend to allow some portion of the Hawking radiation to escape away from but remains entangled with the black hole. We thus glue two copies of Minkowski half space, in which the gravity is totally irrelevant, to the maximally extended black hole in Narayan theory, one copy on each side in the spirit of \cite{almheiri2020replica}, see Figure \ref{Pensrose bath figure}.\\

We couple the Narayan theory to a CFT with a large central charge $c \gg 1$ living in the maximally extended black hole. The action of the Narayan theory coupled to a CFT is 

\begin{equation} \label{Narayan coupled to CFT 1}
    I_{\text{Narayan} + \text{CFT}}[g_{\mu \nu}, \phi, \chi] = I_{\text{Narayan}}[g_{\mu \nu}, \phi] + I_{\text{CFT}}[g_{\mu \nu}, \chi]
 \end{equation}
 where (in Lorentzian signature)

 \begin{equation} \label{Narayan coupled to CFT 2}
     \begin{split}
      I_{\text{Narayan}}[g_{\mu \nu}, \phi] & =  \frac{\phi_0}{16 \pi G_2} \bigg( \int_{M_2} R +2\int_{\partial M_2} K \bigg) \\
 &   +\frac{1}{16 \pi G_2} \bigg[ \int_{M_2} \big(  \phi R + 12 l^{-2} \phi^{-\frac{1}{3}} \big)
 +2\int_{\partial M_2} \phi K - \frac{6}{l} \int_{\partial M_2} \phi^{\frac{2}{3}}\bigg].  
     \end{split}
 \end{equation}
 where $M_2$ denotes the maximally extended black hole in the Narayan theory and, in addition to the renormalized Narayan action (\ref{regularized action}), we add a purely topological term in the first line which provides the extremal entropy of the black hole $S_0 = \frac{\phi_0}{16 \pi G_2}$. We assume the CFT matter does \textbf{NOT} couple to the dilaton $\phi$ in the Narayan theory. We also set the Narayan black hole in our model to be in a global vacuum state, namely the expectation value of the CFT stress-energy tensor is zero throughout, because we can always absorb the trace of the CFT stress-energy tensor $T^{\text{CFT}}_{\mu \nu} = -\frac{2}{\sqrt{-g}}\frac{\delta}{\delta g^{\mu \nu}} I_{\text{CFT}}[g_{\mu \nu}, \chi]$, $T^\mu_{\mu,{\text{CFT}}} = \frac{c}{24\pi} R $, into the extremal value of the dilaton: $\phi_0 \rightarrow \phi_0^{Ren} = \phi_0 + \frac{1}{3} c G_2$ \cite{almheiri2019entropy} . We further couple the above system (\ref{Narayan coupled to CFT 1}) (\ref{Narayan coupled to CFT 2}) to the exactly same CFT living in the baths, i.e. the two Minkowski half spaces, and impose transparent boundary conditions for the CFT at the interface \cite{almheiri2019entropy} \cite{almheiri2020replica}.  We use Poincaré coordinates $\{t, r\}$ (or the null coordinates $x^\pm = t \mp r_*$ whenever at our convenience) for the vacuum Narayan black hole and the natural Minskowskian or Rindler coordinates $\{u, \sigma\}$ (or $y^\pm = u \pm \sigma$) for the flat spaces/baths,
 
\begin{gather}
ds_{\text{in}}^2  =  \frac{l^4}{r^4} \Bigg( 1- \frac{\pi^4 r^4}{\beta^4} \bigg) \bigg[ -dt^2 + {\bigg( 1 - {\frac{\pi^4 r^4}{\beta^4}} \bigg)}^{-2} dr^2 \Bigg] = l^4 \bigg( \frac{1}{r^4} - \frac{\pi^4 }{\beta^4} \bigg) (-dt^2 + dr_*^2),\\
ds_{\text{out}}^2 = \frac{l^4}{\epsilon^4} (-du^2 + d\sigma^2), \quad \epsilon \rightarrow 0.
\end{gather}
Instead of directly gluing the $\sigma = 0$ vertical boundary of half flat space and the $r = 0$ boundary of the maximally extended Narayan black hole together, we choose to glue the two spaces together along the curve called the "physical boundary" near the "absolute" $\sigma = 0$ vertical boundary of half flat space, according to a map  defined on the boundary $\sigma = 0$ from  $ u $ to $f(u)$ : given $f(u)$ on $\sigma = 0$, the physical boundary is identified to be the curve in the Narayan black hole near the $\sigma = 0$ boundary, parameterized in $\{x^+ , x^-\}$ as
\begin{equation} \label{boundary match}
    x^+  = f(u-\epsilon) \approx f(u) - \epsilon f'(u) , \quad   x^- = f(u+\epsilon) \approx f(u) + \epsilon f'(u),
\end{equation}
where the exact form of $f(u)$ needs to be determined by the proper boundary conditions for the metric and the dynamics at the boundary. Admittedly, the gluing map $t = f(u)$ is defined only on the absolute boundary $\sigma = 0$ and the Rindler coordinates $y^\pm$ are defined only in the Minkowski region up until this point, however, we can extend the Rindler coordinates into the Narayan black hole by the following coordinate transformation

\begin{equation} \label{xy transformation}
    x^\pm = f(y^\pm)
\end{equation}
which, in principle, can be done for any gluing maps. Combining the parametric equation of the physical boundary (\ref{boundary match}), the coordinate transformation (\ref{xy transformation}) and the definition of the null Rindler coordinates $y^\pm = u \pm \sigma$ , we see the physical boundary can be exactly identified as the curve $\sigma = -\epsilon$.\\

 The natural boundary conditions for the asymptotically conformally $\text{AdS}_2$ black hole metric (\ref{BH metric in tortoise}) are

\begin{equation} \label{boundary metric 0}
     \quad ds^2_{\text{bd}} \sim \frac{l^4}{\epsilon^4} (-du^2 +d\sigma^2) 
\end{equation}
which matches the flat metric outside the boundary.
Therefore, from the line element on the physical boundary $\sigma = - \epsilon$ and the inverse expansion of (\ref{tortoiseDef}) near the physical boundary
\begin{equation}\label{inverse expansion boundary}
    r = r_* - \frac{\pi ^4 {r_*}^5}{5 \beta^4} + \frac{4 \pi ^8 {r_*}^9}{45 \beta^8} + \dots, 
\end{equation} we have

\begin{equation} \label{boundary metric 01}
    l^4 \bigg(\frac{1}{r^4} - \frac{\pi^4}{\beta^4} \bigg) \Bigg[ - {\bigg(\frac{\partial t}{\partial u} \bigg)}^2 + {\bigg(\frac{\partial r_*}{\partial u} \bigg)}^2  \Bigg] = - \frac{l^4}{\epsilon^4} \longrightarrow r_* \sim \epsilon {\bigg(\frac{\partial t}{\partial u} \bigg)}^{\frac{1}{2}}
\end{equation}
to the leading order in $\epsilon$. Meanwhile, from (\ref{boundary match}) and the definition of null Poincare coordinates $x^\pm = t \mp r_*$ we have at the physical boundary
\begin{equation}\label{boundary legacy}
    t \approx f(u), \quad  r_* \approx \epsilon f'(u) =\epsilon\bigg(\frac{\partial t}{\partial u} \bigg),
\end{equation}
to the leading order in $\epsilon$. Just as in JT gravity,  $t = f(u)$ here also marks the trajectory of the 'boundary particle'. Note that in JT, in principle, $t = f(u)$ can be any real function of $u$ without contradicting the boundary condition $r = \epsilon \partial t/\partial u$ there, as long as it is a solution to the equation of motion dictated by the conservation of ADM energy. However, here in the case of Narayan black hole, by comparing (\ref{boundary metric 01}) and (\ref{boundary legacy}), we see there is a new kinematic (not dynamical) constraint we have to impose on the trajectory of the boundary particle to reconcile (\ref{boundary metric 01}) and (\ref{boundary legacy}), which does not exist in the near $AdS_2$ spacetime, namely 
\begin{equation} \label{constraint tu}
    \frac{\partial t}{\partial u}  = f'(u) \sim 1,
\end{equation}
to the leading order (zeroth order) in $\epsilon$ at the physical boundary $\sigma = -\epsilon$. 
\textbf{At first glance, this 'kinetic constraint' indicates bad choice of gluing the black hole and the baths and may lead to inconsistencies, but we will show it is good for eternal black holes.} As a result, at the physical boundary $\sigma = -\epsilon$, the appropriate boundary condition for the diltaon should be

\begin{equation} \label{dilaton boudnary}
     \phi_{\text{bd}} \sim \frac{l^3}{\epsilon^3}.
\end{equation}
We see dilatons goes to the infinity at the boundary of Narayan black hole which is consistent with freezing gravity in the bath outside the boundary \cite{almheiri2019entropy}\cite{engelsoy2016investigation}.\\

 Now, we would like to derive the correct form of function $f$ for  eternal Narayan black holes, i.e. vacuum Narayan black hole solutions of constant ADM energy, and especially, we want to see if imposing constraint (\ref{constraint tu}) will produce consistent results in our model, so a more rigorous analysis of the boundary dynamics is warranted and presented as follows. \\

For the sake of deriving the ADM energy of our model - a vacuum Narayan black hole joined by two baths, and the equation of motion of the boundary particle which determines the form of gluing function $f$ for eternal Narayan black holes, we need to calculate the on-shell gravitational action of a vacuum Narayan black hole (\ref{Narayan coupled to CFT 2}) and turn it into a boundary integral over boundary time $u$ so that we can extract the ADM energy. As a heads-up for the following calculation, we have to take a closer inspection of the boundary conditions for the metric to keep all the higher-order terms vital to the calculation of relevant boundary quantities, such as the extrinsic curvature and the value of dilaton on the physical boundary.\\

At the physical boundary $\sigma = -\epsilon$, we take the boundary condition for the metric to be exactly that in (\ref{boundary metric 0}) but \textbf{to all orders of $\epsilon$}. Therefore, we have according to (\ref{boundary metric 01})

\begin{equation} \label{pin down r eternal}
  \bigg(\frac{1}{r^4} - \frac{\pi^4}{\beta^4} \bigg) \left[ - {\bigg(\frac{\partial t}{\partial u} \bigg)}^2 + {\bigg(\frac{\partial r_*}{\partial u} \bigg)}^2  \right] \equiv - \frac{1}{\epsilon^4}   
\end{equation}
where $r$ can be expanded in $r_*$ as in (\ref{inverse expansion boundary}) at the physical boundary. We take the following ansatz for the asymptotic expansion of $r_*$ in $\epsilon$  (we denote $\frac{\partial t}{\partial u}$ as $t'$ and so on)
\begin{equation} \label{rstar ansatz}
     r_* ({\sigma = -\epsilon}) =\epsilon  {t'}^{\frac{1}{2}} \left[1+\epsilon  f_1(u) + \epsilon ^2 f_2(u) + \epsilon ^3 f_3(u)+\epsilon ^4 f_4(u) + o(\epsilon^5)\right].
\end{equation}
It then follows that $f_1(u) = 0$ as we plug the above ansatz into (\ref{pin down r eternal}), and due to the kinetic constraint (\ref{constraint tu}) and the relation (\ref{inverse derivative equivalence}), we have
\begin{equation} \label{tu constraint 2ndorder}
    t' = 1 + o(\epsilon^2),
\end{equation}
which then gives $f_2(u) = 0$. Finally, we arrive at the following asymptotic expansion of $r_*$ 

\begin{equation} \label{rstar boundary full}
    r_* ({\sigma = -\epsilon}) = \epsilon  {t'}^{\frac{1}{2}}  \left(1- \frac{\pi ^4  t'^2}{20 \beta ^4} \epsilon ^4 + o(\epsilon^6) \right)
\end{equation}
and of  $r$ in $\epsilon$ 

 \begin{equation} \label{r boundary full}
    r ({\sigma = -\epsilon}) = \epsilon  {t'}^{\frac{1}{2}}  \left(1- \frac{\pi ^4  t'^2}{4 \beta ^4} \epsilon ^4 + o(\epsilon^6) \right)
\end{equation} 

After a lengthy calculation in Appendix \ref{A}, we obtained the extrinsic curvature on any curve where $\sigma$ is constant (\ref{K appendix A})
\begin{equation} \label{K n1 original}
    K = l^{-2} {\left( t'^2 - r_*'^2\right)}^{-\frac{3}{2}} r^{-1} {\bigg(\frac{1}{r^4} - \frac{\pi^4}{\beta^4} \bigg)}^{- \frac{1}{2}} \left( 2t'^3 - 2t'r_*'^2 + t'' r  r_*' - t' r r_*''\right).
\end{equation}
Plugging in (\ref{rstar boundary full}) and (\ref{r boundary full}), we get
the extrinsic curvature at the physical boundary
\begin{equation} \label{K n1  exp}
    K = \epsilon  {t'}^{\frac{1}{2}}  l^{-2} \left[2 + \epsilon ^2 \left(\frac{t''^2}{t'^3}-\frac{t'''}{2 t'^2} \right)+ \epsilon ^4 \left(\frac{\pi ^4 t'^2}{2 \beta ^4}+\frac{21 t''^4}{64 t'^6}-\frac{3 t'''t''^2}{16 t'^5}\right) + o(\epsilon^6)  \right]
\end{equation}
which agrees with the Lorentzian version of the result obtained in equation (38) of \cite{narayan2021aspects} up to an extra term arising from black hole regulation and several opposite signs.\\

We are now ready to compute the on-shell gravitational action of a vacuum Narayan black hole in Lorentzian signature,
\begin{equation} \label{on-shell action n=1 0}
\begin{split}
      I_{\text{Narayan,on-shell}}  & =   \frac{\phi_0}{16 \pi G_2} \bigg( \int_{M_2} R +2\int_{M_2} K \bigg) + \frac{1}{16 \pi G_2} \bigg[ \int_{M_2} \big(  \phi R + 12 l^{-2} \phi^{\frac{1}{3}} \big) +2\int_{\partial M_2} \phi K - \frac{6}{l} \int_{\partial M_2} \phi^{\frac{2}{3}}\bigg]\\
 & =  \frac{1}{16 \pi G_2} \cdot 2 \Biggl\{  \int dt \int_{r({\sigma = -\epsilon})}^{+\infty} dr \; \frac{l^4}{r^4}\Bigg[ \frac{l^3}{r^3}*  \frac{-4 r^2}{l^4}  + 12 l^{-2} {\bigg( \frac{l^3}{r^3} \bigg)}^{\frac{1}{3}}  \Bigg] +2 \int du  \; \frac{l^2 }{\epsilon^2} * \frac{l^3}{{r({\sigma = -\epsilon})}^3} * K \\
& - \frac{6}{l} \int du \; \frac{l^2 }{\epsilon^2}  * \frac{l^2}{{r({\sigma = -\epsilon})}^2} \Biggr\} + \text{topological} \\
& =   \frac{l^3}{8 \pi G_2}  \left[ \int dt 
\,  \left( -\frac{2}{r^4} \Bigg\vert_{r({\sigma = -\epsilon})}^{+\infty} \right) + \frac{2l^2}{\epsilon^2} \int du \, \frac{K}{{r({\sigma = -\epsilon})}^3} - \frac{6}{\epsilon^2} \int du \frac{1}{{r({\sigma = -\epsilon})}^2} \right] +\text{topological} \\
& = \frac{l^3}{8 \pi G_2} \int du \, \left( \frac{2t'}{{r({\sigma = -\epsilon})}^4} + \frac{2 l^2 K}{\epsilon^2 {r({\sigma = -\epsilon})}^3} - \frac{6}{\epsilon^2 {r({\sigma = -\epsilon})}^2}\right) + \text{topological} \\
& =  \frac{l^3}{8 \pi G_2} \int du \, \left[ \frac{1}{\epsilon ^2} \left( \frac{2 t''^2 - t't''' }{ t'^4} \right) + \frac{3 \pi ^4 t'}{\beta ^4}  +\frac{21 t''^4}{32 t'^7} - \frac{3 t''^2 t''' }{8 t'^6} \right] + \text{topological} \\
& =  \frac{l^3}{16 \pi G_2} \int_{\partial M_2} du \, \left(  \frac{3 \pi ^4 t'}{\beta ^4}  +\frac{21 t''^4}{32 t'^7} - \frac{3 t''^2 t''' }{8 t'^6} \right) +\text{topological} 
    \end{split}
\end{equation}
where the overall factor of 2 in the second line accounts for the fact that our black hole has two asymptotic regions and two baths and by the same token, the factor of $\frac{1}{2}$ is restored on the last line since $\partial M_2$ contains both left and right boundaries, we also throw off the $\frac{1}{\epsilon^2}$ divergence in the last line because it is supposed to be formally cancelled by another counterterm \cite{narayan2021aspects}
\begin{equation} \label{epsilon 2 counterterm}
   - \int_{\partial M_2} \left( {\left( \nabla \phi \right)}^2 + \phi \nabla^2 \phi \right) \propto \int \frac{du}{\epsilon ^2} \left( \frac{ t't''' - 2 t''^2 }{ t'^4} \right).
\end{equation}
We see the on-shell gravitational action of a vacuum Narayan black hole is a pure boundary term and thus the ADM energy of a vacuum Narayan black hole of inverse temperature $\beta$ is 
\begin{equation} \label{ADM n=1}
    M(u) = \frac{l^3}{16 \pi G_2} \left(  \frac{3 \pi ^4 t'}{\beta ^4}  +\frac{21 t''^4}{32 t'^7} - \frac{3 t''^2 t''' }{8 t'^6} \right).
\end{equation}
Conservation of ADM energy relates $\frac{dM(u)}{du}$ to the net flux of energy across the physical boundary,
\begin{multline}\label{energy rate of change}
    \frac{dM(u)}{du} =  \frac{l^3}{16 \pi G_2}  \Bigg( \frac{3 \pi ^4}{\beta ^4} - \frac{3 t'''^2}{4 t'^6}-\frac{147 t''^4}{32 t'^8}-\frac{3 t^{(4)} t''}{8 t'^6}+\frac{39 t''' t''^2}{8 t'^7} \Bigg) t'' = T_{y^+ y^+} -T_{y^- y^-}
\end{multline}
This is the equation of motion for the gluing map $f(u) = t(u)$ which, at the same time, must satisfy the kinematic constraint (\ref{constraint tu}) of Narayan black hole joined by flat baths. For eternal black holes, we demand the conservation of its ADM energy
\begin{equation} \label{conservation of ADM}
    \frac{dM(u)}{du} =  T_{y^+ y^+} -T_{y^- y^-} = 0,
\end{equation}
which immediately gives 
\begin{equation}
t'' = 0 \quad \text{or} \quad \frac{3 \pi ^4}{\beta ^4} - \frac{3 t'''^2}{4 t'^6}-\frac{147 t''^4}{32 t'^8}-\frac{3 t^{(4)} t''}{8 t'^6}+\frac{39 t''' t''^2}{8 t'^7} =0.
\end{equation}
Keeping (\ref{tu constraint 2ndorder}) in mind, we can only have $t'' = 0$ because there is no way that derivatives of $t$ that are at least $o(\epsilon^2)$ satisfy $\frac{3 \pi ^4}{\beta ^4} - \frac{3 t'''^2}{4 t'^6}-\frac{147 t''^4}{32 t'^8}-\frac{3 t^{(4)} t''}{8 t'^6}+\frac{39 t''' t''^2}{8 t'^7} =0$. Therefore, at the physical boundary,

\begin{equation} \label{tu quasi-final}
  f(u)  \equiv t(u) = g(\sigma = - \epsilon) u
\end{equation} 
where we set the initial time to zero and $g$ is an undetermined function of $\sigma$ which satisfies $g(\sigma = - \epsilon) \sim 1 + o(\epsilon^2) $. Now that we have obtained the gluing function, we can extend $y^\pm$ from the two Minkowski baths, respectively, to the two asymptotic regions $A_R$ and $A_L$, according to (\ref{tu quasi-final}) at the boundary and (\ref{xy transformation}). We get a simple coordinate transformation in the bulk

\begin{equation} \label{xy transformation exp}
    x^\pm = g(\sigma = - \epsilon) y^\pm,
\end{equation}
which amounts to
\begin{equation}\label{tu quasi-transformation}
    t = g(\sigma = - \epsilon)* u, \quad  r_* = -g(\sigma = - \epsilon) * \sigma.
\end{equation}
To pin down $g(\sigma = - \epsilon)$, we notice that at $\sigma = - \epsilon$, we have according to (\ref{tu quasi-transformation}) and (\ref{rstar boundary full})
\begin{equation}
   r_* ({\sigma = -\epsilon}) = \epsilon * g(\sigma = - \epsilon) = \epsilon  {g(\sigma = - \epsilon)}^{\frac{1}{2}} \left[ 1 +\epsilon ^4 \left(-\frac{\pi ^4 g(\sigma = - \epsilon)^2}{20 \beta ^4} \right) + o(\epsilon^6) \right],
\end{equation}
by self-consistency we have the following expansion of $g(\sigma = - \epsilon)$ in $\epsilon$
\begin{equation} \label{function g}
    g(\sigma = - \epsilon) = 1 - \frac{\pi^4}{10 \beta^4} \, \epsilon^4 + o(\epsilon^8).
\end{equation}
Hence, at the physical boundary $\sigma = -\epsilon$, the asymptotic expansions of $r_*$ and $r$ in the eternal Narayan black hole of inverse temperature $\beta$ are, according to (\ref{rstar boundary full}), (\ref{r boundary full}), (\ref{tu quasi-final}) and (\ref{function g}),
\begin{equation} \label{rstar boundary full eternal}
    r_* ({\sigma = -\epsilon}) = \epsilon  \left( 1 - \frac{\pi ^4 }{10 \beta ^4} \epsilon ^4  + o(\epsilon^8) \right)
\end{equation}
and
\begin{equation} \label{r boundary full eternal}
    r ({\sigma = -\epsilon}) = \epsilon  \left( 1 - \frac{3\pi ^4 }{10 \beta ^4} \epsilon ^4  + o(\epsilon^8) \right).
\end{equation}
Finally, the Rindler coordinates $\{u, \sigma\}$ can be extended into the eternal Narayan black hole as
\begin{equation}\label{tu transformation}
    t = \left(1 - \frac{\pi^4}{10 \beta^4} \, \epsilon^4 + o(\epsilon^8)\right) \, u, \quad  r_* = - \left(1 - \frac{\pi^4}{10 \beta^4} \, \epsilon^4 + o(\epsilon^8)\right) \, \sigma.
\end{equation}

We can now express the metric and the dilaton in an eternal Narayan black hole of inverse temperature $\beta$ in the extended Rindler coordinates,
\begin{gather}
    ds^2_{\text{in}} = l^4 \bigg( \frac{1}{{r}^4} - \frac{\pi^4 }{\beta^4} \bigg) (-dt^2 + dr_*^2) = l^4 \bigg( \frac{1}{{r(\sigma)}^4} - \frac{\pi^4}{\beta^4} \bigg) \left(1 - \frac{\pi^4}{5 \beta^4} \, \epsilon^4 + o(\epsilon^8)\right) (-du^2 + d\sigma^2),\\
    \phi = \frac{l^3}{{r(\sigma)}^3},
\end{gather}
where $r(\sigma)$ is the implicit function of $\sigma$ determined by (\ref{tortoiseDef}),(\ref{tu transformation}) and thus
\begin{equation}\label{r and sigma}
  \left(1 - \frac{\pi^4}{10 \beta^4} \, \epsilon^4 + o(\epsilon^8)\right) \, \sigma = - \frac{\beta}{2 \pi} \biggl[ \arctan{ \Big( \frac{\pi}{\beta}r(\sigma)} \Big) + \arctanh{\Big( \frac{\pi}{\beta}r(\sigma)} \Big) \biggr].
\end{equation}
Our choice of coordinates is depicted in Figure \ref{Pensrose bath figure}.\\

Note that we assume the Narayan black hole coupled to a CFT is in a global vacuum state (Hartle-Hawking state) where $T_{y^\pm y^\pm, \, ds^2_{\text{in}}} \equiv 0$ throughout inside the physical boundary, therefore the CFT stress-energy tensor in an eternal Narayan black hole where the metric were flat, i.e. $ds^2 = \frac{l^4}{\epsilon^4}(-du^2 + d\sigma^2) = \Omega^2 ds^2_{\text{in}}$, would be entirely given by the Weyl anomaly,
\begin{equation} \label{stress-energy would}
    \begin{split}
        T_{y^\pm y^\pm , \, \text{flat}} & = \frac{c}{12 \pi} \Omega^{-1} \partial_{y^\pm}^2 \Omega\\
        & = \frac{c}{12 \pi} {\left[ \frac{1}{\epsilon^2} \left( 1 + \frac{\pi^4}{10 \beta^4} \epsilon^4 +o(\epsilon^8) \right) {\left( \frac{1}{{r(\sigma)}^4} - \frac{\pi^4}{\beta^4} \right)}^{-\frac{1}{2}} \right] }^{-1} \partial_{y^\pm}^2 \left[ \frac{1}{\epsilon^2} \left( 1 + \frac{\pi^4}{10 \beta^4} \epsilon^4 +o(\epsilon^8) \right) {\left( \frac{1}{{r(\sigma)}^4} - \frac{\pi^4}{\beta^4} \right)}^{-\frac{1}{2}} \right]\\
        & = \frac{c}{12 \pi} {\left( \frac{1}{{r \left(\sigma = \frac{1}{2} \left( y^+ - y^-\right) \right)}^4} - \frac{\pi^4}{\beta^4} \right)}^{\frac{1}{2}} \partial_{y^\pm}^2 {\left( \frac{1}{{r \left(\sigma = \frac{1}{2} \left( y^+ - y^-\right) \right)}^4} - \frac{\pi^4}{\beta^4} \right)}^{-\frac{1}{2}}\\
        & = \frac{c}{24 \pi} \cdot \frac{\pi^4 \, {r(\sigma)}^4 + \beta^4 }{\beta^4 \,{r(\sigma)}^2}.
    \end{split}
\end{equation}
At the physical boundary $\sigma = - \epsilon$, according to (\ref{stress-energy would}) and  (\ref{r boundary full eternal}) we have

\begin{equation} \label{stress-energy would bd}
    T_{y^\pm y^\pm , \, \text{flat}} \Bigg\vert_{\sigma = - \epsilon} = \frac{c}{24 \pi \, \epsilon^2} + \frac{ c \, \pi ^3 \, \epsilon ^2}{15 \, \beta ^4} + o(\epsilon^6),
\end{equation}
which, under the transparent boundary conditions imposed on the CFT, means that we have a constant CFT stress-energy tensor  equal to that in (\ref{stress-energy would bd}) across the entire baths,
\begin{equation} \label{stress-energy bath}
    T_{y^\pm y^\pm , \, ds^2_{\text{out}}} =  \frac{c}{24 \pi \, \epsilon^2} + \frac{ c \, \pi ^3 \, \epsilon ^2}{15 \, \beta ^4} + o(\epsilon^6).
\end{equation}
We see from (\ref{stress-energy would bd}) and (\ref{stress-energy bath}) that the flux across the physical boundary is zero, which is consistent with the conservation of ADM energy of an eternal black hole (\ref{conservation of ADM}). Note that the jump of the induced metric $[h_{uu}]$ is zero across the boundary and, since the extrinsic curvature on the outside of the boundary is zero, the jump of extrinsic curvature is $[K_{uu}] = [K] h_{uu}$. As a result, Israel's junction conditions give rise to vanishing total surface stress-energy tensor on the boundary, which adds another layer on the consistency of eternal black hole construction. \\

The ADM energy of an eternal Narayan black hole of inverse temperature $\beta$, according to (\ref{ADM n=1}) and (\ref{tu transformation}), is
\begin{equation} \label{ADM n=1 eternal exp}
    M_{\text{eternal}} = \frac{3 \pi^3 l^3 }{16 G_2 \beta^4}.
\end{equation}\\

\begin{figure}[H]
    \centering
    \includegraphics[width = 0.65\linewidth]{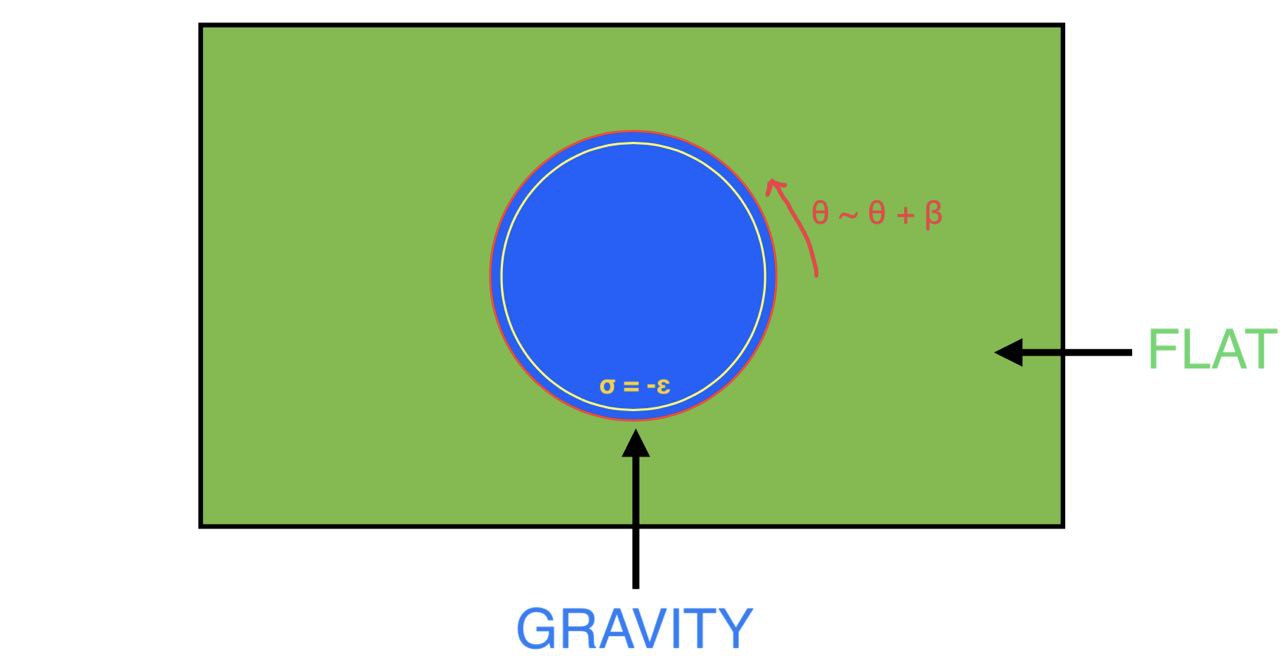}
    \caption{The Euclidean maximally extended Narayan black hole.}
    \label{Euclidean 0}
\end{figure}

 To derive the gravitational entropy of eternal Narayan black holes, we first switch to Euclidean signature and calculate the on-shell action of an Euclidean eternal Narayan black hole of inverse temperature $\beta$. After a Wick rotation $ u \rightarrow  i \tau $, we have
\begin{equation} \label{y Euclidean}
   y = \sigma + i \tau, \quad \bar{y} = \sigma - i \tau.  
\end{equation} 
and the Euclidean black hole by doing analytical continuation, see Figure \ref{Euclidean 0}.

The eternal Narayan black hole metric in Euclidean signature is 

\begin{equation} \label{BH solution E}
ds^2_{\text{in}}  =  \frac{l^4}{r(\sigma)^4} \bigg[ \bigg( 1- \frac{\pi^4 {r(\sigma)}^4}{\beta^4} \bigg) d\theta^2 + {\bigg( 1 - \frac{\pi^4 {r(\sigma)}^4}{\beta^4} \bigg)}^{-1} dr^2 \bigg] = l^4 \bigg( \frac{1}{{r(\sigma)}^4} - \frac{\pi^4}{\beta^4} \bigg) \left(1 - \frac{\pi^4}{5 \beta^4} \, \epsilon^4 + o(\epsilon^8)\right) (d\tau^2 + d\sigma^2).
 \end{equation}
At the physical boundaries $\sigma = -\epsilon$, the tangent and outward-pointing normal vectors are

\begin{equation}
    t^a = \frac{1}{l^2 \sqrt{\frac{1}{r(\sigma = - \epsilon)^4} - \frac{\pi^4}{\beta^4}}} \left(1 + \frac{\pi^4}{10 \beta^4} \, \epsilon^4 + o(\epsilon^8)\right) {\bigg( \frac{\partial}{\partial \tau}\bigg)}^a, \; n^a = \frac{1}{l^2 \sqrt{\frac{1}{r(\sigma = -\epsilon)^4} - \frac{\pi^4}{\beta^4}}} \left(1 + \frac{\pi^4}{10 \beta^4} \, \epsilon^4 + o(\epsilon^8)\right) {\bigg( \frac{\partial}{\partial \sigma}\bigg)}^a.
\end{equation}
The extrinsic curvature with respect to $n^a$ at the physical boundary can now be calculated as follows

\begin{equation}
    \begin{aligned}
   & K = g^{ab} \nabla_a n_b = t^a t^b \nabla_a n_b = \frac{2}{l^2 \, r(\sigma = -\epsilon) \sqrt{\frac{1}{r(\sigma = -\epsilon)^4} - \frac{\pi^4}{\beta^4}}} (1+ o(\epsilon^8))\\
   & =  2 \epsilon l^{-2}  \left(  1 + \frac{ \pi^4 }{5 \beta^4}\epsilon^4 +o(\epsilon^8) \right).
\end{aligned}
\end{equation}
The gravitational part of the on-shell action of our Euclidean eternal black hole of inverse temperature $\beta$ is
\begin{equation} \label{on-shell action n=1}
     \begin{split}
      I_{\text{eternal,on-shell}} & =  - \frac{\phi_0}{16 \pi G_2} \bigg( \int_{M_2} R +2\int_{M_2} K \bigg) - \frac{1}{16 \pi G_2} \bigg[ \int_{M_2} \big(  \phi R + 12 l^{-2} \phi^{\frac{1}{3}} \big) +2\int_{\partial M_2} \phi K - \frac{6}{l} \int_{\partial M_2} \phi^{\frac{2}{3}}\bigg]\\
 & = - \frac{\phi_0}{4 G_2} - \frac{1}{16 \pi G_2} \Biggl\{ \int_0^\beta d\theta \int_{\epsilon  \left( 1 - \frac{3\pi ^4 }{10 \beta ^4} \epsilon ^4  + o(\epsilon^8) \right)}^{\frac{\beta}{\pi}} dr \; \left[  \frac{l^4}{r^4} \left( \frac{l^3}{r^3}*  \frac{-4 r^2}{l^4}  + 12 l^{-2} {\Big( \frac{l^3}{r^3} \Big)}^{\frac{1}{3}}  \right) \right]\\
& +2 \int_0^{\left( 1 + \frac{\pi ^4 }{10 \beta ^4} \epsilon ^4  + o(\epsilon^8) \right) \beta} d\tau  \; \Bigg[\frac{l^2}{\epsilon^2} * \frac{l^3}{{\left[ \epsilon  \left( 1 - \frac{3\pi ^4 }{10 \beta ^4} \epsilon ^4  + o(\epsilon^8) \right) \right]}^3}  * 2  l^{-2} \epsilon \left(  1 + \frac{ \pi^4 }{5 \beta^4}\epsilon^4 +o(\epsilon^8) \right) \Bigg]\\
& - \frac{6}{l} \int_0^{\left( 1 + \frac{\pi ^4 }{10 \beta ^4} \epsilon ^4  + o(\epsilon^8) \right) \beta} d\tau \, \Bigg[ \frac{l^2}{\epsilon^2} *\frac{l^2}{{\left[ \epsilon  \left( 1 - \frac{3\pi ^4 }{10 \beta ^4} \epsilon ^4  + o(\epsilon^8) \right) \right]}^2} \Bigg] \Biggr\}\\
& = - \frac{\phi_0}{4 G_2} - \frac{\beta l^3}{16 \pi G_2} \left[  \left( -\frac{2}{r^4} \Bigg\vert_{\epsilon  \left( 1 - \frac{3\pi ^4 }{10 \beta ^4} \epsilon ^4  \right)}^{\frac{\beta}{\pi}} \right) + \bigg( \frac{4}{\epsilon^4}+\frac{24 \pi^4}{5 \beta^4}\bigg) -  \bigg( \frac{6}{\epsilon^4}+\frac{21 \pi^4}{5 \beta^4}\bigg)+ o(\epsilon^4)\right]
 \end{split}
\end{equation}
\begin{equation*}
\begin{split}
    & = - \frac{\phi_0}{4 G_2}  - \frac{\beta l^3}{16 \pi G_2} \Bigg[ \bigg( \frac{2}{\epsilon^4} + \frac{ 2\pi^4}{5 \beta^4}\bigg) +  \bigg( \frac{4}{\epsilon^4}+\frac{24 \pi^4}{5 \beta^4}\bigg) -  \bigg( \frac{6}{\epsilon^4}+\frac{21  \pi^4}{5 \beta^4}\bigg) + o(\epsilon^4) \Bigg] \\
& = -\frac{1}{4 G_2} \bigg( \phi_0 + \frac{\pi^3 l^3}{4 \beta^3} \bigg), 
\end{split}
   \end{equation*}
where we used the expansion of $r$ in $\sigma = -r_* = -\epsilon$ (\ref{r boundary full eternal}) multiple times in our derivation.
Using the definition of entropy in thermodynamics and the definition of partition function of the gravitational path integral in our case
\begin{equation}
    Z = \int \mathcal{D}g \mathcal{D}\phi \; e^{ - I_{\text{Narayan}}[g_{\mu \nu}, \phi]} Z_{\text{CFT}, g_{\mu \nu}}
\end{equation}

we derive the gravitational entropy of an enternal Narayan black hole of inverse temperature $\beta$
\begin{equation} \label{eternal n=1 grav entropy}
\begin{aligned}
  & S_{\text{grav}} =  (1 - \beta \partial_\beta ) \log Z \\
  & \approx -(1 - \beta \partial_\beta) I_{\text{Narayan}} = \frac{1}{4 G_2} \bigg( \phi_0 + \frac{\pi^3 l^3}{\beta^3}\bigg) = \frac{1}{4 G_2} \bigg( \phi_0 + \phi \big\vert_{\text{horizon}} \bigg),
\end{aligned}
\end{equation}
which matches the normal expectation for the gravitational entropy in a two-dimensional dilaton gravity theory. \\

Meanwhile, we can calculate the gravitational free energy $F_{\text{grav}}$ from (\ref{eternal n=1 grav entropy}) and the ADM energy of the eternal black hole (\ref{ADM n=1 eternal exp})
\begin{equation}
    F_{\text{grav}} = M - TS_{\text{grav}} = \frac{3 \pi^3 l^3 T^4}{16 G_2} - T \cdot \frac{1}{4 G_2} \left( \phi_0 + \pi^3 l^3 T^3\right) = -\frac{1}{4 G_2} \left( \phi_0 T + \frac{1}{4} \pi^3 l^3 T^4  \right) 
\end{equation}and then the gravitational entropy can be recovered 
\begin{equation} \label{n = 1 Bekenstein-Hawking entropy}
   S_{\text{grav}} = - \frac{\partial F_{\text{grav}}}{\partial T} = \frac{1}{4 G_2} \left( \phi_0 +  \pi^3 l^3 T^3  \right) = \frac{1}{4 G_2} \left( \phi_0 + \phi \big\vert_{\text{horizon}} \right)
\end{equation}
 which is exactly (\ref{eternal n=1 grav entropy}).\\

\textbf{We can now conclude that we have established an eternal black hole of inverse temperature $\beta$}.

\newpage

\section{Black hole information problem in Narayan theories }\label{BHIP}
In this section, we set up a version of black hole information problem situated in the eternal Narayan black hole coupled with two baths that we constructed in section \ref{eternal BH + bath}. In subsection 3.1, using (\ref{island formula}), we do the QES calculation in the case of a single interval in eternal black hole in Narayan theory as a warm-up. Then in subsection 3.2, we turn to the case of two intervals in Narayan eternal black hole which exhibits Hawking's information paradox and do the QES calculation again using (\ref{island formula}). Of course, in order to fully justify the results of entropy computations obtained in this section without resorting to holography, we must calculate the nth-Rényi entropy of the region in question and take the $n \rightarrow 1$ limit, which involves the replica trick and the evaluation of the Euclidean gravitational path integral representing the trace of powers of the density matrix. In this section, however, We do not carry out any replica calculation and everything is calculated in Lorentzian signature, but we do plan to explicitly construct replica geometry in section 4 and legitimize the results derived here in subsection 3.1.

\subsection{Single interval in the Narayan eternal black hole}\label{one interval section}
For future convenience, we first develop a coordinate system \{$w^+$, $w^-$\} covering the entirety of the eternal Narayan black hole and the baths, based on the Kruskal coordinates (\ref{Kruskaldef}), (\ref{coordinateAR}) - (\ref{coordinateW}) and the identifications across the physical boundary (\ref{tu transformation}) 
\begin{equation} \label{wDef}
   w^+ = U, \quad w^- = -V, \quad \text{taking }\epsilon \rightarrow 0
\end{equation}
For future reference, we also list the relation between \{$w^+$, $w^-$\} and Rindler coordinates \{$u$ , $\sigma$\} in all parts of Narayan eternal black holes plus the right bath, $B_R$ and the left bath, $B_L$. 

\begin{equation} \label{wDefARBR}
 A_R + B_R:
 \begin{cases}
      w^+ = e^{\frac{2\pi}{\beta}  y^+} = e^{\frac{2\pi}{\beta}  (u + \sigma)},\\
      w^- = e^{\frac{-2\pi}{\beta} y^-} = e^{\frac{-2\pi}{\beta} (u - \sigma)}.
 \end{cases}
\end{equation}

\begin{equation} \label{wDefALBL}
 A_L +B_L:
 \begin{cases}
      w^+ = -e^{\frac{2\pi}{\beta}  y^+} = -e^{\frac{2\pi}{\beta} (u + \sigma)},\\
      w^- = -e^{-\frac{2\pi}{\beta} y^-} = -e^{-\frac{2\pi}{\beta} (u - \sigma)}.
 \end{cases}
\end{equation}

\begin{equation} \label{wDefB}
 BH:
 \begin{cases}
      w^+ = e^{\frac{2\pi}{\beta}  y^+} = e^{\frac{2\pi}{\beta} (u + \sigma)},\\
      w^- = -e^{-\frac{2\pi}{\beta}  y^-} = -e^{-\frac{2\pi}{\beta}  (u - \sigma)}.
 \end{cases}
\end{equation}

\begin{equation} \label{wDefW}
 WH:
 \begin{cases}
      w^+ = -e^{\frac{2\pi}{\beta} y^+} = -e^{\frac{2\pi}{\beta} (u + \sigma)},\\
      w^- = e^{-\frac{2\pi}{\beta} y^-} = e^{-\frac{2\pi}{\beta}  (u - \sigma)}.
 \end{cases}
\end{equation}
In the new coordinates system \{$w^+,w^-$\}, the metric on both sides of the physical boundary can be written as
\begin{equation}
  \begin{aligned}
      & ds^2_{\text{in}} = - l^4 \bigg( \frac{1}{r(\sigma(w^+,w^-))^4} - \frac{\pi^4}{\beta^4}\bigg)  d{y^+} d{y^-} = \frac{\beta^2 l^4}{4 \pi^2} \bigg( \frac{1}{r(\sigma(w^+,w^-))^4} - \frac{\pi^4}{\beta^4}\bigg) \frac{1}{w^+ w^-} d{w^+} d{w^-}\\
      & = \Omega_{\text{in}}^{-2} d{w^+} d{w^-}, \Omega_{\text{in}} = \frac{2 \pi}{\beta l^2} \sqrt{w^+ w^-} {\bigg( \frac{1}{r(\sigma(w^+,w^-))^4} - \frac{\pi^4}{ \beta^4} \bigg)}^{-\frac{1}{2}}
  \end{aligned}  
\end{equation}
and
\begin{equation}
ds^2_{\text{out}} = -  \frac{l^4}{\epsilon^4} d{y^+} d{y^-} =  \frac{\beta^2 l^4}{4 \pi^2 \epsilon^4}  \frac{1}{w^+ w^-} d{w^+} d{w^-} = \Omega_{\text{out}}^{-2} d{w^+} d{w^-}, \ \Omega_{\text{out}} = \frac{2 \pi \epsilon^2}{\beta l^2}  \sqrt{w^+ w^-}.
\end{equation}
The dilaton inside the physical boundary can be expressed as

\begin{equation}
    \phi = \frac{l^3}{{r(\sigma(w^+,w^-))}^3}.
\end{equation}
Note that $r(\sigma(w^+,w^-))$ is a composite function of $w^+$ and $w^-$  by combining the implicit function (\ref{r and sigma}) (now $\epsilon = 0$) and $\sigma = \frac{\beta}{4 \pi}  \log \abs{w^+ w^-}$, namely the inverse function of  
\begin{equation}
    \abs{w^+ w^-} = e^{-2 \biggl[ \arctan{ \Big( \frac{\pi}{\beta} r(\sigma(w^+,w^-))} \Big) + \arctanh{\Big( \frac{\pi}{\beta} r(\sigma(w^+,w^-))} \Big) \biggr]} .
\end{equation}\hfill\break

\begin{figure}[H]
    \centering
    \includegraphics[width = 1\linewidth]{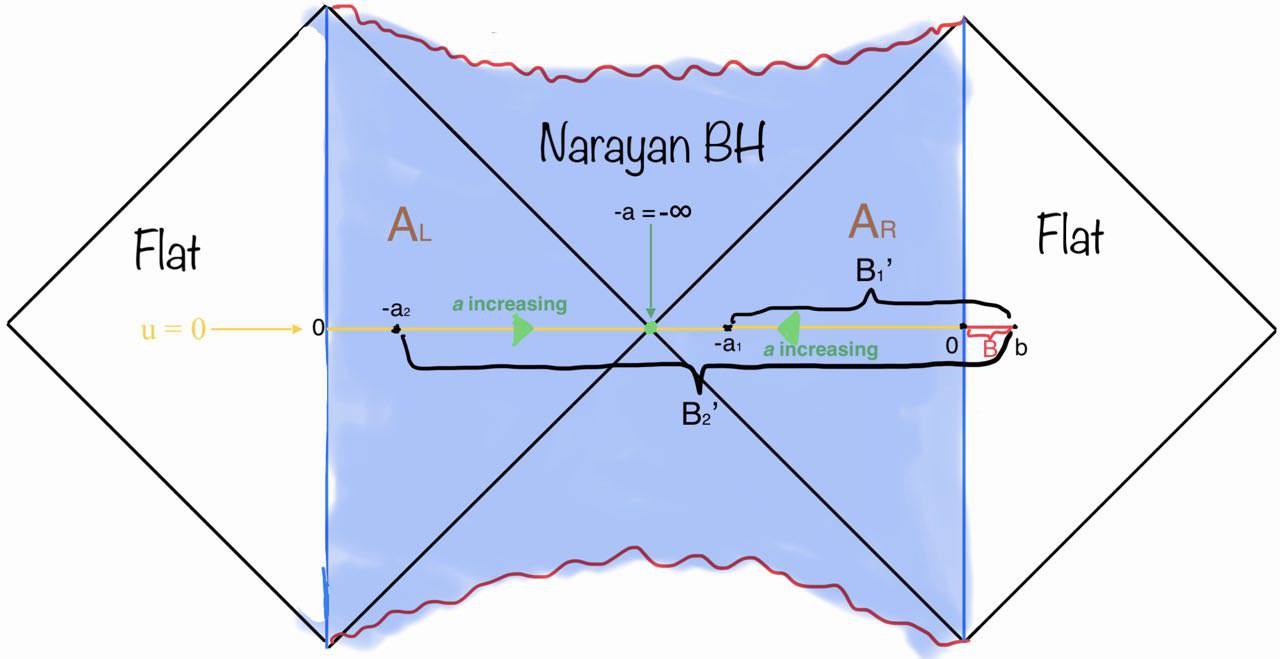}
    \caption{The single interval configuration in Lorentzian signature. We are calculating the true quantum gravity fine-grained entropy of the interval $B = [0 , b]$ (in red). This calculation, however, involves extremizing the generalized entropy of the interval  $B' = [-a , b]$ whose left end point $\sigma = -a$ can move left and right freely on the part of time reflection surface $u = 0$ in the eternal Narayan black hole (painted in blue). Note that, according to our choice of Rindler coordinates, $a \in (0, \infty)$ in both $A_L$ and $A_R$ and $a$ increases in value when it moves closer to the bifurcation surface (throat) where $a = \infty$. Here two possible typical configurations are shown: $B_1' = [-a_1 , b]$ whose left end point lies in $A_R$ and  $B_2' = [-a_2 , b]$ whose left end point lies in $A_L$.}
    \label{single figure}
\end{figure}

We are ready to give the computation of the true quantum gravity fine-grained entropy of the region $B = [0,b]$ in the bath at time $t = u =0$, which is tethered to the boundary of black hole on the left end, see Figure \ref{single figure}. As in JT gravity, the physical meaning of the entropy of the interval $B$ outside the eternal Narayan black hole is the true quantum gravity fine-grained entropy of one side (right side) of the eternal Narayan black hole at $u = 0$. Actually, the calculation involves an interval $B' = [-a,b]$ at $t = u = 0$ sticking into the eternal black hole, with $a,b > 0$, which will be upheld by the analysis in section 4. \\

Note that we assume $c \gg 1$ so that we can ignore the contribution to the total fine-grained entropy and the total stress-energy tensor from gravitons. Therefore, quantum fluctuation from the on-shell solution, i.e. the saddle point in gravitational path integral can thus be ignored. For B', the generalized entropy is the sum of the Bekenstein-Hawking term evaluated at the left end point of $B'$ and the von Neumann entropy of the semi-classical state of the CFT matter on $B'$
\begin{equation} \label{QES1}
    \begin{aligned}
        &  S_{\text{gen}}(B') = S_{\text{grav}}(-a) + S_{vN,\text{CFT}}([-a,b]) \\
        & = \frac{1}{4 G_2} [\phi_0 + \phi (-a)] + \frac{c}{6} \log \Bigg[ \frac{[w^+ (b) - w^+ (-a)][w^- (b) - w^- (-a)]}{\epsilon_{-a,UV} \epsilon_{b,UV} \Omega_{\text{in}} (-a) \Omega_{\text{out}} (b)  } \Bigg]\\
        & = \frac{1}{4 G_2} \Bigg( \phi_0 + \frac{l^3}{{r \Big( \sqrt{w^+ w^-} = e^{-\frac{2 \pi}{\beta} a } \Big)}^3} \Bigg)  + \frac{c}{6} \log \left\{ \frac{ \Big[ e^{\frac{2 \pi \epsilon^2}{\beta} b} \mp e^{-\frac{2 \pi}{\beta} a}\Big] \Big[ e^{\frac{-2 \pi}{\beta} (- b)} \mp e^{\frac{-2 \pi}{\beta} a}\Big]}{\Big( \frac{2 \pi}{\beta l^2} e^{\frac{2 \pi}{\beta} b} \Big) \Big( \frac{2 \pi \epsilon^2}{\beta l^2} e^{- \frac{2 \pi}{\beta} a }\Big) {\Bigg[  \frac{1}{{r \Big( \sqrt{w^+ w^-} = e^{-\frac{2 \pi}{\beta} a } \Big)}^4} - \frac{\pi^4}{\beta^4} \Bigg]}^{-\frac{1}{2}}}\right\}\\
        & = \frac{1}{4 G_2} \Bigg( \phi_0 + \frac{l^3}{{r \Big( \sqrt{w^+ w^-} = e^{-\frac{2 \pi}{\beta} a } \Big)}^3} \Bigg)  + \frac{c}{6} \log  \Big( \cosh{\frac{2 \pi}{\beta} (a + b) \mp 1} \Big) + \frac{c}{12} \log \Bigg[  \frac{1}{{r \Big( \sqrt{w^+ w^-} = e^{-\frac{2 \pi}{\beta} a } \Big)}^4} -\frac{\pi^4}{\beta^4} \Bigg] \\
        & + \frac{c}{6} \log \frac{\beta^2 l^6}{2 \pi^2}, \quad \epsilon = 0,
    \end{aligned}
\end{equation}

where the minus sign in the $\mp$ on the third and the fourth line corresponds to the left end of the interval $\sigma = -a$ lying in the right asymptotic region $A_R$ and the plus sign corresponds to $\sigma = -a$  lying in the left asymptotic region $A_L$. We also absorbed $\epsilon$ and the UV divergence $\epsilon_{-a,UV} $ into the Newton constant $G_2$ and dropped $\epsilon_{b,UV} $ since it is constant and thus does not feed into the extremization of $S_{\text{gen}}(B')$ below.\hfill\break

$S_{\text{gen}}(B')$ possibly can have more than one stationary points on the time reflection Cauchy slice $u = 0$, which is different from the situation in JT. 
The quantum extremal surface (QES) of $S_{\text{gen}}(B')$, by its definition, lies at the stationary point of $S_{\text{gen}}(B')$ where the value of $S_{\text{gen}}(B')$ there is smaller than that at any other stationary point of  $S_{\text{gen}}(B')$.\\

The extremization goes as

\begin{equation} \label{extremality 1}
    \begin{aligned}
        & 0 = \partial_a S_{\text{gen}}(B') \equiv \partial_a S_{\text{gen}}([-a,b]) \\
        & = \frac{3 l^3}{4 G_2 r^4} \frac{dr}{d\sigma} \bigg\vert_{\sigma = -a}  + \frac{c}{6} \cdot \frac{\sinh{\frac{2 \pi}{\beta} (a + b)} \cdot \frac{2\pi}{\beta}}{\cosh{\frac{2 \pi}{\beta} (a + b)} \mp 1} - \frac{c}{12} \cdot \frac{r^4 r_0^4}{r_0^4 - r^4} \cdot \Big( - \frac{4}{r^5}\Big) \cdot \frac{dr}{d\sigma} \bigg\vert_{\sigma = -a}\\
        & = \frac{c}{3} \left( \frac{\pi}{\beta} \cdot  \frac{\sinh{\frac{2 \pi}{\beta} (a + b)}}{ \Big[ \cosh{\frac{2 \pi}{\beta} (a + b) \mp 1}  \Big]} - \frac{1}{r(\sigma = -a)} \right) - \frac{3 \pi^4 l^3}{4 G_2 \beta^4}  \bigg( \frac{\beta^4}{\pi^4 {r(\sigma = -a)}^4} -1 \bigg).
    \end{aligned}
\end{equation}
As a result, the location of the QES, $\sigma = -a$,  implicitly satisfies following condition
\begin{equation}
  \begin{cases} \label{QES1 location}
   \coth{\frac{\pi}{\beta} (a+b)} - \frac{\beta}{\pi r(\sigma = -a)} = \frac{9 \pi^3 l^3}{4 c G_2 \beta^3} \bigg[ {\bigg( \frac{\beta}{\pi r(\sigma = -a)} \bigg)}^4 - 1 \bigg], \; \text{if} \; \sigma = - a \;\text{is in} \, A_R, \\
   \tanh{\frac{\pi}{\beta} (a+b)} - \frac{\beta}{\pi r(\sigma = -a)} = \frac{9 \pi^3 l^3}{4 c G_2 \beta^3} \bigg[ {\bigg( \frac{\beta}{\pi r(\sigma = -a)} \bigg)}^4 - 1 \bigg], \; \text{if} \;\sigma = - a \; \text{is in}  \, A_L.
\end{cases}  
\end{equation}
Alternatively, the above equations for "$a$" can be transformed into equations for $\frac{\beta}{\pi r}$ by use of (\ref{r and sigma})
\begin{equation} \label{QES1 location 2}
\begin{cases}
     \coth{\left( \frac{1}{2} \arctan{\frac{1}{x}}+ \frac{1}{2} \arctanh{\frac{1}{x}}   + \frac{\pi b}{\beta}\right)}  - x  - \frac{9 \pi^3 l^3}{4 c G_2 \beta^3} \left( x^4 -1 \right) = 0,\; \text{if} \;\sigma = - a \; \text{is in} \, A_R,\\
 \tanh{\left( \frac{1}{2} \arctan{\frac{1}{x}}+ \frac{1}{2} \arctanh{\frac{1}{x}}   +\frac{\pi b}{\beta}\right)}  - x  - \frac{9 \pi^3 l^3}{4 c G_2 \beta^3} \left( x^4 -1 \right) = 0,\; \text{if} \;\sigma = - a \; \text{is in} \, A_L,
\end{cases}
   x = \frac{\beta}{\pi r} \in [ 1,+ \infty). 
\end{equation}

\begin{figure}[H]
\centering
\begin{minipage}[t]{.45\textwidth}
    \centering
    \includegraphics[width=1.0\linewidth]
    {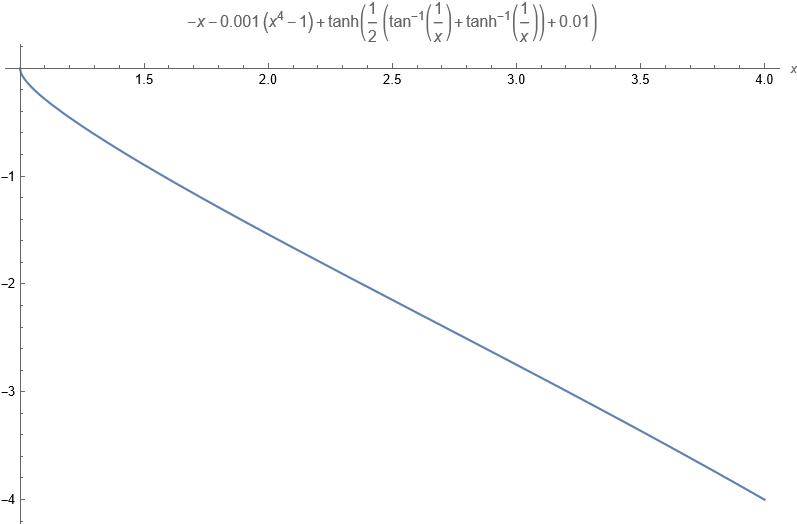}
  \caption{Plotting $\frac{c}{3} \partial_a S_{\text{gen}}(B')$ against $x = \frac{\beta}{\pi r}$ if the left end point of the interval $B'$, $\sigma = - a $, lies in $A_L$ and when $\frac{9 \pi^3 l^3}{4 c G_2 \beta^3}$ takes the very small value of 0.001 and $ \frac{\pi b}{\beta}$ takes the very small value of 0.01. As expected from the range of $\tanh$, there is no stationary point of $S_{\text{gen}}(B')$ in $A_L$. }
  \label{QES1 Tanh}
\end{minipage}\hfill
\begin{minipage}[t]{.45\textwidth}
  \centering
  \includegraphics[width=\linewidth]
  {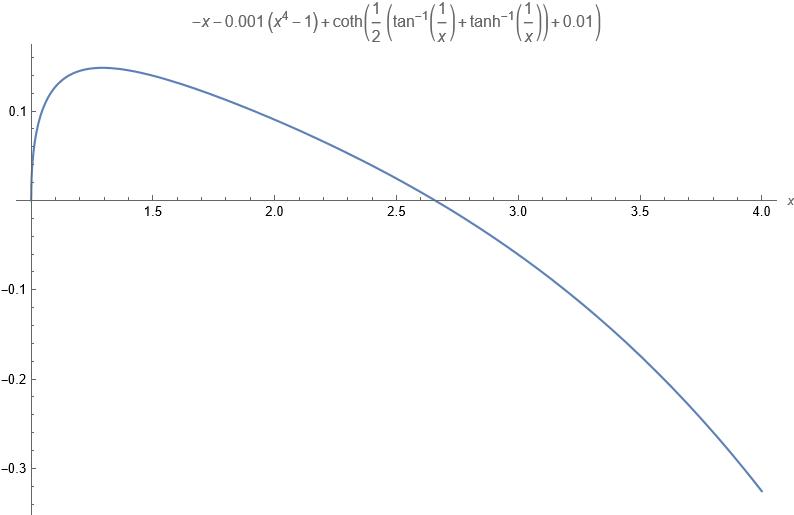}
    \caption{Plotting $\frac{c}{3} \partial_a S_{\text{gen}}(B')$ against $x = \frac{\beta}{\pi r}$ if the left end point of the interval $B'$, $\sigma = - a $, lies in $A_R$ and when $\frac{9 \pi^3 l^3}{4 c G_2 \beta^3}$ takes the very small value of 0.001 and $ \frac{\pi b}{\beta}$ takes the very small value of 0.01. There is another stationary point of $S_{\text{gen}}(B')$ other than the bifurcation surface (throat) at $x = \frac{\beta}{\pi r} \approx 2.6559$, showing a local minimum of $S_{\text{gen}}(B')$ there. Since $\partial_a S_{\text{gen}}(B')$ is always negative when $x>1 \Leftrightarrow a < \infty \Leftrightarrow r>\frac{\beta}{\pi}$ in $A_L$, and, $\partial_a S_{\text{gen}}(B')$ is positive in $A_R$ in the vicinity of the bifurcation surface at $x=1 \Leftrightarrow a = \infty \Leftrightarrow r=\frac{\beta}{\pi}$  as shown in this Figure, we see the bifurcation surface of the event horizon is not an extremum (Figure \ref{single figure} is helpful to see this). As a result, the QES lies at the local minimum $x = \frac{\beta}{\pi r} \approx 2.6559$. }
    \label{QES1 Coth}
\end{minipage}
\end{figure}

\begin{figure}[H]
\centering
\begin{minipage}[t]{.45\textwidth}
    \centering
    \includegraphics[width=1.0\linewidth]
    {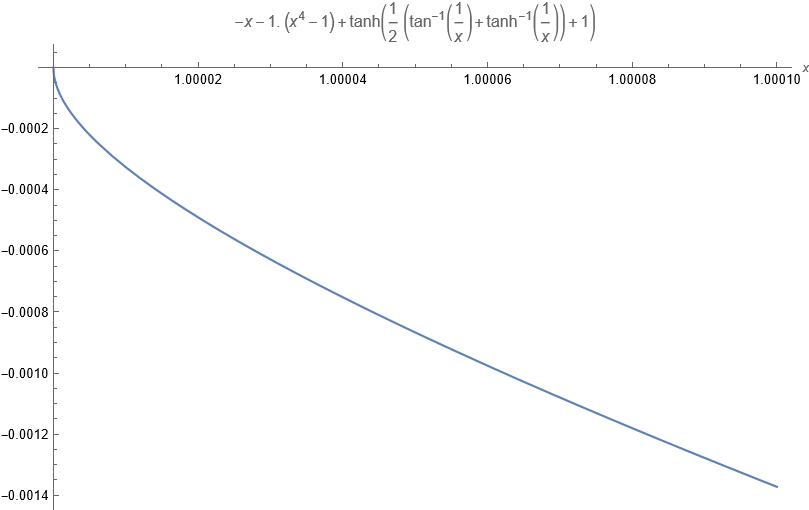}
  \caption{Plotting $\frac{c}{3} \partial_a S_{\text{gen}}(B')$ against $x = \frac{\beta}{\pi r}$ if the left end point of the interval $B'$, $\sigma = - a $, lies in $A_L$ and when $\frac{9 \pi^3 l^3}{4 c G_2 \beta^3}$ and $ \frac{\pi b}{\beta}$ both take the value 1. Still, as expected from the range of $\tanh$, there is no stationary point of $S_{\text{gen}}(B')$ in $A_L$. }
  \label{Normal2}
\end{minipage}\hfill
\begin{minipage}[t]{.45\textwidth}
  \centering
  \includegraphics[width=\linewidth]
  {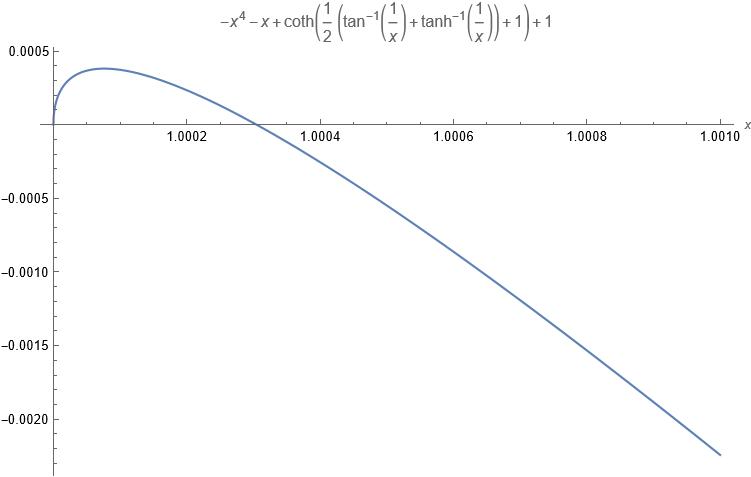}
    \caption{Plotting $\frac{c}{3} \partial_a S_{\text{gen}}(B')$ against $x = \frac{\beta}{\pi r}$ if the left end point of the interval $B'$, $\sigma = - a $, lies in $A_R$ and  when $\frac{9 \pi^3 l^3}{4 c G_2 \beta^3}$ and $ \frac{\pi b}{\beta}$ both take the value 1. There is another stationary point other than the bifurcation surface (throat) at $x = \frac{\beta}{\pi r} \approx 1.0003$, showing a local minimum of $S_{\text{gen}}(B')$ there, which is still very close to the bifurcation surface (throat) of the event horizon. By the same reason mentioned in the caption of Figure \ref{QES1 Coth}, we see the bifurcation surface of the event horizon is not an extremum. As a result, the QES lies at the local minimum $x = \frac{\beta}{\pi r} \approx 1.0003$, though it is very close to the bifurcation surface. }
    \label{Normal}
\end{minipage}
\end{figure}

It is easy to see $x = 1 \Leftrightarrow r = \frac{\beta}{\pi} \Leftrightarrow  a = + \infty$ is a solution to both equations in (\ref{QES1 location 2}), therefore the bifurcation sphere of the event horizon (or throat) is always a stationary point of $S_{\text{gen}}(B')$. Since we require 
\begin{equation}
    \frac{9 \pi^3 l^3}{4 c G_2 \beta^3} = \frac{9l^3}{4 c G_2 r_0^3}>0, \ \frac{\pi b}{\beta} = \frac{b}{r_0}>0
\end{equation}
there is no stationary point of $S_{\text{gen}}(B')$ in the left asymptotic region $A_L$, as expected from the range of the function $\tanh$. From numerical analysis, the right derivative of 
 the function
 \[
 \coth{\left( \frac{1}{2} \arctan{\frac{1}{x}}+ \frac{1}{2} \arctanh{\frac{1}{x}}   + \frac{\pi b}{\beta}\right)}  - x  - \frac{9 \pi^3 l^3}{4 c G_2 \beta^3} \left( x^4 -1 \right)
 \]
 at $x = 1$ is always positive infinite, so there is always another stationary point of $S_{\text{gen}}(B')$ in the right asymptotic region $A_R$ whose location depends on the values of the dimensionless parameters involved, namely $\frac{9 \pi^3 l^3}{4 c G_2 \beta^3} = \frac{9l^3}{4 c G_2 r_0^3}$ and $\frac{\pi b}{\beta} = \frac{b}{r_0}$. Further numerical analysis shows, unless $\frac{9 \pi^3 l^3}{4 c G_2 \beta^3} = \frac{9l^3}{4 c G_2 r_0^3}$ and $\frac{\pi b}{\beta} = \frac{b}{r_0}$ both take very small values ($\ll 1$), the other stationary point is extremely close to the throat $x = 1 \Leftrightarrow r = \frac{\beta}{\pi} \Leftrightarrow  a = + \infty$ (see Figures \ref{QES1 Tanh}, \ref{QES1 Coth} and Figures \ref{Normal2}, \ref{Normal} on the next page), meaning that the QES always exists and  it lies in $A_R$ --- almost exactly at the bifurcation sphere (throat) of the event horizon when at least one of $\frac{9 \pi^3 l^3}{4 c G_2 \beta^3} = \frac{9l^3}{4 c G_2 r_0^3}$ and $\frac{\pi b}{\beta} = \frac{b}{r_0}$ does not take very small value ($\ll 1$).\\
 
 In the general case where at least one of $\frac{9 \pi^3 l^3}{4 c G_2 \beta^3} = \frac{9l^3}{4 c G_2 r_0^3}$ and $\frac{\pi b}{\beta} = \frac{b}{r_0}$ does not take very small value ($\ll 1$),  the QES lies extremely close to the bifurcation surface (throat), we have 
\begin{equation} \label{one-side BH entropy}
     S(B) \approx S_{\text{gen}}([-\infty,b]) = \frac{1}{4 G_2} \left( \phi_0 + \frac{\pi^3 l^3}{\beta^3} \right) + \frac{\pi c}{3 \beta} \cdot b + \frac{c}{3} \log{\frac{\beta l^3}{\pi}} +\frac{\pi c}{24} - \frac{c}{12} \log{2}.
\end{equation}

We see, from (\ref{one-side BH entropy}) and (\ref{n = 1 Bekenstein-Hawking entropy}), that the true entropy of one side of the black hole $S(B)$ is dominated by the Bekenstein-Hawking term associated with classical gravitational entropy of the one-side black hole if $b$ is small ($b \ll 1$). Also, due to the existence of a time-like Killing vector outside the event horizon of eternal Narayan black hole, this one-side black hole configuration is invariant under time translation and the general QES at $u \neq 0$ is related by a time translation. 

\newpage

\subsection{Two intervals in the Narayan eternal black hole}
\begin{figure}[H]
    \centering
    \includegraphics[width = 1\linewidth]{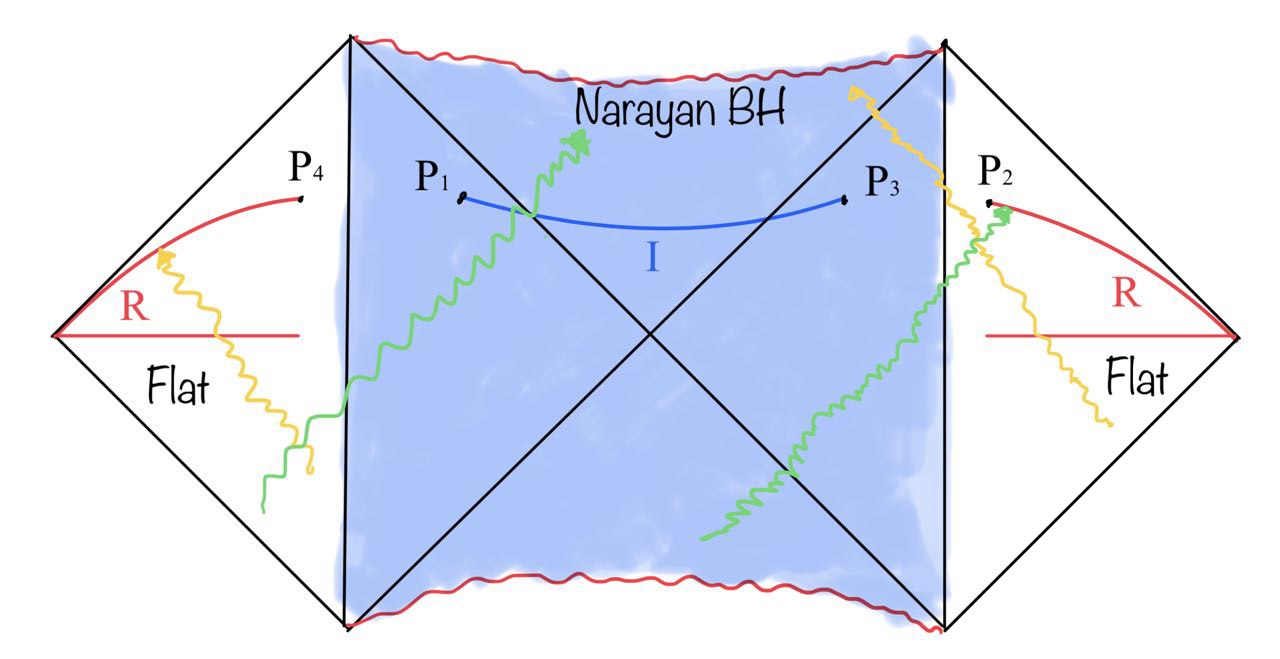}
    \caption{The two interval configuration in Lorentzian signature. Here we visualize two instants in time: $u = 0$ and late time. The two particles of the same  color are entangled as a Hawking pair.} 
    \label{two figure}
\end{figure}

We now move on to formally set up our version of the black hole information paradox in the eternal Narayan black hole configuration constructed in subsection \ref{eternal BH + bath}. For the initial state, we choose a pure thermofield double state for the black hole plus radiation. This initial state can be prepared by an Euclidean gravitational path integral which will be explored in section 4, while the corresponding Lorentzian geometry in this setup is shown in Figure \ref{two figure}. \\

In Figure \ref{two figure}, the Rindler ($u,\sigma$) coordinates of the points are
\begin{equation} \label{P1-P4 coordinates}
    P_1 = (-u_{-a} , -a), \quad P_2 = (u_{b} , b), \quad P_3 = (u_{-a} , -a), \quad P_4 = (-u_{b} , b).
\end{equation}

The radiation region $R$ is
\begin{equation}
R = [P_4 , i^0_L) \cup [P_2 , i^0_R),    
\end{equation}

and the island region $I$ is
\begin{equation}
I = [P_1, P_3].
\end{equation}

The CFT state on the full Cauchy slice is pure, so we have
\begin{equation}
    S_{\text{CFT}} (I \cup R) = S_{\text{CFT}} ([P_4, P_1] \cup [P_3, P_2]).
\end{equation}
which clearly depends on the CFT itself. For simplicity, we assume the CFT matter in our model is $c$ free Dirac fermions \cite{casini2005entanglement}. The entanglement entropy of the two intervals combined ($[P_4, P_1] \cup [P_3, P_2]$) on the metric $ds^2 = \Omega^{-2} d{w^+} d{w^-}$ is
\begin{equation} \label{fermions casini}
    S_{\text{fermions}} = \frac{c}{6} \log \Bigg[ \frac{{\abs{x_{21} x_{32} x_{34} x_{41}}}^2}{\abs{x_{24} {x_{31}}}^2 \Omega_1 \Omega_2 \Omega_3 \Omega_4}\Bigg],
\end{equation}
where all the UV divergences are dropped. \hfill\break

Plugging in all the configuration data, (\ref{fermions casini}) gives the following entanglement entropy in the limit $\epsilon \rightarrow 0$
\begin{multline}\label{sgen_island_1}
   S_{{\text{fermions}}} (I \cup R)   =  \frac{c}{6} \log \left\{ l^8 \bigg[ \bigg( \frac{\beta}{\pi r(\sigma = -a)}^4 \bigg) -1 \bigg] \cross e^{\frac{-4 \pi}{\beta} (a + 2b - u_b)} {\big( e^{\frac{2 \pi}{\beta} (-a + u_{-a})} + e^{\frac{2 \pi}{\beta} (b - u_{b})}\big)}^2 \right. \\
\left. \cross  \frac{{\big( -1 + e^{\frac{2 \pi}{\beta} (a + b + u_{-a} - u_b)} \big)}^2 {\big( -1 + e^{\frac{2 \pi}{\beta} (a + b - u_{-a} + u_b)} \big)}^2 {\big( 1 + e^{\frac{2 \pi}{\beta} (a + b + u_{-a} + u_b)} \big)}^2}{16 {\big( 1 + e^{\frac{4 \pi}{\beta}  u_{-a}} \big)}^2 {\big( 1 + e^{\frac{4 \pi}{\beta}  u_b} \big)}^2 } \right\}.
\end{multline}
The generalized entropy of the radiation region $R$ plus the island region $I$ is thus
\begin{equation} \label{sgen_island_2}
   S_{\text{gen}} (I \cup R)   =  2(\phi_0 + \phi (-a)) + S_{{\text{fermions}}} (I \cup R)   =  2 \bigg( \phi_0 + \frac{l^3}{{r(\sigma = -a)}^3} \bigg) + S_{{\text{fermions}}} (I \cup R) 
\end{equation} \hfill\break

Next, we run an analysis of the generalized entropy of the radiation part in both scenarios where an island is either absent or present.

\paragraph{When island is absent}

We first calculate the generalized entropy in the simpler scenario where there is no island, which corresponds to the Hawking saddle where there is no conical singularities,
 \begin{equation} \label{generalized entropy no island 0}
    \begin{aligned}
        S^{\text{no island}}_{\text{gen}} & = S_{\text{gen}}(R) =  S_{\text{fermions}} (R) =  S_{\text{fermions}}([P_2 , P_4])\\
        & = \frac{c}{6} \log \Bigg[ \frac{[w^+ (P_2) - w^+ (P_4)][w^- (P_2) - w^- (P_4)]}{\epsilon_{P_2,UV} \epsilon_{P_4,UV} \Omega_{\text{out}}  (P_2) \Omega_{\text{out}} (P_4)} \Bigg]\\
        & = \frac{c}{6} \log \Bigg[ \frac{\Big( e^{\frac{2 \pi}{\beta} (u_b + b)} + e^{\frac{2 \pi}{\beta} (-u_b + b)}\Big) \Big( e^{-\frac{2 \pi}{\beta} (u_b - b)} + e^{-\frac{2 \pi}{\beta} (-u_b - b)}\Big)}{{\big( \frac{2 \pi}{\beta l^2} \big)}^2 e^{\frac{4 \pi b}{\beta}}} \Bigg]\\
        & = \frac{c}{3} \log \bigg[ \frac{\beta l^2}{\pi} \cosh{ \bigg( \frac{2 \pi}{\beta} u_b} \bigg) \bigg], \quad \epsilon = 0.
    \end{aligned}
\end{equation} \hfill\break
Obviously, the generalized entropy in this scenario is automatically an extreme value since there is no QES at all to extremalize with. \\

\begin{figure}[H]
    \centering
    \includegraphics[width = 0.75\linewidth]{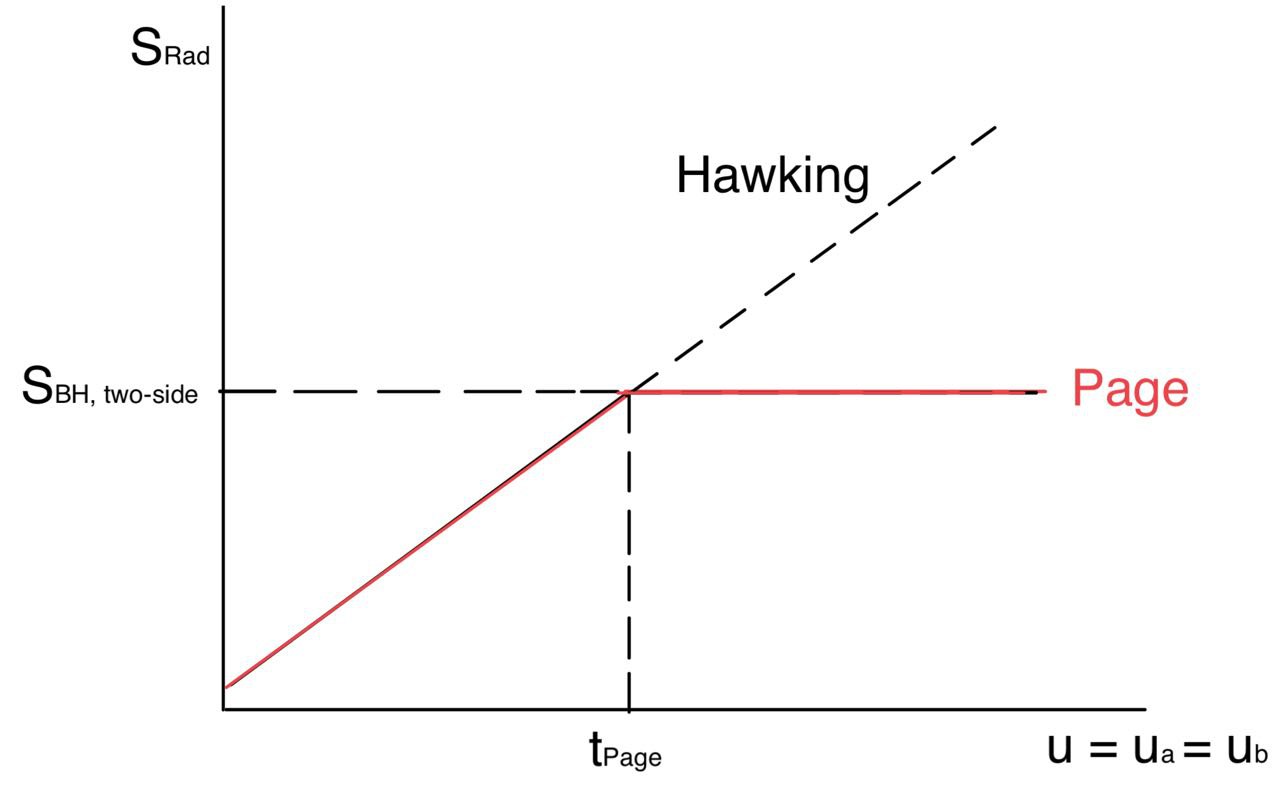}
    \caption{Page curve for the fine-grained entropy of the radiation in our model. The oblique dash line represents the linear growth originally derived by Hawking while the horizontal dash line represents the thermodynamic/coarse-grained entropy of the two-side eternal Narayan black hole. The page curve for our model traces along the minimum of the two. The true fine-grained entanglement entropy of the two-side eternal Narayan black hole also follows the same Page curve.}
    \label{page figure}
\end{figure}

We start to collect Hawking radiation in region $R$ in Figure \ref{two figure} from $u_b  = 0$. The radiation region $R$, as a function of $u_b$, moves upwards in both the baths to the left and right of the black hole. According to (\ref{generalized entropy no island 0}), the generalized entropy of the radiation without an island, now equal to the von Neumann entropy of the matter, grows monotonically with the time $u_b$. Especially, at late times, the generalized entropy of the radiation (\ref{generalized entropy no island 0}) can be approximated as

\begin{equation} \label{QES1 late}
   S^{\text{no island}}_{\text{gen}}  = S_{\text{gen}}(R) \approx \frac{c}{3} \cdot \frac{2 \pi}{\beta} \cdot u_b,  \quad u_b \gg 0,
\end{equation}
showing an linear growth with $u_b$, see Figure \ref{page figure}.  This growth of radiation entropy, matching Hawking's original calculation, can be explained as follows. Initially, the radiation modes on the left and the right are entangled and in pairs. As time moves on, in more and more pairs of entangled modes, only one mode reaches and registers itself in the radiation region $R$ while the other falls into the black hole, therefore the entanglement entropy of the radiation continues to increase. However, if this growth continued forever, the entanglement entropy of the radiation would exceed the thermodynamic/coarse-grained entropy of the two-side eternal Narayan black hole, which is equal to double the Bekenstein-Hawking entropy of the one-side eternal Narayan black hole, causing a contradiction and posing our version of black hole information paradox. It calls for some large non-perturbative corrections to show up as early as around the page time, whose representation in Lorentzian geometry is discussed as follows.

\paragraph{When island is present}

We turn into the more interesting scenario where there is an island corresponding to one of the replica wormhole saddles in Euclidean path integral, see Figure \ref{Replica trick}. According to (\ref{sgen_island_1}) and (\ref{sgen_island_2}), when $u_{-a} = u_b =0$ we have in the limit $\epsilon \rightarrow 0$
\begin{equation}
    \begin{aligned}
        S^{\text{island}}_{\text{gen}} & = S_{\text{gen}}(I \cup R) \\
        & = 2 \bigg( \phi_0 + \frac{l^3}{{r(\sigma = -a)}^3} \bigg) + \frac{c}{6} 
        \log \Biggl\{ \frac{1}{16} \; l^8 \; {\sinh{\frac{2 \pi}{\beta} (a + b)}}^4 \; \bigg[ {\bigg( \frac{\beta}{\pi r(\sigma = -a)} \bigg)}^4 - 1 \bigg] \Biggr\}
    \end{aligned}
\end{equation}
Extremalizing the generalized entropy gives  
\begin{equation}
    \begin{aligned}
         0 & = \partial_a S^{\text{island}}_{\text{gen}} = S_{\text{gen}}(I \cup R) = \partial_a S_{\text{gen}}([P_4, P_1] \cup [P_3, P_2]) \\
        & = \frac{6 l^3}{{r(\sigma = -a)}^4} \frac{dr}{d\sigma} \bigg\vert_{\sigma = -a}  + \frac{4 \pi c}{3 \beta} \coth{\frac{2 \pi}{\beta} (a + b)} + \frac{2 c}{3} \cdot \frac{ {\bigg( \frac{\beta}{\pi r(\sigma = -a)} \bigg)}^4  }{ {\bigg( \frac{\beta}{\pi r(\sigma = -a)} \bigg)}^4 - 1 } \cdot \frac{1}{r(\sigma = -a)}  \frac{dr}{d\sigma} \bigg\vert_{\sigma = -a}\\
        \end{aligned}
\end{equation}
\begin{equation*}
     = \left[ \frac{6 l^3}{{r(\sigma = -a)}^4} + \frac{2 c}{3 r(\sigma = -a)} \cdot \frac{ {\bigg( \frac{\beta}{\pi r(\sigma = -a)} \bigg)}^4  }{ {\bigg( \frac{\beta}{\pi r(\sigma = -a)} \bigg)}^4 - 1 } \right]\bigg( -1 + {\frac{\pi r(\sigma = -a)}{\beta}}^4 \bigg) \bigg\vert_{\sigma = -a} + \frac{4 \pi c}{3 \beta} \coth{\frac{2 \pi}{\beta} (a + b)}.
\end{equation*}
Consequently, the location $\sigma = -a$ of the QES's satisfies the following identity,
\begin{equation} \label{QES20}
     \frac{6}{l} \bigg[ {\bigg( \frac{l}{r(\sigma = -a)} \bigg)}^4 - {\bigg(\frac{\pi l}{\beta} \bigg)}^4 \bigg] + \frac{2 c}{3 r(\sigma = -a)} = \frac{4 \pi c}{3 \beta} \coth{\frac{2 \pi}{\beta} (a + b)}.
\end{equation} \hfill\break
Note again that $r$ and $\sigma$ is related by (\ref{r and sigma}) while $\epsilon = 0$, and that as $a$ goes up from $0$ to $\infty$, $r(\sigma = -a)$ goes up, monotonically, from $0$ to $\frac{\beta}{\pi}$.\hfill\break

There \textbf{always exist} solutions to (\ref{QES20}) at $u_{-a} = u_b =0$. Here is why :  As $a$ goes up from $0$ to $\infty$, the LHS of (\ref{QES20}) decreases monotonically from $\infty$ to $\frac{2 \pi c}{3 \beta}$  and the RHS of (\ref{QES20}) goes monotonically down from $ \frac{4 \pi c}{3 \beta} \coth{\frac{2 \pi}{\beta} b} = \frac{4 \pi c}{3 \beta} \Big( \frac{\beta}{2 \pi b} + \frac{2 \pi b}{3 \beta} + \dots \Big) < \infty$, when $b$ is small but not zero,  to $\frac{4 \pi c}{3 \beta} > \frac{2 \pi c}{3 \beta}$. Therefore, there is at least one value that $a \in [0, \infty)$ evaluates to at which both sides of (\ref{QES20}) are equal. It is natural to infer that the end points of the island are located somewhere very near the event horizon ($a = \infty, r= r_0 = \frac{\beta}{\pi}$). The fact that the solutions $\sigma = -a$ to (\ref{QES20}) at $u_{-a} = u_b =0$  always exist means, even if the replica wormhole saddle is not in dominance so early, there is \textbf{always an island solution in existence} at $u_{-a} = u_b =0$ irrelevant of the value of the parameters in (\ref{QES20}) so long as $b > 0$, which is different from the situation in JT gravity where the existence of the island at $u = 0$ depends on the value of parameters $b$ and the boundary condition for the dilaton $\phi$ \cite{almheiri2020replica}.\\

Meanwhile, at any given time $u_{-a}$ and $u_b$ the extremity condition $\partial_a S^{\text{island}}_{\text{gen}} = 0$ leads to the following identity satisfied by $\sigma = -a$
\begin{multline} \label{QES2late}
    \frac{6}{l} \bigg[ {\bigg( \frac{l}{r(\sigma = -a)} \bigg)}^4 - {\bigg(\frac{\pi l}{\beta} \bigg)}^4 \bigg] + \frac{2 c}{3 r(\sigma = -a)}  = \\
    \frac{ \pi c}{3 \beta} \Big[ \coth{\frac{ \pi}{\beta} (a + b + u_{-a} - u_b)} + \coth{\frac{ \pi}{\beta} (a + b - u_{-a} + u_b)} +\tanh{\frac{ \pi}{\beta} (a + b - u_{-a} - u_b)} +\tanh{\frac{ \pi}{\beta} (a + b + u_{-a} + u_b)}\Big].
\end{multline} \hfill\break

Following the same reasoning as under (\ref{QES20}) for $u_{-a} = u_b =0$, one can easily show there always exists an island at any timelike separated $u_{-a}$ and $u_b$ prior to extremalizing along temporal direction $u$. Therefore for any given $u_b$, in principle, one can locate the islands by Maximin algorithm \cite{wall2014maximin} \cite{akers2020quantum} --- extremalizing the generalized entropy in temporal coordinate $u_{-a}$ after solving (\ref{QES2late}) which extremalizes the generalized entropy in $a$ at every $u_{-a}$. Furthermore, at late times, it is valid that the entanglement entropy can be approximated by doubling the modified single-interval answer, where we restore time dependence in (\ref{QES1}), in the OPE limit, namely

\begin{equation}
\begin{aligned}
   S_{{\text{fermions}}} (I \cup R) & \approx 2S_{{\text{fermions}}} ([P_1 , P_2]]) = \frac{c}{3} \log  \Big[ \cosh{\frac{2 \pi}{\beta} (a + b)} - \cosh{\frac{2 \pi}{\beta} (u_b - u_{-a})} \Big] \\
   & + (\text{terms independent of the temporal coordinate} \ u),  
\end{aligned}
\end{equation}
we see the extremalization in temporal direction further mandates that $u_{-a} = u_b$ at the end points of the island at late times.\hfill\break

According to the QES prescription, the true fine-grained full entropy for the radiation in two-integral configuration is

\begin{equation} \label{island}
    S_{\text{Rad}} = \min \Bigl\{  S^{\text{no island}}_{\text{gen}} \, , \,  S^{\text{island}}_{\text{gen}} \Bigr\}.
\end{equation}
As we have seen in (\ref{QES1 late}), the entropy in the absence of island grows linearly with time at late times, hence island not only always exists but dominates the entropy at late times. This suggests that we should regard the island, which is inside the black hole, as a subsystem of the outgoing Hawking radiation at late times. This amazing new insight leads to the observation that the entanglement across the event horizon of the entangled pairs no longer contributes to the entanglement entropy of outgoing radiation, for the two quanta in each of those entangled pairs now belong to the same system and thus purify each other. As a result, a unitary Page curve for the fine-grained entropy of the radiation is produced as shown in Figure \ref{page figure}. At the same time, because the CFT state on the full Cauchy slice at any given time is pure, the fine-grained matter entropy of the radiation is always equal to that of the two-side black hole at any given time, plus the Bekenstein-Hawking term is always the same in both fine-grained full entropy of the radiation and of the black hole, we conclude the fine-grained full entropy of the two-side eternal Narayan black hole is equal to that of the radiation at any given time, 
\begin{equation}
    S_{\text{BH, two-side}} = S_{\text{Rad}}.
\end{equation}
More specifically, we are essentially extremizing two always equal quantities, namely (see Figure \ref{two figure})

\begin{equation}
    S_{\text{gen,\,BH}}^{\text{trivial}}\equiv S_{\text{gen}}([P_4, P_2])  =  S_{\text{gen}}(R) \equiv S_{\text{gen,\,rad}}^{\text{no island}}
\end{equation}
before the Page time and 
\begin{equation}
S_{\text{gen,\,BH}}^{\text{disconnected}}\equiv 
S_{\text{gen}} ([P_4, P_1] \cup [P_3, P_2]) = S_{\text{gen}}(I \cup R)  \equiv S_{\text{gen,\, rad}}^{\text{island}}
\end{equation}
since the Page time, over the locations of the same set of points (no points to extremize over before the Page time, extremizing over the locations of $P_1$ and $P_3$ since the Page time), according to the QES prescription.\\

Therefore, the fine-grained full entropy of the two-side eternal Narayan black hole follows the same Page curve as the true fine-grained entropy of the radiation does in Figure \ref{page figure}, which shows the eternal Narayan black hole and the radiation remain perfectly are entangled all the time and thus unitarity is preserved. \\

\textbf{Our new version of black hole information problem in the eternal Narayan black holes is thus resolved.}

\newpage

\section{The “Replica geometry" of eternal black holes in Narayan theories}\label{rep}
In this section we try to build the replica geometry of eternal Narayan black holes on the basis of a hypothesis and then attempt to recover the extremity condition of the QES in one-interval case constructed in subsection \ref{one interval section} by taking the limit $n \rightarrow 1$. In subsection 4.1, we first review the replica trick for calculating the true quantum gravity von Neumann entropy of a system coupled to gravity and show that the total action of a gravity theory coupled with matter QFTs equals to the generalized entropy up to a constant term. We then specify more details about applying the replica trick to the theory at issue, that is, the Narayan theory plus a CFT. Then, in subsection 4.2, given the modes of boundary particle, we discuss the "conformal welding problem" of joining the gravitational region and the flat region by finding holomorphic maps of both inside and outside the boundary onto the complex z plane that are compatible at the boundary, following \cite{almheiri2020replica}. We also make a hypothesis of the replica geometry of the covering manifold and derive the replica equation of motion for the boundary mode in the orbifold, generalizing (\ref{energy rate of change}). We then make an attempt to extract the extremity condition for the single interval case from this replica equation of motion. Although this effort proves to be only partially successful in the end, the same method/algorithm deriving the bulk action with conical singularities can be used in more general settings. Furthermore, there is still more interesting stuff worth being discussed in the wake of our failure of recovering the extremity condition in the one-interval case (\ref{QES1 location}), which will be further explored in section 5.

\subsection{The replica trick produces the generalized entropy} \label{general replica formalism}
In this section we review the ideas in \cite{lewkowycz2013generalized} \cite{faulkner2013quantum} \cite{dong2016deriving} \cite{dong2018entropy} for proving the holographic formula for the von Neumann entropy and generalize them to the case the system in question is coupled with gravity. We also show that the generalized entropy appears when we evaluate the $n \sim 1$ total action based on simple assumptions and how it gives the von Neumann entropy of a subregion within a gravitationally non-dynamic region.\\

\begin{figure}[H]
    \centering
    \includegraphics[width = 0.8\linewidth]{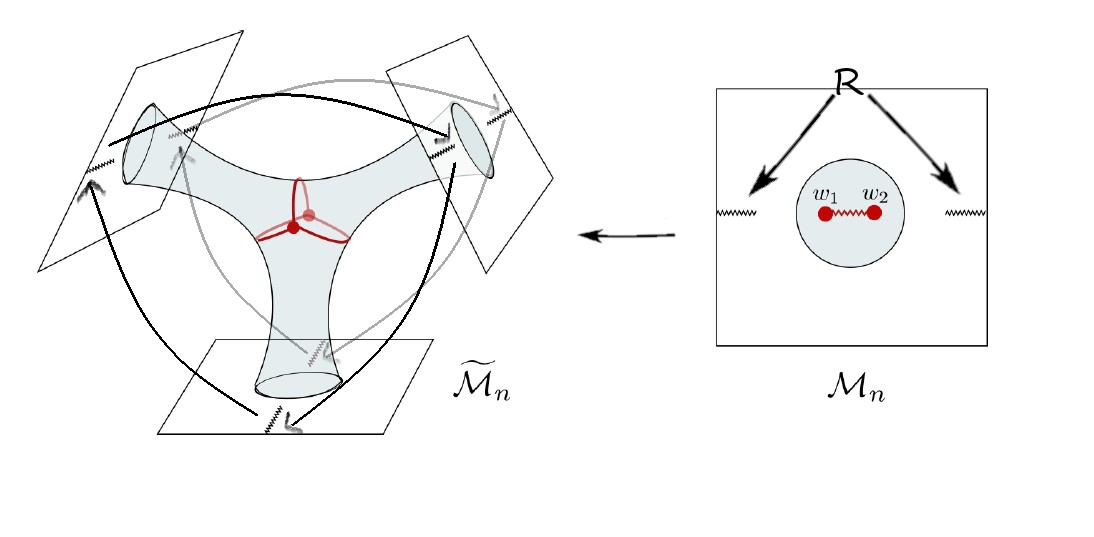}
    \caption{Here are the two representations of the same Euclidean replica geometry. $\mathcal{M}_n$ on the right is the orbifold of the two-interval scenario in the eternal Narayan black hole. The two intervals $R$ denote the radiation region in the two flat baths. We are interested to calculate $R$'s entanglement entropy using replica tricks. The replica/covering manifold $\Tilde{\mathcal{M}}_n$ on the left, here displayed as $\Tilde{\mathcal{M}}_3$, is the product of  cyclically gluing $n$ copies of $\mathcal{M}_n$ along $R$'s. In the center of $\Tilde{\mathcal{M}}_3$, there is a replica wormhole geometry straddling glued $\mathcal{M}_3$', which can also be represented as two fixed points $w_1$ and $w_2$ on the ends of a branch cut in a disk at the center of $\mathcal{M}_3$. This figure is adapted from Figure 9 of \cite{almheiri2020replica}.}
    \label{Replica trick}
\end{figure}

Suppose there is a subregion $R$ inside a region where gravity is non-dynamic and one can associate a density matrix $\rho_R$, which we assume can be expressed as an Euclidean path integral,  with the quantum state of $R$ in full quantum gravity. Then, the von Neumann or fine-grained entropy of the subregion $R$ can be calculated by evaluating and analytically continuating its nth-Renyi entropy to $n \rightarrow 1$
\begin{equation} \label{n-Renyi}
    S[\rho_R] = \lim_{n \rightarrow 1} \frac{1}{1 - n} \text{Tr} \left( {\hat{\rho}_R}^n \right) = \lim_{n \rightarrow 1} \frac{1}{1 - n} \frac{ Z_n}{Z_1^n}= -\frac{\partial}{\partial n} \left( \frac{\log Z_n }{n} \right) \Bigg\vert_{n = 1} = \lim_{n \rightarrow 1} \frac{1}{1-n} \delta \left( \frac{\log Z_n }{n} \right)
\end{equation}
where $\hat{\rho}_R = \frac{{\rho}_R}{{\text{Tr}[\rho_R]}}$ is the normalized state of $R$ and $Z_n = Z[\Tilde{\mathcal{M}}_n] = \text{Tr} \left( \hat{\rho}_R^n \right)$ can be viewed as the partition function of the theory on a topologically non-trivial manifold $\Tilde{\mathcal{M}}_n$ which is the product of cyclically identifying n copies of the original theory along the subregion $R$, see Figure \ref{Replica trick} for a typical example. \\

The covering manifold (this name will be justified shortly) $\Tilde{\mathcal{M}}_n$ involved in the calculation of the nth-Renyi entropy has its geometry totally fixed in the non-gravitational region that the subregion in question - $R$ is a part of. Meanwhile, since it is gravitational path integral involved in the evaluation of $Z_n$, we have to consider any geometry with any topology that obeys the appropriate boundary conditions characterizing the fundamental constructs of our model. Therefore, by assuming the gravitational sector of the theory being characterized the metric $g$ and the dilaton $\phi$ and adding minimally coupled matter now collectively denoted by $\chi$, $Z_n$ can be expressed as 
\begin{equation} \label{Zn1}
\begin{split}
   Z_n & = \int \mathcal{D}g \mathcal{D}\phi \mathcal{D}\chi \; e^{ - I_{\text{total}}[\Tilde{\mathcal{M}}_n; g , \phi, \chi]} = \int \mathcal{D}g \mathcal{D}\phi \mathcal{D}\chi  \; e^{ - \left(I_{\text{grav}}[\Tilde{\mathcal{M}}_n; g , \phi] + I_{\text{mat}}[\Tilde{\mathcal{M}}_n; g , \chi] \right)}\\
   & =  \int \mathcal{D}g \mathcal{D} \phi  \; e^{ -I_{\text{grav}}[\Tilde{\mathcal{M}}_n; g , \phi]} \int \mathcal{D} \chi \; e^{ -I_{\text{mat}}[\Tilde{\mathcal{M}}_n; g , \chi]} = \int \mathcal{D}g \mathcal{D} \phi  \; e^{ -I_{\text{grav}}[\Tilde{\mathcal{M}}_n; g , \phi]} Z_{\text{mat}} [\Tilde{\mathcal{M}}_n; g]\\
   & = \sum_{i, \text{saddles}} e^{ - \left( I_ {\text{grav}}[\Tilde{\mathcal{M}}_n; g_i , \phi_i] -\log Z_{\text{mat}} [\Tilde{\mathcal{M}}_n; g_i] \right)}
\end{split}
\end{equation}
where it is presupposed that the entanglement entropy of matter is large for kinematic reasons such as a CFT with a large central charge $c \gg 1$, and thus the contribution to the total stress-energy tensor from the gravitons can be ignored. Subsequently, we evaluate the integral over geometries on semi-classical saddle points, namely classical solutions of metric and dilaton plus quantum fluctuations of matter QFT on the background of those classical geometries excluding the quantum aspects of gravity brought by graviton fluctuations.\\

\textbf{Unlike in \cite{almheiri2020replica}, we do \textbf{NOT} assume that all the off-shell geometries on $\Tilde{\mathcal{M}}_n$ in the above gravitational path integral necessarily have replica $\mathbb{Z}_n$ symmetry.} In the spirit of \cite{lewkowycz2013generalized}, for on-shell geometries - the saddle points - we assume replica $\mathbb{Z}_n$ symmetry extends into the bulk, arising from the Einstein equations and the $\mathbb{Z}_n$ symmetry of the boundary conditions. Note that there are some off-shell geometries bearing $\mathbb{Z}_n$ symmetry in the gravitational path integral since any geometry satisfying the boundary conditions is included in the gravitational path integral. The $\mathbb{Z}_n$ symmetry allows us to consider everything on the orbifold $\mathcal{M}_n = \Tilde{\mathcal{M}}_n / \mathbb{Z}_n$ when it applies. When there is $\mathbb{Z}_n$ symmetry on $\Tilde{\mathcal{M}}_n$,  there are conical singularities on $\mathcal{M}_n$ with opening angle $2 \pi /n$ - the fixed points of the $\mathbb{Z}_n$ action - because on the covering manifold $\Tilde{\mathcal{M}}_n$ the geometry is smooth at these points. We can add codimensional-two cosmic branes of tension
\begin{equation} \label{cosmic brane tension}
     T_n = \frac{1}{4 G_N} \left( 1 - \frac{1}{n} \right)
\end{equation}
to enforce the conical singularities. This permits us to express the gravitational part of the action by
\begin{equation} \label{cosmic brane}
    \frac{1}{n}  I_ {\text{grav}}[\Tilde{\mathcal{M}}_n] = I_ {\text{grav}}[\mathcal{M}_n] + T_n \int_{\Sigma_{d-2}} \sqrt{\gamma}
\end{equation}
where $\gamma$ is the determinant of the induced transverse metric on the codimensional-2 cosmic brane. Note that in (\ref{cosmic brane}) we integrate the gravitational Lagrangian across the singularity where there is a $\delta$-function for the Ricci curvature scalar and the resulting contribution to the action from the neighborhood around this singularity can generally be evaluated by Gauss-Bonnet theorem. This contribution to the action coming from conical singularity is then cancelled by the cosmic brane term, consistent with the non-singular geometry in the covering manifold $\Tilde{\mathcal{M}}_n$. Also, the position of cosmic branes in the gravitational region is determined by Einstein equations from the on-shell geometry on $\Tilde{\mathcal{M}}_n$. They can not get into the non-gravitational region due to diffeomorphism symmetry. We then insert twist operators $\mathcal{T}_n$ at all the conical singularities in the gravitational region and the endpoints of the subregion $R$ in the non-gravitational region to account for the partition function of the matter sector.\\

We then need to evaluate $\delta \left( \log Z_n/n \right)$ in (\ref{n-Renyi}). When $n = 1$, we have the original solution to the problem on the manifold $\mathcal{M}_1 = \Tilde{\mathcal{M}}_1$ and thus
\begin{equation}\label{logZ1}
   \log Z_1 = - I_ {\text{grav}}[\Tilde{\mathcal{M}}_1] + \log  Z_{\text{mat}}[\Tilde{\mathcal{M}}_1].
\end{equation}
Meanwhile, when $n \sim 1$  we have generalized entropy coming out of the total action on any geometry on $\Tilde{\mathcal{M}}_n$ with replica $\mathbb{Z}_n$ symmetry

\begin{equation} \label{action to sgen}
    \begin{split}
      & I_{\text{total}}[\Tilde{\mathcal{M}}_n] = I_ {\text{grav}}[\Tilde{\mathcal{M}}_n]  - \log  Z_{\text{mat}}[\Tilde{\mathcal{M}}_n] \\
    & = n\left( I_ {\text{grav}}[\mathcal{M}_n] + T_n \int_{\Sigma_{d-2}} \sqrt{\gamma} \right) - (1 - n) \cdot \frac{1}{1-n} \log \frac{Z_{\text{mat}}[\Tilde{\mathcal{M}}_n]}{{Z_{\text{mat}}[\mathcal{M}_n]}^n} - n \log Z_{\text{mat}}[\mathcal{M}_n]\\
    & = n \left( I_ {\text{grav}}[\mathcal{M}_n] - \log Z_{\text{mat}}[\mathcal{M}_n]\right) + nT_n \int_{\Sigma_{d-2}} \sqrt{\gamma} + (n - 1)S_{\text{mat}}[R;\mathcal{M}_n] \\
  & = n \left( I_ {\text{grav}}[\mathcal{M}_1] - \log Z_{\text{mat}}[\mathcal{M}_1]\right) + n\underbrace{\left[ \left( I_ {\text{grav}}[\mathcal{M}_n] - \log Z_{\text{mat}}[\mathcal{M}_n]\right) - \left( I_ {\text{grav}}[\mathcal{M}_1] - \log Z_{\text{mat}}[\mathcal{M}_1]\right) \right]}_{ = 0 \ \text{as} \ \delta I_{\text{total}} \ \text{around the} \ n = 1 \ \text{original solution}} \\
    & + \frac{n - 1}{4 G_N} \sum_{\alpha, \text{Cosmic Branes}} \text{Area} (w_\alpha) + (n - 1)S_{\text{mat}}[R;\mathcal{M}_1] + o {(n-1)}^2 \\
 & =  (n - 1)\cdot \left( \sum_{\alpha, \text{Cosmic Branes}} \text{Area} (w_\alpha) + S_{\text{mat}}[R;\mathcal{M}_1] \right) - n \log Z_1 + o {(n-1)}^2  \\
 & = (n-1) S_{\text{gen}} (w)  - n \log Z_1  + o {(n-1)}^2     
 \end{split}   
\end{equation}

 where we used (\ref{cosmic brane}) on the second line and the replica formula for entanglement entropy of matter in a QFT on a fixed geometry on the third line. Note  also that on the fifth line of (\ref{action to sgen}) we consider the cosmic branes and twist fields as living on the original $n=1$ solution on $\mathcal{M}_1$ since these terms are already of order $(n-1)$. Next, we can express $\log Z_n/n,\ n \sim 1$ as 
\begin{equation} \label{action to sgen2}
    \begin{split}
     \frac{\log Z_n }{n} & = \frac{1}{n} \log \left[ \sum_{i, \text{saddles}} e^{ - \left( I_ {\text{grav}}[\Tilde{\mathcal{M}}_n; g_i , \phi_i] -\log Z_{\text{mat}} [\Tilde{\mathcal{M}}_n; g_i] \right)} \right]  \\
     & = \frac{1}{n} \log \left[ \sum_{i, \text{saddles}} e^{ - \left( I_ {\text{grav}}[\Tilde{\mathcal{M}}_n; g_i , \phi_i] -\log Z_{\text{mat}} [\Tilde{\mathcal{M}}_n; g_i] \right)} + \int_{\substack{\text{off-shell} \; g \; \text{and} \; \phi \\ \text{also with replica} \\ \mathbb{Z}_n \; \text{symmetry}}} \mathcal{D}g \mathcal{D}\phi e^{ - \left( I_ {\text{grav}}[\Tilde{\mathcal{M}}_n] -\log Z_{\text{mat}} [\Tilde{\mathcal{M}}_n] \right)} \right] \\
     & = \frac{1}{n} \log \left[ \sum_{i, \text{saddles}} e^{ -\left[ (n-1) S_{\text{gen}} (w_i)  - n \log Z_1  \right]} + \int_{\substack{\text{off-shell} \; g \; \text{and} \; \phi \\ \text{also with replica} \\ \mathbb{Z}_n \; \text{symmetry}}}  \mathcal{D}g \mathcal{D}\phi e^{ - \left[ (n-1) S_{\text{gen}} (w)  - n\log Z_1  \right]} \right]\\
     & = \frac{1}{n} \log \left( \sum_{i, \text{saddles}} e^{ - \left[ (n-1) S_{\text{gen}} (w_i)  - n\log Z_1  \right]} \right) \  \text{where} \ \partial_{w_{i_\alpha}}S_{\text{gen}} (w_i)=0, \ \forall w_{i_\alpha} \ \text{for the} \ i \text{-th saddle},\\
     & = \frac{1}{n} \log \left( e^{ - \left[ (n-1) S_{\text{gen}} (w)  - n \log Z_1  \right]} \right) = (1-n) S_{\text{gen}} (w) + \log Z_1 + o {(n-1)}^2 
    \end{split}
\end{equation}
where
\begin{equation}
     S_{\text{gen}} (w) = \min_{i, \text{saddles}} S_{\text{gen}} (w_i).
\end{equation}
We see, on the third line of (\ref{action to sgen2}), in the process of extremizing the total action on the covering manifold $\Tilde{\mathcal{M}}_n$, we need to equivalently extremize the generalized entropy over the position of cosmic 
branes, $w$  (we must include the off-shell geometries also with the replica symmetry to continuously extremize over $w$, but the contribution to the logarithm from these off-shell geometries is zero so they only play an auxiliary role here), and finally identify the solution where $S_{\text{gen}}$ reaches its global minimum (when $n \sim 1$, it really means $n > 1$ ) as the leading saddle point. Also, note that on the fourth line of (\ref{action to sgen2}), as already accounted for in (\ref{action to sgen}), one saddle in the gravitational path integral can lead to multiple conical singularities on the orbifold. Finally, combining (\ref{n-Renyi}), (\ref{logZ1}) and (\ref{action to sgen2}), we have
\begin{equation} \label{sgen prescription}
  S[\rho_R] =   \lim_{n \rightarrow 1} \frac{1}{1-n} \delta \left( \frac{\log Z_n }{n} \right) =  S_{\text{gen}} (w),
\end{equation}
which manifestly gives the quantum extremal surface prescription of \cite{ryu2006holographic} \cite{hubeny2007covariant} \cite{engelhardt2015quantum} and substantiates the results obtained in (\ref{extremality 1}) and (\ref{island}) without any help from holography.\\

We proceed to specify the application of (\ref{sgen prescription}) to our theory, an eternal Narayan black hole in the Narayan theory coupled to a CFT matter theory with $c \gg 1$. \\

As specified in subsection 2.3, the same CFT theory lives in the Narayan black hole where the gravity is dynamic and flat baths where the gravity is non-dynamic, we impose transparent boundary conditions at the physical boundary. We thus have the following Euclidean full action on any smooth geometry 
\begin{equation}
   I_ {\text{total}} = - \frac{\phi_0}{16 \pi G_2} \left( \int_{M_2} R +2\int_{\partial M_2} K \right) - \frac{1}{16 \pi G_2} \left[ \int_{M_2} \big(  \phi R + 12 l^{-2} \phi^{\frac{1}{3}} \big) +2\int_{\partial M_2} \phi K - \frac{6}{l} \int_{\partial M_2} \phi^{\frac{2}{3}}\right] + \log Z_{\text{CFT}}[g].
\end{equation}
Instead of working on a smooth geometry on the covering manifold $\Tilde{\mathcal{M}}_n$, as mentioned earlier, we choose to quotient $\Tilde{\mathcal{M}}_n$ by $Z_n$ and work on the orbifold $\mathcal{M}_n = \Tilde{\mathcal{M}}_n / \mathbb{Z}_n$ for our convenience. As a result, the total gravitational action on $\mathcal{M}_n$ contains extra terms involving cosmic branes 

\begin{equation}\label{action cb}
    \begin{split}
      & I_ {\text{grav}}[\mathcal{M}_n] =  \frac{1}{n} I_ {\text{grav}}[\Tilde{\mathcal{M}}_n] \\  
    & = - \frac{\phi_0}{16 \pi G_2} \left( \int_{\mathcal{M}_n} R +2\int_{\partial\mathcal{M}_n} K \right) - \frac{1}{16 \pi G_2} \left[ \int_{\mathcal{M}_n} \big(  \phi R + 12 l^{-2} \phi^{\frac{1}{3}} \big) +2\int_{\partial \mathcal{M}_n} \phi K  - \frac{6}{l} \int_{\partial \mathcal{M}_n} \phi^{\frac{2}{3}}\right] \\& + \frac{1}{4 G_2} \left( 1 - \frac{1}{n} \right) \sum_{\alpha, \text{Cosmic Branes}} \left[ \phi_0 + \phi(w_\alpha) \right].
     \end{split}
\end{equation} 
In some senses, the above action can be regarded as a new gravity theory and we then add n copies of CFT on the orbifold $\mathcal{M}_n$. In addition, we place twist fields at the position of the cosmic branes, $w_\alpha$.\\

We define an interior complex coordinate $w$ so that the metric on $\mathcal{M}_n$ in the gravitational region inside the boundary can be written in conformal gauge 
\begin{equation} \label{metric conformal gauge}
    ds^2_{\text{in}} = e^{2\rho} dw d\Bar{w},  \quad \abs{w} \leq 1
\end{equation}
where the boundary between the gravitational and non-gravitational region lies at $\abs{w} = 1$, $w = e^{i \frac{2\pi}{\beta}\theta} $.
There are conical singularities at certain $w_\alpha$ with an opening angle $2 \pi /n$. By varying the action (\ref{action cb}) with respect to the dilaton, the function $\rho$ in the conformal factor in (\ref{metric conformal gauge}) is subject to the following dilaton equation of motion 
\begin{equation} \label{dilaton EOM}
    - 4 \partial_w \partial_{\Bar{w}} \rho + + 2l^{-2} e^{2\rho} \phi^{-\frac{2}{3}} = 4\pi \left( 1 - \frac{1}{n} \right) \sum_{\alpha}  \delta^{(2)} (w - w_\alpha).
\end{equation}
The above equation is coupled with the Einstein equation derived by varying (\ref{action cb}) with respect the metric
\begin{gather} \label{Einstein Eq}
    \partial_w^2 \phi - 2 \partial_w \phi \partial_w \rho = - 8 \pi G_2 \langle T^{\text{CFT}}_{ww} \rangle ,\\
    \partial_w \partial_{\Bar{w}} \phi - 3l^{-2} \phi^{\frac{1}{3}} e^{2\rho} = 8 \pi G_2 \langle T^{\text{CFT}}_{w \Bar{w}} \rangle,\\
    \partial_{\Bar{w}}^2 \phi - 2 \partial_{\Bar{w}} \phi \partial_{\Bar{w}} \rho = - 8 \pi G_2 \langle T^{\text{CFT}}_{{\Bar{w}}{\Bar{w}}} \rangle.
\end{gather} 
Once we impose the above Einstein equations and dilaton EOM, the cosmic brane action terms in (\ref{action cb}) are cancelled by the contribution from the singular Ricci curvature scalar which is, in turn, caused by the singular part in $\rho$.\\

\begin{figure}[H]
    \centering
    \includegraphics[width = 0.8\linewidth]{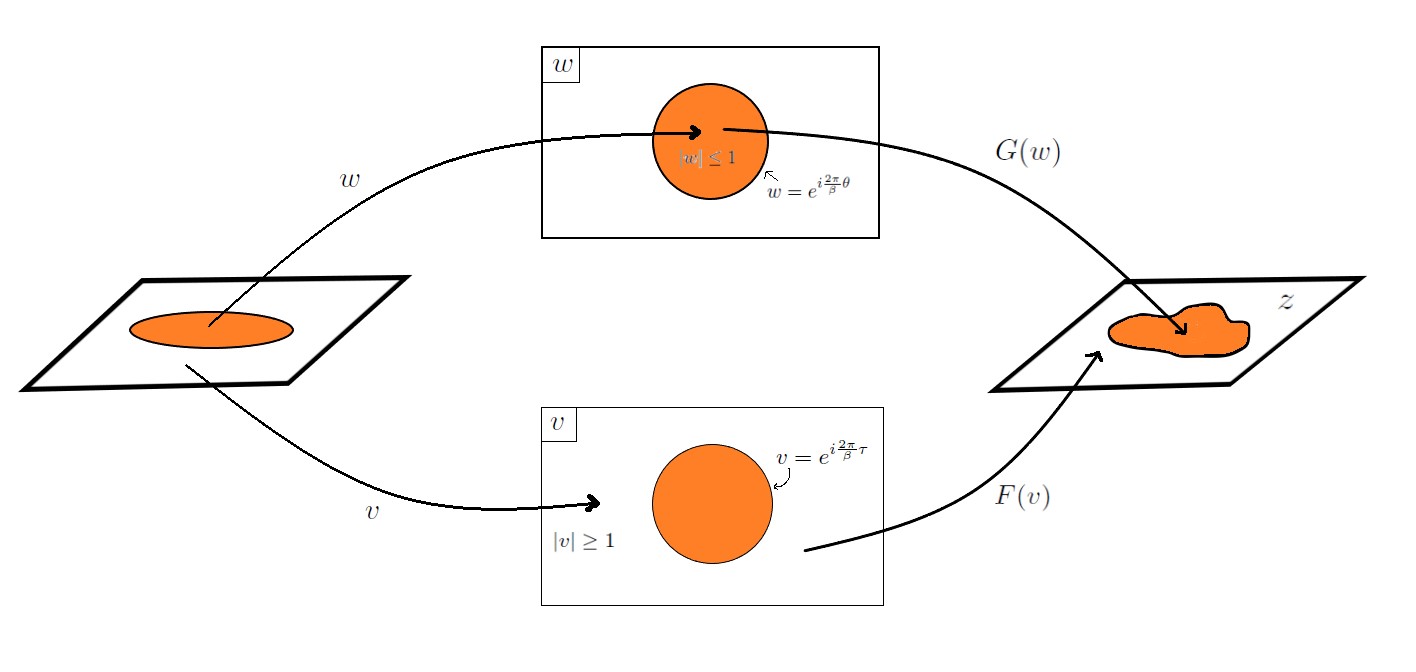}
    \caption{The conformal welding problem involves two disks, one parameterized by $\abs{w} \leq 1$ and another by $\abs{v} \geq 1$ with their boundaries glued according to a given function $\theta(\tau)$ where $w = e^{i \frac{2\pi}{\beta}\theta}$ and $v = e^{i \frac{2\pi}{\beta}\tau}$. What we need to find is holomorphic maps $G$ from the $\abs{w} \leq 1$ disk to the inside of and $F$ from the $\abs{v} \geq 1$ disk to the outside of some region on the complex $z$ plane so that $G$ and $F$ are compatible at the boundary. It is not required that the $G$ and $F$ are holomorphic at the boundary, however.}
    \label{conformal welding figure}
\end{figure}

Now that we have set up the coordinate system in the gravitational region, we need to do the same for the flat space outside the boundary. We consider a finite temperature configuration where $\tau \sim \tau + \beta$. We define the coordinate $v = e^{\frac{2\pi}{\beta} y}$, $y = \sigma + i \tau$ in the flat space such that
\begin{equation}
     ds^2_{\text{out}} = \frac{l^4}{\epsilon^4} dy d\Bar{y}, \quad \abs{v} \geq 1.
\end{equation}
At the boundary $\abs{w} = \abs{v} = 1$, we have $w = e^{i \frac{2\pi}{\beta}\theta (\tau)} $ and $v =e^{i \frac{2\pi}{\beta}\tau}$. Much noteworthy here is that, different from the $n = 1$ original solution where $\theta(\tau) \equiv \tau$ when we take $\epsilon$ to 0 after finishing boundary renormalization, we generally can \textbf{NOT} directly extend the coordinate $v$ on the outside to a holomorphic coordinate inside the boundary. Nevertheless, we can find another coordinate $z$ so that there are holomorphic maps from both the interior $\abs{w} \leq 1$ and the exterior $\abs{v} \geq 1$ to the complex $z$ plane, see Figure \ref{conformal welding figure}. The problem of finding this coordinate map $z$ covering the whole orbifold $\mathcal{M}_n$ can be framed as the “conformal welding problem", following \cite{almheiri2020replica}.  For given data of $\theta (\tau)$ at the boundary, we need to find two functions $G$ and $F$ such that
\begin{equation} \label{conformal welding}
    \begin{split}
       & z = G (w), \quad \text{for} \quad \abs{w} \leq 1\\
       & z = F (v), \quad \text{for} \quad \abs{v} \geq 1\\
       & G(e^{i \frac{2\pi}{\beta}\theta (\tau)}) = F (e^{i \frac{2\pi}{\beta}\tau}), \quad \text{for} \quad \abs{w} = \abs{v} =  1.
    \end{split}
\end{equation}
We require the functions $G$ and $F$ are holomorphic in their respective domains but they do not have to be holomorphic at the boundary.  $G$ and $F$ depend non-locally on $\theta (\tau)$ at the boundary and, respectively, they map the inside disk $\abs{w} \leq 1$ and the outside disk $\abs{v} \geq 1$ to the inside and outside of some irregular region on the complex $z$ plane. Put into the context of replica geometry of Narayan black holes, $\theta(\tau)$ represents the boundary mode in the  Narayan theory, in parallel with that in the nearly-$AdS_2$ gravity \cite{maldacena2016conformal} \cite{engelsoy2016investigation} \cite{jensen2016chaos}. Note again that, in the original $n = 1$ theory developed in subsection \ref{eternal BH + bath}, the boundary mode $\theta(\tau) = \tau$ when $\epsilon = 0$ such that the physical boundary $\sigma = -\epsilon$ is identical with the true boundary $\sigma = 0$ after the action is renormalized. \\

\subsection{A partially successful attempt to construct replica geometry for one-interval case in eternal Narayan black holes} \label{attempt}

\begin{figure}[H]
    \centering
    \includegraphics[width = 1.0\linewidth]{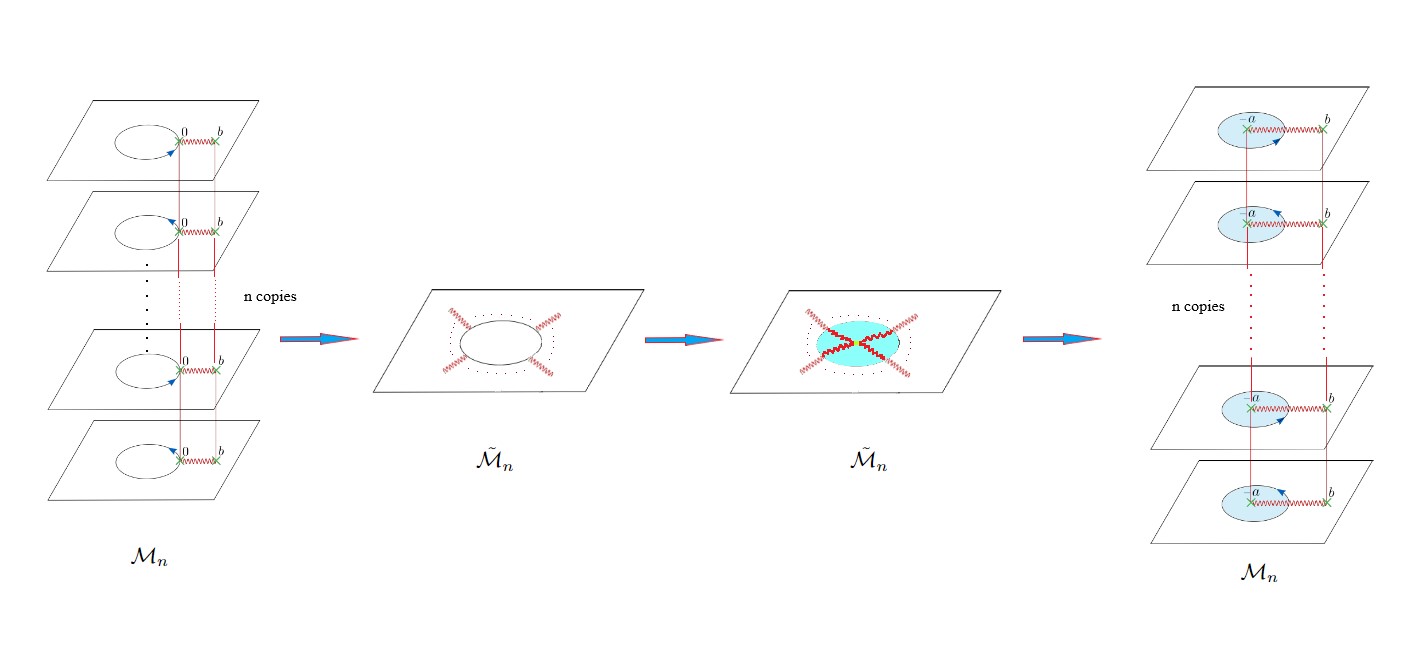}
    \caption{From left to right: We have n copies of $\mathcal{M}_n$. On each copy, there is a flat space field theory outside the boundary shown as a unit circle. In the one-interval scenario, we want to calculate the true fine-grained entropy of an interval in the exterior with one of its end tethered to the boundary. The disk is hollow in $\mathcal{M}_n$ and will be filled in with gravitational configurations subject to the boundary conditions on the unit circle. We then glue together these n copies along the n intervals, each now shown as a branch cut on each $\mathcal{M}_n$. The blue arrows on each  $\mathcal{M}_n$ indicate how to go around the branch cuts. The n boundaries on n copies of $\mathcal{M}_n$ are now connected into a single long circle n times as long as the original one on the covering manifold $\Tilde{\mathcal{M}}_n$. According to the hypothesis we made below (\ref{branch points}), we fill in the same standard Narayan black hole metric of inverse temperature $\beta$ in the interior of this elongated boundary on $\Tilde{\mathcal{M}}_n$ which is perfectly smooth and has a topology of disk. Then, moving to the farthest right, the same replica wormhole geometry on $\Tilde{\mathcal{M}}_n$  we just filled in can also be described by n copies of, again, orbifold $\mathcal{M}_n$. However, the point at the center  of replica geometry on $\Tilde{\mathcal{M}}_n$ is mapped to a branch point inside the boundary on every copy of $\mathcal{M}_n$, which now stands for a conical singularity with an opening angle of $2 \pi/n$. This makes the branch cut on each $\mathcal{M}_n$ extend into the interior, anchoring one of its end point to the conical singularity. This figure is based on figure 11 in \cite{almheiri2020replica}.}
    \label{Mn}
\end{figure}

For a given boundary mode $\theta(\tau)$,  we have already defined the conformal welding problem of matching the inside and outside of the boundary in a compatible way at the boundary. We then need to deal with the problem of deriving the equation of motion for the $\theta(\tau)$ on the orbifold $\mathcal{M}_n$. It is important to notice that the  on-shell gravitational action for $\mathcal{M}_n$ is no longer the Euclidean version of that in (\ref{on-shell action n=1 0}) for the $n=1$ original solution, namely 
\begin{equation} \label{n = 1 Euclidean action}
  - \frac{l^3}{16 \pi G_2} \int_{\partial \mathcal{M}_n} d\tau \, \left( \frac{3 \pi ^4 \theta '(\tau )}{\beta ^4}+\frac{21 \theta ''(\tau )^4}{32 \theta '(\tau )^7}-\frac{3 \theta''' (\tau ) \theta ''(\tau )^2}{8 \theta '(\tau )^6} \right).
\end{equation}
since we need to take into account the conical singularities on $\mathcal{M}_n$. When $n \sim 1$, these conical singularities lead to small deviations in the interior metric from the  $n = 1$ Narayan black hole metric below
\begin{equation} \label{n = 1 Euclidean metric}
   ds^2_{\text{in}, \mathcal{M}_1} = e^{2\rho_0} dw d\Bar{w}= \frac{\beta^2 l^4}{4 \pi^2} \bigg( \frac{1}{r {(\abs{w})}^4} - \frac{\pi^4}{\beta^4}\bigg) \frac{1}{\abs{w}^2} d{w} d{\Bar{w}}
\end{equation}
where we define $w = e^{ \frac{2\pi}{\beta} (\gamma + i \theta)}$ and $y$ was defined in (\ref{y Euclidean}), $r(\abs{w})$ is the inverse function of
\begin{equation} \label{inverse r Euclidean}
    \abs{w} = e^{- \left[ \arctan{  \frac{\pi}{\beta} r(\abs{w})}  + \arctanh{ \frac{\pi}{\beta}r(\abs{w})}  \right]} .
\end{equation}
Therefore, in order to figure out the exact form of the metric perturbation when $n \sim 1$ in the case of a single interval in eternal Narayan black hole studied in section \ref{one interval section}, we need to seek the geometry of the replica wormhole on the nth-covering manifold, $\Tilde{\mathcal{M}}_n$, which in turn is the n-fold cover of the orbifold $\mathcal{M}_n$ where there are branch points at $-a$ and $b$ and a branch cut in between, see Figure \ref{Mn}. Obviously, due to the existence of the branch points and the branch cut, it is more convenient to introduce different coordinates on the inside and outside just like the initial setup in the conformal welding problem in the previous section. Likewise, we define $w = e^{\frac{2\pi}{\beta} (\gamma + i \theta)}, \, \abs{w} < 1$ for the interior and  $v = e^{ \frac{2\pi}{\beta} y}= e^{ \frac{2\pi}{\beta} (\sigma + i \tau)}, \, \abs{v} > 1$ for the exterior, on the basis of (\ref{conformal welding}). We write the coordinates of the branch points as
\begin{equation} \label{branch points}
    w = A = e^{-\frac{2\pi}{\beta} a}, \quad v = B = e^{\frac{2\pi}{\beta} b}.
\end{equation}
following the convention in \cite{almheiri2020replica}. \\

We now make the following important hypothesis, which is in part inspired by $AdS_2$ geometry in JT gravity studied in \cite{almheiri2020replica}, that the leading saddle of the gravitational action on $\Tilde{\mathcal{M}}_n$, $I_ {\text{grav}}[\Tilde{\mathcal{M}}_n]$, is still an Euclidean Narayan black hole of the same temperature $\beta$ which is totally smooth on $\Tilde{\mathcal{M}}_n$ when we fill with it the inside of the boundary on $\Tilde{\mathcal{M}}_n$. In other words, the geometry of the so-called "replica wormhole" in one-interval case is nothing but a Narayan black hole of the temperature $\beta$ but is "n times as large as" the original Narayan black hole solution on $\mathcal{M}_1$. \\

We then use the following map from $\mathcal{M}_n$ to $\Tilde{\mathcal{M}}_n$ to uniformize n copies of the inner disk
\begin{equation} \label{uniformize}
    \Tilde{w} = {\left( \frac{w-A}{1-Aw} \right)}^{1/n}
\end{equation}
which maps the boundary in each copy of $\mathcal{M}_n$ to the elongated boundary in $\Tilde{\mathcal{M}}_n$ and the branch point at $A$ in each copy of $\mathcal{M}_n$ to 0 in the unit disk in $\Tilde{\mathcal{M}}_n$ where the metric is smooth there. We define a new auxiliary coordinate $w_n$ on the unit disk in $\Tilde{\mathcal{M}}_n$
\begin{equation} \label{wn}
   \Tilde{w} = \frac{w_n - A}{1- A w_n}, \quad \text{or} \quad w_n = \frac{\Tilde{w} + A}{1+ A \Tilde{w}} = \frac{{\left( \frac{w-A}{1-Aw} \right)}^{1/n} + A}{1+ A {\left( \frac{w-A}{1-Aw} \right)}^{1/n}},
\end{equation}
we see $w_n$ becomes $w$ when $n$ goes to 1. 
Then, the Euclidean Narayan black hole metric on $\Tilde{\mathcal{M}}_n$, the hypothesized leading saddle point of $I_ {\text{grav}}[\Tilde{\mathcal{M}}_n]$, takes the following form in $\{ \Tilde{w}, \Bar{\Tilde{w}}\}$ coordinates
\begin{equation} \label{metric in covering manifold}
\begin{split}
    ds^2_{\text{in}, \Tilde{\mathcal{M}}_n} & = \frac{\beta^2 l^4}{4 \pi^2} \bigg( \frac{1}{r_n (\abs{w_n})^4} - \frac{\pi^4}{\beta^4}\bigg) \frac{1}{\abs{w_n}^2} d{w_n} d{\Bar{w_n}}\\
    & = \frac{\beta^2 l^4 }{4 \pi^2} \bigg( \frac{1}{r_n \left( \abs{ \frac{\Tilde{w} +A}{1 + A\Tilde{w}}} \right)^4} - \frac{\pi^4}{\beta^4}\bigg) \frac{\left( 1 - A^2 \right)^2}{\abs{\frac{\Tilde{w} + A}{1+ A \Tilde{w}}}^2 \abs{1 + A \Tilde{w}}^4} d{\Tilde{w}} d{\Bar{\Tilde{w}}}
\end{split}
\end{equation}
where, similar to (\ref{inverse r Euclidean}),
\begin{equation}\label{inverse rn Euclidean}
    \abs{w_n} = \abs{\frac{\Tilde{w} + A}{1+ A \Tilde{w}}} = e^{- \left( \arctan{ \frac{\pi}{\beta} r_n}  + \arctanh{ \frac{\pi}{\beta} r_n}  \right)} 
\end{equation}
and $ds^2_{\text{in}, \Tilde{\mathcal{M}}_n} = ds^2_{\text{in}, \mathcal{M}_1}$ when $n = 1$, as expected. \\

Going back onto $\mathcal{M}_n$, from a combination of (\ref{n = 1 Euclidean metric}), (\ref{uniformize}), (\ref{metric in covering manifold}) and the previous definition $w = e^{\frac{2\pi}{\beta} (\gamma + i \theta)}$ we have when $n \sim 1$
\begin{equation} \label{replica metric}
    ds^2_{\text{in}, \mathcal{M}_n} = e^{2\rho} dw d\Bar{w} = e^{2 \delta\rho} ds^2_{\text{in}, \mathcal{M}_1}, \ \rho = \rho_0 + \delta \rho,
\end{equation}
where $\delta \rho$ goes as (we shall always only keep the $o(n-1)$ order for our purpose)
\begin{equation} \label{replica metric boundary}
   \delta \rho = U_1 (\theta) + \gamma^2 U_2 (\theta) + \gamma^4 U_3 (\theta) + o(\gamma^6), \quad \text{as} \ \abs{w} \rightarrow 1 \,(\gamma \rightarrow 0)
\end{equation}
and 
\begin{equation} \label{U1exp}
 U_1(\theta) = (n-1) \left[ 1+\frac{A \,e^{-\frac{2 i \pi  \theta }{\beta }} \left(1 - e^{\frac{4 i \pi  \theta }{\beta }}\right) \log \left(\frac{e^{\frac{2 i \pi  \theta }{\beta }}-A}{1 - A e^{\frac{2 i \pi  \theta }{\beta }}}\right)}{1 - A^2} \right],
 \end{equation}
 \begin{equation}  \label{U2exp}
 \begin{split}
  U_2(\theta) & = \frac{(n-1) A\pi ^2  e^{-\frac{2 i \pi  \theta }{\beta }}}{3 \left(A^2-1\right) \beta ^2 \left(A-e^{\frac{2 i \pi  \theta }{\beta }}\right)^2 \left(-1+A e^{\frac{2 i \pi  \theta }{\beta }}\right)^2} \cross\\
 & \Bigg[ -2 \left(A^2-1\right) e^{\frac{2 i \pi  \theta }{\beta }} \left(-2 A^2 e^{\frac{2 i \pi  \theta }{\beta }} \left(1+e^{\frac{4 i \pi  \theta }{\beta }}\right)+A \left(6 e^{\frac{4 i \pi  \theta }{\beta }}+e^{\frac{8 i \pi  \theta }{\beta }}+1\right)-2 e^{\frac{2 i \pi  \theta }{\beta }} \left(1+e^{\frac{4 i \pi  \theta }{\beta }}\right)\right)\\
 & -2 \left(-1+e^{\frac{4 i \pi  \theta }{\beta }}\right) \left(A^2 \left(-e^{\frac{2 i \pi  \theta }{\beta }}\right)+A e^{\frac{4 i \pi  \theta }{\beta }}+A-e^{\frac{2 i \pi  \theta }{\beta }}\right)^2 \log \left(\frac{A-e^{\frac{2 i \pi  \theta }{\beta }}}{-1+A e^{\frac{2 i \pi  \theta }{\beta }}}\right) \Bigg],
 \end{split}
 \end{equation}
 
 \begin{equation} \label{U3exp}
 U_3(\theta) = \frac{2 (n-1) \pi ^4 \left(A^2-1\right)^2 e^{\frac{4 i \pi  \theta }{\beta }} \left(A^4 e^{\frac{4 i \pi  \theta }{\beta }}+A^2 \left(-12 e^{\frac{4 i \pi  \theta }{\beta }}+5 e^{\frac{8 i \pi  \theta }{\beta }}+5\right)+e^{\frac{4 i \pi  \theta }{\beta }}\right)}{5 \beta ^4 \left(A-e^{\frac{2 i \pi  \theta }{\beta }}\right)^4 \left(-1+A e^{\frac{2 i \pi  \theta }{\beta }}\right)^4}.
 \end{equation}
 Clearly, the metric perturbation functions $U_1$, $U_2$ and $U_3$, are all of order 1 in $(n-1)$ and depend on $A$, the location of the conical singularity in $\mathcal{M}_n$, and thus also on the moduli of the Riemann surface.\\

 The on-shell gravitational action on $\mathcal{M}_n$, after a long derivation in Appendix \ref{B}, is then (\ref{replica action in appendix B})
 \begin{equation} \label{action orbifold}
     \begin{split}
        I_{\text{grav}}[\mathcal{M}_n] & = -\frac{l^3}{16 \pi G_2} \int_{\partial \mathcal{M}_n} d\tau \Bigg[ \left(\frac{3 \pi ^4 \theta '(\tau )}{\beta ^4}+\frac{21 \theta ''(\tau )^4}{32 \theta '(\tau )^7}-\frac{3 \theta''' (\tau ) \theta ''(\tau )^2}{8 \theta '(\tau )^6} \right)\\
        & +\left( \frac{11 \pi ^4 \theta '(\tau ) U_1 (\theta (\tau ))}{3 \beta ^4}+\theta '(\tau ) U_2''(\theta (\tau ))+\frac{5 \theta ''(\tau ) U_2'(\theta (\tau ))}{2 \theta '(\tau )}+\frac{27 \theta ''(\tau )^2 U_2(\theta (\tau ))}{4 \theta '(\tau )^3}-\frac{50}{3} \theta '(\tau ) U_3 (\theta (\tau )) \right.\\
        & -\frac{\theta ''(\tau ) U_1'''(\theta (\tau ))}{6 \theta '(\tau )}-\frac{3 \theta ''(\tau )^2 U_1''(\theta (\tau ))}{8 \theta '(\tau )^3}+\frac{\theta ''(\tau )^3 U_1'(\theta (\tau ))}{\theta '(\tau )^5}+\frac{63 \theta ''(\tau )^4 U_1(\theta (\tau ))}{32 \theta '(\tau )^7}-\frac{3 \theta''' (\tau ) U_2(\theta (\tau ))}{\theta '(\tau )^2}\\
        & \left.-\frac{1}{12} \theta '(\tau ) U_1^{(4)}(\theta (\tau ))-\frac{3 \theta'''(\tau ) \theta ''(\tau ) U_1'(\theta (\tau ))}{4 \theta '(\tau )^4}-\frac{9 \theta'''(\tau ) \theta ''(\tau )^2 U_1 (\theta (\tau ))}{8 \theta '(\tau )^6} \right)\Bigg].
     \end{split}
 \end{equation}
 By varying (\ref{action orbifold}) with respect to $\theta(\tau)$ and following the same variation of the CFT part in Appendix A of \cite{almheiri2020replica} where they did a variation of the exterior metric localized at the boundary in lockstep with the variation of $\theta(\tau)$  maintaining the same boundary condition, we get the following equation of motion for the boundary mode $\theta(\tau)$
 \begin{equation} \label{boundary mode EOM 1}
     \begin{split}
       &  \frac{l^3}{16 \pi  G_2} \left[ \left(\frac{3 \pi ^4 \theta ''(\tau )}{\beta ^4}-\frac{147 \theta ''(\tau )^5}{32 \theta '(\tau )^8}-\frac{3 \theta ^{(4)}(\tau ) \theta ''(\tau )^2}{8 \theta '(\tau )^6}+\frac{39 \theta'''(\tau ) \theta ''(\tau )^3}{8 \theta '(\tau )^7}-\frac{3 \theta'''(\tau )^2 \theta ''(\tau )}{4 \theta '(\tau )^6} \right)\right.\\
       & + \frac{11 \pi ^4 \theta '(\tau )^2 U_1'(\theta (\tau ))}{3 \beta ^4}+\frac{11 \pi ^4 \theta ''(\tau ) U_1(\theta (\tau ))}{3 \beta ^4}+\frac{7}{2} \theta ''(\tau ) U_2''(\theta (\tau ))-\frac{50}{3} \theta '(\tau )^2 U_3'(\theta (\tau ))-\frac{50}{3} \theta ''(\tau ) U_3(\theta (\tau ))\\
       & \frac{17 \theta ''(\tau )^3 U_1''(\theta (\tau ))}{8 \theta '(\tau )^4}-\frac{97 \theta ''(\tau )^4 U_1'(\theta (\tau ))}{32 \theta '(\tau )^6}-\frac{441 \theta ''(\tau )^5 U_1(\theta (\tau ))}{32 \theta '(\tau )^8}+\frac{17 \theta ''(\tau )^2 U_2'(\theta (\tau ))}{4 \theta '(\tau )^2}-\frac{81 \theta ''(\tau )^3 U_2(\theta (\tau ))}{4 \theta '(\tau )^4}\\
       & -\frac{5 \theta ''(\tau )^2 U_1'''(\theta (\tau ))}{24 \theta '(\tau )^2}-\frac{3 \theta'''(\tau ) \theta ''(\tau ) U_1''(\theta (\tau ))}{2 \theta '(\tau )^3}+\theta '(\tau )^2 U_2'''(\theta (\tau ))-\frac{\theta'''(\tau ) U_2'(\theta (\tau ))}{2 \theta '(\tau )}+\frac{39 \theta'''(\tau ) \theta ''(\tau ) U_2(\theta (\tau ))}{2 \theta '(\tau )^3}\\
       &-\frac{\theta'''(\tau )  U_1'''(\theta (\tau ))}{6 \theta '(\tau )}-\frac{3 \theta'''(\tau )^2 U_1'(\theta (\tau ))}{4 \theta '(\tau )^4}+\frac{39 \theta''' (\tau ) \theta ''(\tau )^2 U_1'(\theta (\tau ))}{8 \theta '(\tau )^5}+\frac{117 \theta'''(\tau ) \theta ''(\tau )^3 U_1(\theta (\tau ))}{8 \theta '(\tau )^7}-\frac{1}{4} \theta ''(\tau ) U_1^{(4)}(\theta (\tau ))\\
      & -\frac{3 \theta ^{(4)}(\tau ) \theta ''(\tau ) U_1'(\theta (\tau ))}{4 \theta '(\tau )^4}+\frac{9 \theta'''(\tau )^2 \theta ''(\tau ) U_1(\theta (\tau ))}{4 \theta '(\tau )^6}-\frac{3 \theta ^{(4)}(\tau ) U_2 (\theta (\tau ))}{\theta '(\tau )^2} \left. -\frac{1}{12} \theta '(\tau )^2 U_1^{(5)}(\theta (\tau ))-\frac{9 \theta ^{(4)}(\tau ) \theta ''(\tau )^2 U_1(\theta (\tau ))}{8 \theta '(\tau )^6} \right]\\
      & = i\left[ T_{yy} \left( e^{i \frac{2\pi}{\beta} \tau}\right) -  T_{\Bar{y} \Bar{y}} \left( e^{i \frac{2\pi}{\beta} \tau}\right) \right] = T_{\tau \sigma}.
 \end{split}
\end{equation}
 The physical stress-energy tensor on the RHS of (\ref{boundary mode EOM 1}) can be calculated using the conformal anomaly in the setting of conformal welding problem. Note that there is an ambiguity under $SL(2,C)$ transformation of $z$ for both $G$ and $F$ maps that we can use to map the twist operator at $w=A$ to $z=0$ and the twist operator at $v=B$ to $z=\infty$. So for any original maps $G$ and $F$ to the complex $z$ plane, we can define a new $G'$ and a new $F'$ such that 
 \begin{equation} \label{new GF}
 G'(w) = \frac{G(w) - A}{B - G(w)}, \quad F'(v) = \frac{F(v) - A}{B - F(v)}.
 \end{equation}
 From now on, we assume $G$ and $F$ by default map $w = A$ to $z = 0$ and $v = B$ to $z=\infty$.
 Provided this choice of maps $G$ and $F$, we have on and outside the boundary 
 \begin{equation} \label{T anomaly 1}
     T_{yy} \left(e^{i \frac{2\pi}{\beta}y} \right) = {\left( \frac{dF \left(e^{i \frac{2\pi}{\beta}y} \right)}{dy} \right)}^2 T_{zz}\left( F \left(e^{i \frac{2\pi}{\beta}y} \right) \right) - \frac{c}{24\pi} \biggl\{ F \left(e^{i \frac{2\pi}{\beta}y} \right) , y \biggr\}
 \end{equation}
 and a similar expression for $T_{\Bar{y} \Bar{y}}$.
 Even though the $z$ coordinate covers the entire complex plane in a holomorphic way in the sense of conformal welding problem, see the discussion following (\ref{conformal welding}), there are still twist operators at the branch points $z=0$ and $z=\infty$ and a branch cut between them. So we further define the standard mapping $\Tilde{z} = z^{\frac{1}{n}}$ to remove them. The vacuum expectation value of the stress-energy tensor vanishes on the $\Tilde{z}$ plane due to conformal symmetry, so back on the $z$-plane we have
 \begin{equation}
     T_{zz} (z) = -\frac{c}{24\pi} \{ z^{\frac{1}{n}},z \} = -\frac{c}{48\pi} \left( 1 - \frac{1}{n^2} \right) \frac{1}{z^2}.
 \end{equation}
 By inverting the conformal welding map to the $v-$plane, we finally obtain the expression for $T_{yy}$
 \begin{equation} \label{Tyy}
     T_{yy}(y) = \frac{4 \pi^2}{\beta^2} e^{\frac{4\pi}{\beta}y} \left[ F'(v)^2 T_{zz}\left( F(v) \right) - \frac{c}{24\pi}  \{ F , v\}+ \frac{c}{48\pi} \right]
 \end{equation}
 and a similar one for $T_{\Bar{y} \Bar{y}}$ by taking the complex conjugation of (\ref{Tyy}).\\

 Putting it all together, the equation of motion when $n \sim 1$ is 
 \begin{equation} \label{eom final 0}
     \begin{split}
       &  \frac{3 l^3}{2 c  G_2} \left[ \left(\frac{3 \pi ^4 \theta ''(\tau )}{\beta ^4}-\frac{147 \theta ''(\tau )^5}{32 \theta '(\tau )^8}-\frac{3 \theta ^{(4)}(\tau ) \theta ''(\tau )^2}{8 \theta '(\tau )^6}+\frac{39 \theta'''(\tau ) \theta ''(\tau )^3}{8 \theta '(\tau )^7}-\frac{3 \theta'''(\tau )^2 \theta ''(\tau )}{4 \theta '(\tau )^6} \right)\right.\\
       & + \frac{11 \pi ^4 \theta '(\tau )^2 U_1'(\theta (\tau ))}{3 \beta ^4}+\frac{11 \pi ^4 \theta ''(\tau ) U_1(\theta (\tau ))}{3 \beta ^4}+\frac{7}{2} \theta ''(\tau ) U_2''(\theta (\tau ))-\frac{50}{3} \theta '(\tau )^2 U_3'(\theta (\tau ))-\frac{50}{3} \theta ''(\tau ) U_3(\theta (\tau ))\\
       & \frac{17 \theta ''(\tau )^3 U_1''(\theta (\tau ))}{8 \theta '(\tau )^4}-\frac{97 \theta ''(\tau )^4 U_1'(\theta (\tau ))}{32 \theta '(\tau )^6}-\frac{441 \theta ''(\tau )^5 U_1(\theta (\tau ))}{32 \theta '(\tau )^8}+\frac{17 \theta ''(\tau )^2 U_2'(\theta (\tau ))}{4 \theta '(\tau )^2}-\frac{81 \theta ''(\tau )^3 U_2(\theta (\tau ))}{4 \theta '(\tau )^4}\\
       & -\frac{5 \theta ''(\tau )^2 U_1'''(\theta (\tau ))}{24 \theta '(\tau )^2}-\frac{3 \theta'''(\tau ) \theta ''(\tau ) U_1''(\theta (\tau ))}{2 \theta '(\tau )^3}+\theta '(\tau )^2 U_2'''(\theta (\tau ))-\frac{\theta'''(\tau ) U_2'(\theta (\tau ))}{2 \theta '(\tau )}+\frac{39 \theta'''(\tau ) \theta ''(\tau ) U_2(\theta (\tau ))}{2 \theta '(\tau )^3}\\
       &-\frac{\theta'''(\tau )  U_1'''(\theta (\tau ))}{6 \theta '(\tau )}-\frac{3 \theta'''(\tau )^2 U_1'(\theta (\tau ))}{4 \theta '(\tau )^4}+\frac{39 \theta''' (\tau ) \theta ''(\tau )^2 U_1'(\theta (\tau ))}{8 \theta '(\tau )^5}+\frac{117 \theta'''(\tau ) \theta ''(\tau )^3 U_1(\theta (\tau ))}{8 \theta '(\tau )^7}\\
      \end{split}
 \end{equation}
 \begin{equation*}
     \begin{split}
          & -\frac{1}{4} \theta ''(\tau ) U_1^{(4)}(\theta (\tau ))-\frac{3 \theta ^{(4)}(\tau ) \theta ''(\tau ) U_1'(\theta (\tau ))}{4 \theta '(\tau )^4}+\frac{9 \theta'''(\tau )^2 \theta ''(\tau ) U_1(\theta (\tau ))}{4 \theta '(\tau )^6}-\frac{3 \theta ^{(4)}(\tau ) U_2 (\theta (\tau ))}{\theta '(\tau )^2}\\
       & \left. -\frac{1}{12} \theta '(\tau )^2 U_1^{(5)}(\theta (\tau ))-\frac{9 \theta ^{(4)}(\tau ) \theta ''(\tau )^2 U_1(\theta (\tau ))}{8 \theta '(\tau )^6} \right] =   i\frac{4\pi^2}{\beta^2} e^{i\frac{4 \pi}{\beta} \tau} \left[ -\frac{1}{2} \left( 1- \frac{1}{n^2} \right) \frac{F'\left(  e^{i\frac{2 \pi}{\beta} \tau}\right)^2}{F\left(  e^{i\frac{2 \pi}{\beta} \tau}\right)^2} - \{ F , e^{i\frac{2 \pi}{\beta} \tau} \} \right] + c.c.
     \end{split}
 \end{equation*}
 We now have an EOM entirely written on the orbifold $\mathcal{M}_n$, when $n \sim 1$. Note that $\theta (\tau + \beta) = \theta(\tau) + \beta$. The time reflection symmetry of the twist fields insertions allows us to choose a solution satisfying $\theta(-\tau) = -\theta(\tau)$ which automatically obeys $\theta(0) = 0, \; \theta''(0) = 0$, and as a result we should also impose $\theta(\beta/2) = \beta/2$ and $\theta''(\beta/2) = 0$. This EOM can be considerably simplified thanks to perturbative analysis and gauge-fixing, as we show below. \\

 We now expand perturbatively
 \begin{equation}\label{theta exoansion 1}
     \theta(\tau) = \tau + \delta \theta (\tau),
 \end{equation}
 hence
  \begin{equation}\label{theta derivative exoansion 1}
     \theta'(\tau) = 1 + \delta \theta' (\tau),
 \end{equation}
 where $\delta \theta(\tau)$ is of the order $o(n - 1)$ and all higher derivatives of $\theta(\tau)$ with respect to $\tau$ are at least of the order $o(n - 1)$.\\

 Keeping all the terms up to the order $o(n - 1)$ and noting that all $U_i (\theta(\tau))$ (\ref{U1exp}) (\ref{U2exp})(\ref{U3exp}) and their derivative are already at least of the order $o(n-1)$, only a few terms on the LHS of (\ref{eom final 0}) survive. We have
 \begin{equation} \label{eom LHS n-1}
 \begin{split}
  \text{LHS} & = \frac{3 l^3}{2 c  G_2} \left(\frac{3 \pi ^4 \theta ''(\tau )}{\beta ^4}  + \frac{11 \pi ^4 \theta '(\tau )^2 U_1'(\theta (\tau ))}{3 \beta ^4} -\frac{50}{3} \theta '(\tau )^2 U_3'(\theta (\tau )) +\theta '(\tau )^2 U_2'''(\theta (\tau )) -\frac{1}{12} \theta '(\tau )^2 U_1^{(5)}(\theta (\tau )) \right)  \\
  & = \frac{3 l^3}{2 c  G_2} \left(\frac{3 \pi ^4}{\beta^4}  \delta\theta'' (\tau) + \frac{11 \pi ^4 U_1'(\tau)}{3 \beta ^4} -\frac{50}{3} U_3'(\tau) + U_2'''(\tau) -\frac{1}{12} U_1^{(5)}(\tau) \right) + o(n-1)^2.
 \end{split}
\end{equation}
 Meanwhile, it is shown in Appendix \ref{B} of \cite{almheiri2020replica} that if $\theta (\tau)$ is expanded in Fourier series
 \begin{equation}
     \theta (\tau) = \tau + \sum_{m=-\infty}^\infty c_m e^{i \frac{2 m \pi}{\beta}\tau},
 \end{equation} then for any given $\delta \theta (\tau)$, one can always gauge-fix the ambiguities in the Fourier expansions of both functions $G$ and $F$
 \begin{equation} \label{G F expansion}
     G(w) = w + \sum_{l=0}^\infty g_l w^l, 
     \quad F(v) = v + \sum_{l= - \infty}^2 f_l v^l,
 \end{equation}
 such that there is a unique solution where
 \begin{equation}
     g_0 = f_1 = f_2 = 0 \quad \text{and} \quad f_l = ic_{l-1}\ \text{for} \ l \leq 0, \quad  g_l = -ic_{l-1} \ \text{for} \ l > 0.
 \end{equation}
 On this account, after a little bit of algebra one can show the second term on the RHS of (\ref{eom final 0}) becomes
 \begin{equation} \label{vanishing sch}
     e^{i\frac{4 \pi}{\beta} \tau} \{F , e^{i\frac{2 \pi}{\beta} \tau}\} = - \left( \delta \theta' (\tau) + \frac{\beta^2}{4\pi^2} \delta \theta''' (\tau) \right)_- = i\frac{2\pi}{\beta} \sum_{l - -\infty}^{-2} l(l^2 - 1) c_l \, e^{i\frac{2  \pi l}{\beta} \tau}
 \end{equation}
 where the minus subscript indicates that the expression inside the parenthesis is projected onto the modes of negative frequency. \\

 At the same time, note that $F$ is already multiplied by a factor of $(n-1)$ in the first term on the RHS of (\ref{eom final 0}), we can thus set $F = \frac{v - A}{B - v}$ per our convention below (\ref{new GF}), ignoring the effect of conformal welding. We have
 \begin{equation}
  \frac{1}{2} \left( 1- \frac{1}{n^2} \right) \frac{F'\left(  e^{i\frac{2 \pi}{\beta} \tau}\right)^2}{F\left(  e^{i\frac{2 \pi}{\beta} \tau}\right)^2} =  \left( n -1\right) \cdot \frac{(A-B)^2}{\left( A - e^{i\frac{2\pi}{\beta} \tau} \right)^2 \left( B - e^{i\frac{2\pi}{\beta} \tau} \right)^2}.  
 \end{equation}

 Ultimately, the perturbed EOM (\ref{eom final 0}) can be simplified to
 \begin{equation} \label{eom final symplified}
   \begin{split}
     &  \frac{9 l^3 \pi ^4}{2 c  G_2 \beta^4}   \delta\theta'' (\tau)  + \left[ - i \left(\delta \theta''' (\tau) + \frac{4\pi^2}{\beta^2}\delta \theta' (\tau) \right)_- + c.c.\right]\\
     & =  - \frac{3 l^3}{2 c  G_2} \left( \frac{11 \pi ^4 U_1'(\tau)}{3 \beta ^4} -\frac{50}{3} U_3'(\tau) + U_2'''(\tau) -\frac{1}{12} U_1^{(5)}(\tau) \right) + \frac{4\pi^2 (n-1)}{\beta^2}  \left[ - \frac{i (A-B)^2 e^{i\frac{4\pi}{\beta} \tau}}{\left( A - e^{i\frac{2\pi}{\beta} \tau} \right)^2 \left( B - e^{i\frac{2\pi}{\beta} \tau} \right)^2} + c.c. \right].
   \end{split}  
 \end{equation}

 We see, unlike in JT gravity (see (3.27) of \cite{almheiri2020replica}) where all the terms on the LHS of the perturbative EOM (\ref{eom final symplified}) are the functionals of $\delta \theta'''(\tau) + \frac{4\pi^2}{\beta^2}\delta \theta' (\tau)$, we actually have a term proportional to $\delta \theta ''(\tau)$ on the LHS here. As a result, when we look at the $k = 1$ Fourier mode of the LHS of (\ref{eom final symplified}), the second term vanishes because of (\ref{vanishing sch}) but the Fourier mode of  $\delta \theta ''(\tau)$ survives and we can \textbf{NOT} extract the extremity condition just from the $k = 1$ Fourier mode of the RHS of (\ref{eom final symplified}) as in \cite{almheiri2020replica}. Actually, we need to directly solve $\delta \theta (\tau)$ before we Fourier transform (\ref{eom final symplified}) and then plug it back in an attempt to recover the extremity condition (\ref{QES1 location}).\\

 Instead of going through it, we now show, by cutting corners, that the hypothesis about the leading saddle point of the replica geometry on $\Tilde{\mathcal{M}}_n$ is not the complete story because we can only recover from it one of two the stationary point of the $S_{\text{gen}}(B') \equiv S_{\text{gen}}([-a,b])$ --- the one at the bifurcation surface (throat) of the event horizon of an eternal Narayan black hole (located at $x=1 \Leftrightarrow a = \infty \Leftrightarrow r=\frac{\beta}{\pi}$ in the context of subsection \ref{one interval section}), not the true QES. So the leading point we hypothesized earlier is actually the subleading saddle point.\\

 As shown in \cite{almheiri2020replica}, there is actually a shortcut to derive the on-shell replica gravitational action from a given saddle point. We made a hypothesis earlier in this subsection that the leading saddle of the gravitational action on $\Tilde{\mathcal{M}}_n$ is still an Euclidean Narayan black hole of the same temperature $\beta$ but "n times as large as" the original Narayan black hole solution on $\mathcal{M}_1$. As a result, according to the on-shell gravitational action of the original $n = 1$ Narayan black hole (\ref{n = 1 Euclidean action}), we can immediately write down the gravitational action on $\Tilde{\mathcal{M}}_n$
\begin{equation}\label{Igrav alt}
   I_{\text{grav}}[\Tilde{\mathcal{M}}_n] = - \frac{l^3}{16 \pi G_2} \int_{\partial \mathcal{M}_n} d\tau \, \left( \frac{3 \pi ^4 \theta_n '(\tau )}{\beta ^4}+\frac{21 \theta_n ''(\tau )^4}{32 \theta_n '(\tau )^7}-\frac{3 \theta_n''' (\tau ) \theta_n ''(\tau )^2}{8 \theta_n '(\tau )^6} \right)
\end{equation}
  where $\theta_n (\tau )$, the counterpart of $\theta(\tau)$ on the boundary of $\Tilde{\mathcal{M}}_n$, replaces $\theta(\tau)$ in (\ref{n = 1 Euclidean action}). Using the expression of $\theta_n$ at the boundary of $\Tilde{\mathcal{M}}_n$ (\ref{theta n bd}) , we can now return to the orbifold $\mathcal{M}_n$ and obtain the on-shell replica gravitational action on $\mathcal{M}_n$ 
  \begin{equation} \label{action alt}
      \begin{split}
       I_{\text{grav}}[\mathcal{M}_n] & =  \frac{1}{n} I_{\text{grav}}[\Tilde{\mathcal{M}}_n]\\
       & = -  \frac{l^3}{16 \pi G_2} \int_{\partial \mathcal{M}_n} d\tau \, \left[\left( \frac{3 \pi ^4 \theta '(\tau )}{\beta ^4}+\frac{21 \theta ''(\tau )^4}{32 \theta '(\tau )^7}-\frac{3 \theta''' (\tau ) \theta ''(\tau )^2}{8 \theta '(\tau )^6} \right) + \left(-\frac{3 \pi ^4 \theta '(\tau ) U_1(\theta (\tau ))}{\beta ^4}+ \frac{3 \theta ''(\tau )^2 U_1''(\theta (\tau ))}{8 \theta '(\tau )^3}\right. \right.\\
       & -\frac{3 \theta ''(\tau )^3 U_1'(\theta (\tau ))}{2 \theta '(\tau )^5} \left.\left. +\frac{3 \theta'''(\tau ) \theta ''(\tau ) U_1'(\theta (\tau ))}{4 \theta '(\tau )^4}+\frac{63 \theta ''(\tau )^4 U_1(\theta (\tau ))}{32 \theta '(\tau )^7}-\frac{9 \theta''' (\tau ) \theta ''(\tau )^2 U_1 (\theta (\tau ))}{8 \theta '(\tau )^6}\right)\right].
      \end{split}
\end{equation}
We see, unlike the situation in JT (see Chapter 3 and Appendix A of \cite{almheiri2020replica}), the $o(n-1)$ order perturbation terms in (\ref{action alt}), namely terms containing $U_i(\theta (\tau )), i = 1,2,3$ and its derivatives, are \textbf{different}, at least formally, from those in the same on-shell replica gravitational action on $\mathcal{M}_n$ we obtained earlier (\ref{action orbifold}).\\

However, the two results (\ref{action orbifold}) and (\ref{action alt}) for the same on-shell replica gravitational action $I_ {\text{grav}}[\mathcal{M}_n]$ derived through different methods but from the same hypothesis --- the leading saddle point of $I_ {\text{grav}}[\Tilde{\mathcal{M}}_n]$ is still an Euclidean Narayan black hole of the same temperature  $\beta$ --- must be equal, if the replica metric we used to derive (\ref{action orbifold}) and (\ref{action alt}) is the true on-shell replica metric. (The true on-shell replica metric here referes to the replica metric not only taking the form in (\ref{replica metric}) - (\ref{U3exp}) but $A = e^{-\frac{2\pi}{\beta} a}$ inside its metric perturbation functions --- $U_i(\theta), i =1,2,3$ --- also evaluating to the exact right value for the on-shell replica geometry). \\

 One important observation is that, if the conical singularity/branch point on $\mathcal{M}_n$ lies at $A = e^{-\frac{2\pi}{\beta} a} = 0 \Leftrightarrow a=\infty$ (see \ref{branch points}), we have 
\begin{equation} \label{A = 0 aftermath}
    U_1(\theta) = n - 1, \ U_2(\theta) = 0, \ U_3(\theta) = \frac{2 \pi^4}{5 \beta^4} (n - 1)
\end{equation}
by setting $A = 0$ in (\ref{U1exp}), (\ref{U2exp}) and (\ref{U3exp}).
Plugging (\ref{A = 0 aftermath}) into (\ref{action orbifold}), we get 
\begin{equation} \label{recover part 1}
    \begin{split}
        I_{\text{grav}}[\mathcal{M}_n] & = -\frac{l^3}{16 \pi G_2} \int_{\partial \mathcal{M}_n} d\tau \bigg[ \left(\frac{3 \pi ^4 \theta '(\tau )}{\beta ^4}+\frac{21 \theta ''(\tau )^4}{32 \theta '(\tau )^7}-\frac{3 \theta''' (\tau ) \theta ''(\tau )^2}{8 \theta '(\tau )^6} \right)+ (n-1) \cdot \left( \frac{11 \pi ^4 \theta '(\tau )}{3 \beta ^4}  \right. \\
        & \left.- \frac{50}{3} \theta '(\tau ) \cdot \frac{2 \pi^4}{5 \beta^4} +\frac{63 \theta ''(\tau )^4 }{32 \theta '(\tau )^7} -\frac{9 \theta'''(\tau ) \theta ''(\tau )^2 }{8 \theta '(\tau )^6} \right)\bigg]\\
        & = -\frac{l^3}{16 \pi  G_2} \int_{\partial \mathcal{M}_n} d\tau \bigg[ \left(\frac{3 \pi ^4 \theta '(\tau )}{\beta ^4}+\frac{21 \theta ''(\tau )^4}{32 \theta '(\tau )^7}-\frac{3 \theta''' (\tau ) \theta ''(\tau )^2}{8 \theta '(\tau )^6} \right)+ (n-1) \cdot \left( - \frac{3 \pi ^4 \theta '(\tau )}{ \beta ^4}  \right. \\
        & \left.+\frac{63 \theta ''(\tau )^4 }{32 \theta '(\tau )^7} -\frac{9 \theta'''(\tau ) \theta ''(\tau )^2 }{8 \theta '(\tau )^6} \right)\bigg].
     \end{split}
\end{equation}
On the other hand, again plugging (\ref{A = 0 aftermath}) into (\ref{action alt}), we get the result finally matching that in (\ref{recover part 1}) 
 \begin{equation} \label{recover part 2}
      \begin{split}
       I_{\text{grav}}[\mathcal{M}_n] & =  \frac{1}{n} I_{\text{grav}}[\Tilde{\mathcal{M}}_n]\\
       & = -\frac{l^3}{16 \pi  G_2} \int_{\partial \mathcal{M}_n} d\tau \bigg[ \left(\frac{3 \pi ^4 \theta '(\tau )}{\beta ^4}+\frac{21 \theta ''(\tau )^4}{32 \theta '(\tau )^7}-\frac{3 \theta''' (\tau ) \theta ''(\tau )^2}{8 \theta '(\tau )^6} \right)+ (n-1) \cdot \left( - \frac{3 \pi ^4 \theta '(\tau )}{ \beta ^4}  \right. \\
        & \left.+\frac{63 \theta ''(\tau )^4 }{32 \theta '(\tau )^7} -\frac{9 \theta'''(\tau ) \theta ''(\tau )^2 }{8 \theta '(\tau )^6} \right)\bigg].
      \end{split}
\end{equation}\\

Note that if $A \neq 0$, the $o(n-1)$ order perturbation terms in (\ref{action orbifold}) and (\ref{action alt}) are formally different and they do not equal. Otherwise, there would be a non-trivial new equation by equating the $o(n-1)$ order perturbation terms in (\ref{action orbifold}) and (\ref{action alt}) when $A \neq 0$, namely

\begin{equation}\label{OOPArt}
\begin{split}
  \frac{11 \pi ^4 \theta '(\tau ) U_1 (\theta (\tau ))}{3 \beta ^4}+\theta '(\tau ) U_2''(\theta (\tau )) -\frac{50}{3} \theta '(\tau ) U_3 (\theta (\tau )) - \frac{1}{12} \theta '(\tau ) U_1^{(4)}(\theta (\tau )) = -\frac{3 \pi ^4 \theta '(\tau ) U_1(\theta (\tau ))}{\beta ^4}  \\
  \Rightarrow   \frac{20 \pi ^4 \theta '(\tau ) U_1 (\theta (\tau ))}{3 \beta ^4}+\theta '(\tau ) U_2''(\theta (\tau )) -\frac{50}{3} \theta '(\tau ) U_3 (\theta (\tau )) - \frac{1}{12} \theta '(\tau ) U_1^{(4)}(\theta (\tau )) = 0.
\end{split}
\end{equation}

We could directly solve the boundary mode $\theta(\tau)$ from (\ref{OOPArt}) and this solution would not involve $B = e^{2\pi/b}$ --- the other end point of interval in the flat space, since the gravitational action does not involve the flat space outside the boundary and thus no $B$ goes into the metric perturbation functions $U_i(\theta(\tau)), i = 1,2,3$. Hence, this $\theta(\tau)$ solution clashes with the the perturbation EOM of the $\theta(\tau)$ (\ref{eom final 0}) where there is $B$ going into it through the twist operator at $B$ when we calculate the stress tensor and consequently the exterior holomorphic map $F$ on the RHS of (\ref{eom final 0}). As a result, we conclude the two expressions, (\ref{action orbifold}) and (\ref{action alt}), for the same on-shell replica gravitational action $I_ {\text{grav}}[\mathcal{M}_n]$ match only if $A=0$. \\

Therefore, we see from the identical results in (\ref{recover part 1}) and (\ref{recover part 2}) when $A = 0$ that, even without explicitly solving the EOM of the boundary mode $\theta(\tau)$ (\ref{eom final symplified}), the saddle point we chose in our central hypothesis (we filled in an Euclidean Narayan black hole of the same temperature $\beta$ inside the boundary of covering manifold $\Tilde{\mathcal{M}}_n$) mandates the conical singularity/branch point on the orbifold $\mathcal{M}_n$ to be exactly at $w = A = e^{-\frac{2\pi}{\beta} a} = 0 \Leftrightarrow a=\infty$. The perturbed equation of motion for the boundary mode $\theta(\tau)$ associated with the hypothesized saddle point is thus further simplified from (\ref{eom final symplified}) to 
\begin{equation}
    \frac{9 l^3 \pi ^4}{2 c  G_2 \beta^4}   \delta\theta'' (\tau)  + \left[ - i \left(\delta \theta''' (\tau) + \frac{4\pi^2}{\beta^2}\delta \theta' (\tau) \right)_- + c.c.\right] = \frac{4\pi^2 (n-1)}{\beta^2}  \left[ - \frac{i B^2}{ \left( B - e^{i\frac{2\pi}{\beta} \tau} \right)^2} + c.c. \right]
\end{equation}
after we plugged $A=0$ into (\ref{eom final symplified}), which we will not try to solve in this article. Going back to Lorentzian signature, this conical singularity/branch point translates into $a=\infty \Leftrightarrow r =\beta/\pi$ at the bifurcation surface (throat) of the event horizon at $u=0$, \textbf{so we just recovered one of the two stationary points --- the stationary point other than, but generally still very close to, the QES in the one-interval scenario (see the analysis below (\ref{QES1 location 2}))}.\\

Based on the arguments in subsection \ref{general replica formalism}, the results in subsection \ref{one interval section} show that there are actually two saddle points of the Euclidean Narayan action leading to two stationary points of the generalized entropy in the one-interval case --- the bifurcation surface(throat) of the event horizon and the true QES. The saddle point we earlier 
 hypothesized to be the leading saddle point, namely a standard Euclidean Narayan black hole of the same temperature $\beta$ but larger, is actually the subleading saddle point. To recover the other stationary point, the true QES, we need to find and fill in the leading saddle point of Euclidean Narayan action. This leading saddle point has inverse temperature $\beta$ and, at the same time, satisfying the boundary conditions for the metric (\ref{boundary metric 0}) and dilaton (\ref{dilaton boudnary}) as the replica wormhole geometry. We have yet found an answer to that so far. It is an interesting open problem and some thoughts are shared in the next section.

\newpage

\section{Conclusion and Discussion}\label{condis}
In this article, we first built an eternal black hole on the basis of maximally extended Narayan black hole geometry coupled with two flat baths and then we established a new version of black hole information paradox in the setting of the same CFT living on an eternal vacuum Narayan black hole joined to flat baths on both sides. We derived the extremity conditions and analyzed the existence and the locations of the QES/island solutions in both one-interval and two-interval scenarios and found that there are both similarities and remarkable differences from the JT gravity case. In the two-interval setting which constitutes our version black hole information paradox, we found that, just as in JT gravity, the Hawking saddle is exponentially suppressed in the Euclidean path integral as the time goes on and the entropy continues to grow inside the boundary. Therefore, another saddle of a different topology, in this case the replica wormhole, dominates because the associated generalized entropy finally saturates at that of two copies of single interval entropy at late times and the contribution in the Euclidean path integral stops to diminish for long periods. Although we provided solid justifications for the QES prescription we used to write down the extremity condition and calculate the fined-grained entropy of the radiation and the two-side eternal Narayan black hole, for the one-interval scenario we only partially successfully constructed the replica geometry on the covering manifold (the subleading saddle point), having recovered one of the two stationary point, but did not recover the extremity condition itself and the other stationary point --- the true QES. It is intriguing to guess what the other solution of the Narayan action satisfying the boundary conditions and leading to the true QES is.  Here we share some thoughts on this open problem.\\

Fundamentally different from the JT gravity where a constant curvature $R = - 2$ Euclidean $AdS_2$ saddle point is dictated by the JT action on $\mathcal{M}_1$ by varying with respect to the dilaton $\phi$ in the bulk,
\begin{equation} \label{JT action}
    I_{\text{JT}}[\mathcal{M}_1] = - \frac{1}{16 \pi G_2} \left[\int_{\mathcal{M}_1} \phi (R + 2) + 2\phi_b\int_{\partial \mathcal{M}_1} (K + 1) \right] +(\text{topological}),
\end{equation} 
there are likely more than one classical solutions to the Narayan action on $\mathcal{M}_1$
 \begin{equation} \label{Narayan action section 5}
 I_{\text{Narayan}}[\mathcal{M}_1] = -\frac{1}{16 \pi G_2} \bigg[ \int_{\mathcal{M}_1} \big(  \phi R + 12 l^{-2} \phi^{-\frac{1}{3}} \big)+2\int_{\partial \mathcal{M}_1} \phi K - \frac{6}{l} \int_{\partial \mathcal{M}_1} \phi^{\frac{2}{3} }\bigg] +(\text{topological})
 \end{equation} are of inverse temperature $\beta$ and also satisfying the boundary conditions for both the metric (\ref{boundary metric 0}) and the dilaton (\ref{dilaton boudnary}). To account for the true QES in the one-interval scenario, we need to find a classical solution other than the Narayan black hole satisfying all the boundary conditions and can be shown to have inverse temperature $\beta$. On top of it, the on-shell gravitational action of this solution on $\mathcal{M}_1$, $I_{\text{Narayan, the other solution}}[\mathcal{M}_1]$, should be no smaller than the on-shell gravitational action of the Narayan black hole solution (\ref{on-shell action n=1}) dominating the Narayan black hole saddle so that we still have the correct gravitational entropy for the Narayan black hole (\ref{eternal n=1 grav entropy}). It is an interesting open problem to find this solution, as it concerns the evaluation of the saddle points of Euclidean path integral in a theory other than JT gravity, which is more complicated than JT gravity but is relatively simple given that it incorporates JT gravity as its simplest example. Furthermore, being able to evaluate nonperturbative quantities and effects in other 2D gravity theories may bring new insights to studying quantum gravity in 2D.\\

 We now move on to highlight some similarities and differences in regard to the eternal black holes in Narayan theory and JT and some other possiblilities in future developments. \\

 In section \ref{narayan} we imposed a kinetic constraint for the boundary mode (\ref{constraint tu}) and constructed a consistent eternal black hole on the basis of it, the same construction can be generalized to Narayan theory with $d_i >3$ without fundamental obstacles despite an algebraic mess. However, unlike JT where there is no inconsistency on the boundary conditions to begin with, it is problematic when a black hole joined by two baths is not eternal (say an evaporating black hole) in the Narayan theory and the flux across the boundary is not infinitesimal, say of order 1: if we insist the kinetic constraint (\ref{constraint tu}) be imposed: there is no solution to (\ref{energy rate of change}) satisfying (\ref{constraint tu}). Therefore, if we intend to study the non-eternal Narayan black holes, we need to find a new way to join the baths to the black hole, which may involve baths with a curved but non-dynamic metric (possibly conformaly flat) or a natural but a different way than (\ref{xy transformation}) to extend the Rindler coordinates on flat bath into the Narayan black hole. It is challenging but interesting to design a more general formalism of joining baths to asymptotically conformally $AdS_2$ geometries as it allows for evaporating black holes which can lead to interesting dynamics, providing more insights to the black hole information paradox in 2D. \\

 In section \ref{BHIP}, some interesting results were presented. In one-interval scenario in eternal Narayan black hole, like in the JT construction, the QES always exists and lies somewhere to the right of the bifurcation surface (throat) of the event horizon, albeit generally even closer to the bifurcation surface. In the physically significant two-interval scenario, although at late times a Page curve is produced just as in JT gravity, we found that real solutions to extremity condition for islands always exist, independent on the parameters and even before Page time, which is different than the same scenario in JT.

\section*{Acknowledgement}
The author is grateful to his Master's thesis supervisors Professor Umut Gursoy and Professor Stefan Vandoren for helpful discussions and enlightening comments. The author also wants to thank Prof. Daniel Grumiller from TU Wien for pointing out a mistake at the very beginning of section 2 in the first submitted version of this article about the most general action of 2D dilaton gravity.
\newpage

\appendix
\section{Calculation of the extrinsic curvature in an  n = 1 vacuum Narayan black hole }\label{A}
In this appendix we calculate the extrinsic curvature on any curve of constant $\sigma$ in a vacuum Narayan black hole of inverse temperature $\beta$, which will be used in subsection \ref{eternal BH + bath} to compute the on-shell gravitational action of a vacuum Narayan black hole in Lorentzian signature. (As to Euclidean vacuum Narayan black hole, we get the same result up to several opposite signs.)\\

For convenience, we choose to set up the calculation in the tortoise coordinates  $\{ t, r_* \}$ and name the curve of constant value of $\sigma$ in question $\Sigma$. As the Rindler coordinates $\{ u, \sigma \}$ are extended from the baths to the vacuum Narayan black hole, even though we do not know as of yet the exact form of this coordinate extension, $\Sigma$ is by definition a $u$ coordinate line. We adopt the convention that the derivatives with respect to $u$ is denoted by a prime mark 'and the derivatives with respect to $\sigma$ is denoted by dot notation and from (\ref{xy transformation}) we have one important observation
\begin{equation}
    \begin{split}
       & t' = \frac{\partial t}{\partial x^+} \frac{d x^+}{d y^+ } \frac{dy^{+}}{du}  + \frac{\partial t}{\partial x^-} \frac{d x^- }{d y^- } \frac{dy^{-}}{du}  = \frac{1}{2} \left( \frac{d x^+ }{d y^+ } + \frac{d x^- }{d y^- } \right),\\
       & \dot{t} = \frac{\partial t}{\partial x^+} \frac{d x^+}{d y^+ } \frac{dy^{+}}{d\sigma}  + \frac{\partial t}{\partial x^-} \frac{d x^- }{d y^- } \frac{dy^{-}}{d\sigma}  = \frac{1}{2} \left( \frac{d x^+ }{d y^+ } - \frac{d x^- }{d y^- } \right),\\
       & r_*' = \frac{\partial r_*}{\partial x^+} \frac{d x^+}{d y^+ } \frac{dy^{+}}{du}  + \frac{\partial r_*}{\partial x^-} \frac{d x^- }{d y^- } \frac{dy^{-}}{du}  = \frac{1}{2} \left( \frac{d x^+ }{d y^+ } - \frac{d x^- }{d y^- } \right),\\
       & \dot{r_*} = \frac{\partial r_*}{\partial x^+} \frac{d x^+}{d y^+ } \frac{dy^{+}}{d\sigma}  + \frac{\partial r_*}{\partial x^-} \frac{d x^- }{d y^- } \frac{dy^{-}}{d\sigma}  = \frac{1}{2} \left( \frac{d x^+ }{d y^+ } + \frac{d x^- }{d y^- } \right),
    \end{split}
\end{equation}
therefore,
\begin{equation} \label{inverse derivative equivalence}
    t' = \dot{r_*}, \quad \dot{t} = r_*'.
\end{equation}
So now we have the following expression of the tangent vector $t^a$ and outwardly pointing normal vector $n^a$ on $\Sigma$ according to the metric of vacuum Narayan black holes of inverse temperature $\beta$ (\ref{BH metric in tortoise})

\begin{equation} \label{tangent 0}
    t^a = \frac{1}{l^2 \sqrt{\left( \frac{1}{r^4} - \frac{\pi^4}{\beta^4} \right) \left( t'^2 - r_*'^2\right)}} {\left( \frac{\partial}{\partial u} \right)}^a = \frac{t'}{l^2 \sqrt{\left( \frac{1}{r^4} - \frac{\pi^4}{\beta^4} \right) \left( t'^2 - r_*'^2\right)}} {\left( \frac{\partial}{\partial t} \right)}^a + \frac{r_*'}{l^2 \sqrt{\left( \frac{1}{r^4} - \frac{\pi^4}{\beta^4} \right) \left( t'^2 - r_*'^2\right)}} {\left( \frac{\partial}{\partial r_*} \right)}^a 
\end{equation}
and
\begin{equation} \label{normal 0}
    n^a  = - \frac{r_*'}{l^2 \sqrt{\left( \frac{1}{r^4} - \frac{\pi^4}{\beta^4} \right) \left( t'^2 - r_*'^2\right)}} {\left( \frac{\partial}{\partial t} \right)}^a - \frac{t'}{l^2 \sqrt{\left( \frac{1}{r^4} - \frac{\pi^4}{\beta^4} \right) \left( t'^2 - r_*'^2\right)}} {\left( \frac{\partial}{\partial r_*} \right)}^a,
\end{equation}
\begin{equation} \label{normal co 0}
    n_a  = \frac{l^2 \, r_*' \sqrt{\frac{1}{r^4} - \frac{\pi^4}{\beta^4}}}{ \sqrt{   t'^2 - r_*'^2}} {\left( dt \right)}_a - \frac{l^2 \, t' \sqrt{\frac{1}{r^4} - \frac{\pi^4}{\beta^4}}}{ \sqrt{   t'^2 - r_*'^2}}  {\left( dr_*\right)}_a. 
\end{equation}   
 Meanwhile, the inverse metric and the non-zero Christoffel symbols in tortoise coordinates are

\begin{equation} \label{inverse metric tortoise}
    g^{tt} = - g^{r_* r_*} = - l^{-4} {\left( \frac{1}{r^4} - \frac{\pi^4}{\beta^4} \right)}^{-1}, \quad g^{tr_*} = g^{r_*t} = 0,
\end{equation}
and 
\begin{equation} \label{Christoffel 0}
    \Gamma^t_{t r_*} = \Gamma^t_{r_* t} = \Gamma^{r_*}_{t t} = \Gamma^{r_*}_{r_* r_*} = \frac{2 \beta ^4}{\pi ^4 r^5-\beta ^4 r} * \frac{dr}{dr_*} = - \frac{2}{r}. 
\end{equation} 
The extrinsic curvature $K$ with respect to $n^a$ on $\Sigma$ is thus
\begin{equation} \label{K appendix A}
    \begin{split}
        K & = g^{\mu \nu} \nabla_\mu n_\nu = g^{\mu \nu} \left(\partial_\mu n_\nu - \Gamma^\rho_{\mu \nu} n_\rho \right) =g^{t t} \partial_t n_t + g^{r_* r_*} \partial_{r_*} n_{r_*} \\
        & = l^{-4} {\left( \frac{1}{r^4} - \frac{\pi^4}{\beta^4} \right)}^{-1} \left[ -\left( \frac{\partial u}{\partial t} n_t' +  \frac{\partial \sigma}{\partial t} \dot{n_t} \right) + \left( \frac{\partial u}{\partial r_*} n_{r_*}' +  \frac{\partial \sigma}{\partial {r_*}} \dot{n_{r_*}} \right) \right]\\
        & =  l^{-4} {\left( \frac{1}{r^4} - \frac{\pi^4}{\beta^4} \right)}^{-1} \left[ - \frac{t'}{t'^2 - r_*'^2} \cdot \frac{\partial}{\partial u} \left( \frac{l^2 \, r_*' \sqrt{\frac{1}{r^4} - \frac{\pi^4}{\beta^4}}}{ \sqrt{   t'^2 - r_*'^2}} \right) - \frac{-r_*'}{t'^2 - r_*'^2} \cdot \frac{\partial}{\partial \sigma} \left( \frac{l^2 \, r_*' \sqrt{\frac{1}{r^4} - \frac{\pi^4}{\beta^4}}}{ \sqrt{   t'^2 - r_*'^2}} \right) \right]\\
        & + l^{-4} {\left( \frac{1}{r^4} - \frac{\pi^4}{\beta^4} \right)}^{-1} \left[ \frac{- r_*'}{t'^2 - r_*'^2} \cdot \frac{\partial}{\partial u} \left( \frac{- l^2 \, t' \sqrt{\frac{1}{r^4} - \frac{\pi^4}{\beta^4}}}{ \sqrt{   t'^2 - r_*'^2}} \right) + \frac{t'}{t'^2 - r_*'^2} \cdot \frac{\partial}{\partial \sigma} \left( \frac{-l^2 \, t' \sqrt{\frac{1}{r^4} - \frac{\pi^4}{\beta^4}}}{ \sqrt{   t'^2 - r_*'^2}} \right) \right]\\
        & =  - l^{-2} {\left( \frac{1}{r^4} - \frac{\pi^4}{\beta^4} \right)}^{-1} \cross\\
        & \left\{ \frac{ t'}{t'^2 - r_*'^2} \cdot \left[ \frac{r_*'' \sqrt{\frac{1}{r^4} - \frac{\pi^4}{\beta^4}}}{\sqrt{t'^2 - r_*'^2}} + \frac{r_*' \cdot \left( -\frac{4}{r^5} \, \frac{dr}{dr_*} \right) \cdot r_*'}{2 \sqrt{\left( \frac{1}{r^4} - \frac{\pi^4}{\beta^4} \right) \left( t'^2 - r_*'^2\right)}} - \frac{r_*' \sqrt{\frac{1}{r^4} - \frac{\pi^4}{\beta^4}}}{2 \, {\left( t'^2 - r_*'^2 \right)}^{\frac{3}{2}}} \cdot \left( 2t't'' -2r_*'r_*'' \right) \right] \right.\\
        & -\frac{ r_*'}{t'^2 - r_*'^2} \cdot \left[ \frac{t'' \sqrt{\frac{1}{r^4} - \frac{\pi^4}{\beta^4}}}{\sqrt{t'^2 - r_*'^2}} + \frac{r_*' \cdot \left( -\frac{4}{r^5} \, \frac{dr}{dr_*} \right) \cdot t'}{2 \sqrt{\left( \frac{1}{r^4} - \frac{\pi^4}{\beta^4} \right) \left( t'^2 - r_*'^2\right)}} - \frac{r_*' \sqrt{\frac{1}{r^4} - \frac{\pi^4}{\beta^4}}}{2 \, {\left( t'^2 - r_*'^2 \right)}^{\frac{3}{2}}} \cdot \left( 2t'r_*'' -2r_*'t'' \right) \right]\\
        & -\frac{ r_*'}{t'^2 - r_*'^2} \cdot \left[ \frac{t'' \sqrt{\frac{1}{r^4} - \frac{\pi^4}{\beta^4}}}{\sqrt{t'^2 - r_*'^2}} + \frac{r_*' \cdot \left( -\frac{4}{r^5} \, \frac{dr}{dr_*} \right) \cdot t'}{2 \sqrt{\left( \frac{1}{r^4} - \frac{\pi^4}{\beta^4} \right) \left( t'^2 - r_*'^2\right)}} - \frac{t' \sqrt{\frac{1}{r^4} - \frac{\pi^4}{\beta^4}}}{2 \, {\left( t'^2 - r_*'^2 \right)}^{\frac{3}{2}}} \cdot \left( 2t't'' -2r_*'r_*'' \right) \right]\\
        & \left. \frac{ t'}{t'^2 - r_*'^2} \cdot \left[ \frac{r_*'' \sqrt{\frac{1}{r^4} - \frac{\pi^4}{\beta^4}}}{\sqrt{t'^2 - r_*'^2}} + \frac{t' \cdot \left( -\frac{4}{r^5} \, \frac{dr}{dr_*} \right) \cdot t'}{2 \sqrt{\left( \frac{1}{r^4} - \frac{\pi^4}{\beta^4} \right) \left( t'^2 - r_*'^2\right)}} - \frac{t' \sqrt{\frac{1}{r^4} - \frac{\pi^4}{\beta^4}}}{2 \, {\left( t'^2 - r_*'^2 \right)}^{\frac{3}{2}}} \cdot \left( 2t'r_*'' -2t'r_*'' \right) \right] \right\}\\
        & = - l^{-2} {\left( \frac{1}{r^4} - \frac{\pi^4}{\beta^4} \right)}^{- \frac{1}{2}} \cross \left[ \frac{2 \left( t' r_*'' - t'' r_*' + t' r^{-1} r_*'^2 - t'^3 r^{-1} \right)}{{\left( t'^2 - r_*'^2 \right)}^{\frac{3}{2}}} + \frac{r_*'^2 \left( t' r_*'' - t'' r_*' \right) - t'^2 \left( t' r_*'' - t'' r_*' \right)  }{{\left( t'^2 - r_*'^2 \right)}^{\frac{5}{2}}}\right]\\
        & = l^{-2} r^{-1}  {\left( t'^2 - r_*'^2\right)}^{-\frac{3}{2}} {\bigg(\frac{1}{r^4} - \frac{\pi^4}{\beta^4} \bigg)}^{- \frac{1}{2}} \left( 2t'^3 - 2t'r_*'^2 + t'' r  r_*' - t' r r_*''\right).
    \end{split}
\end{equation}

In Euclidean signature, K takes the following form
\begin{equation}
    K = l^{-2} r^{-1}  {\left( t'^2 + r_*'^2\right)}^{-\frac{3}{2}} {\bigg(\frac{1}{r^4} - \frac{\pi^4}{\beta^4} \bigg)}^{- \frac{1}{2}} \left( 2t'^3 + 2t'r_*'^2  + t' r r_*''- t'' r  r_*'\right).
\end{equation}

\newpage

\section{An attempt to derive the on-shell replica gravitational action }\label{B}
In this appendix we offer a detailed derivation of the $n \sim 1$ on-shell replica gravitational action based on our hypothesis about the leading saddle point of replica gravitational path integral above Figure \ref{Mn}. Although the results of this appendix only account for the saddle point we hypothesized, the method used here is more general and can be easily applied to other cases without significant changes.\\

 The replica gravitational action on the orbifold $\mathcal{M}_n$ is
\begin{equation}\label{action cb in Appendix B}
    \begin{split}
      & I_ {\text{grav}}[\mathcal{M}_n] =  \frac{1}{n} I_ {\text{grav}}[\Tilde{\mathcal{M}}_n] \\  
    & = - \frac{\phi_0}{16 \pi G_2} \left( \int_{\mathcal{M}_n} R +2\int_{\mathcal{M}_n} K \right) - \frac{1}{16 \pi G_2} \left[ \int_{\mathcal{M}_n} \big(  \phi R + 12 l^{-2} \phi^{\frac{1}{3}} \big) +2\int_{\partial \mathcal{M}_n} \phi K  - \frac{6}{l} \int_{\partial \mathcal{M}_n} \phi^{\frac{2}{3}}\right] \\& + \frac{1}{4 G_2} \left( 1 - \frac{1}{n} \right) \sum_{\alpha, \text{Cosmic Branes}} \left[ \phi_0 + \phi(w_\alpha) \right],
     \end{split}
\end{equation} 
To compute the two boundaries terms in (\ref{action cb in Appendix B}), we need to calculate the extrinsic curvature $K$ and the value of the dilaton $\phi$ at the physical boundary $\sigma = -\epsilon$.\\

We start with the expansion of the replica metric near the boundary, (\ref{replica metric}) and (\ref{replica metric boundary}),
\begin{equation}
\begin{split}
   ds^2_{\text{in}, \mathcal{M}_n}  & = \frac{\beta^2 l^4}{4 \pi^2} \bigg( \frac{1}{r (\abs{w})^4} - \frac{\pi^4}{\beta^4}\bigg) \frac{1}{\abs{w}^2} d{w} d{\Bar{w}} \left[  1 + 2U_1 (\theta) + 2\gamma^2 U_2 (\theta) + 2\gamma^4 U_3 (\theta) + o(\gamma^6) + o(n-1)^2 \right]\\
   & = l^4\left( \frac{1}{r \left(e^{\frac{2\pi}{\beta}\gamma}\right)^4} - \frac{\pi^4}{\beta^4}\right)\left( d\gamma^2 + d\theta^2 \right) \left[  1 + 2U_1 (\theta) + 2\gamma^2 U_2 (\theta) + 2\gamma^4 U_3 (\theta) + o(\gamma^6) + o(n-1)^2 \right]\\
\end{split}
\end{equation}
where $w = e^{\frac{2\pi}{\beta} (\gamma + i \theta)}$ with $\abs{w} < 1\Leftrightarrow \gamma<0$, $r$ satisfies (\ref{inverse r Euclidean}) and $U_i (\theta)$'s are of the form (\ref{U1exp}) (\ref{U2exp}) and (\ref{U3exp}), $\ i = 1,2,3$.\\

Here the boundary conditions for the metric is still 
\begin{equation}
    ds^2_{\text{bd},\mathcal{M}_n} = \frac{l^4}{\epsilon^4} \left( d\tau^2 +d\sigma^2\right)
\end{equation}
which gives rise to
\begin{equation} \label{pin down gamma}
     \left( \frac{1}{r \left(e^{\frac{2\pi}{\beta}\gamma}\right)^4} - \frac{\pi^4}{\beta^4}\right)\left( \gamma'^2 + \theta'^2 \right) \left[  1 + 2U_1 (\theta) + 2\gamma^2 U_2 (\theta) + 2\gamma^4 U_3 (\theta) + o(\gamma^6) + o(n-1)^2 \right] = \frac{1}{\epsilon^4}
\end{equation}
in parallel to (\ref{pin down r eternal}) at the physical boundary $\sigma = - \epsilon$ (we still mark the derivative with respect to the Rindler time $\tau$ with ').
In the same spirit of (\ref{rstar ansatz}), we make another ansatz for $\gamma$ at the boundary
\begin{equation} \label{gamma ansatz}
    \gamma_{\text{bd}} (\tau) = -\epsilon {\theta '(\tau )}^{\frac{1}{2}} \left( 1 + f_{\gamma 0}(\tau )+\epsilon  f_{\gamma 1}(\tau )+\epsilon ^2 f_{\gamma 2}(\tau )+\epsilon ^3 f_{\gamma 3}(\tau )+\epsilon ^4 f_{\gamma 4}(\tau ) + o\left( \epsilon \right)^5\right)
\end{equation}
and plug it into (\ref{pin down gamma}). The solution, to the first order of $(n-1)$, is
\begin{equation}
    f_{\gamma 0}(\tau ) = \frac{1}{2} U_1(\theta (\tau )), 
\end{equation}
\begin{equation}
    f_{\gamma 2}(\tau ) = \frac{1}{2} U_2(\theta (\tau )) \theta '(\tau ),  
\end{equation}
\begin{equation}
    f_{\gamma 4}(\tau ) = -\frac{\pi ^4 \theta '(\tau )^2}{20 \beta ^4} + \left(-\frac{\pi ^4 \theta '(\tau )^2 U_1(\theta (\tau ))}{8 \beta ^4}+\frac{1}{8} \theta ''(\tau ) U_2'(\theta (\tau ))+\frac{3 \theta ''(\tau )^2 U_2(\theta (\tau ))}{16 \theta '(\tau )^2}+\frac{1}{2} \theta '(\tau )^2 U_3(\theta (\tau ))\right),
\end{equation}
\begin{equation}
     f_{\gamma 1}(\tau ) =  f_{\gamma 3}(\tau ) = 0,
\end{equation}
where $ U_i'(\theta (\tau ))$ means the derivative of $ U_i \left(\theta (\tau ) \right)$ with respect to $\theta(\tau)$.\\

We see from the above equations that the replica wormhole geometry we hypothesized leads to $o(n-1)$ perturbations in every order of $\epsilon$ in (\ref{gamma ansatz}), even the zeroth order, 1, receives $o(n-1)$ perturbation. \\

We can now proceed to calculate the function $r$ at the boundary through the following expansion of $r$ in $\gamma$ near the boundary
\begin{equation}
    r = -\gamma + \frac{\pi ^4 {\gamma}^5}{5 \beta^4} + \frac{4 \pi ^8 {r_*}^9}{45 \beta^8} + \dots,  
\end{equation}
which is essentially the same thing as (\ref{inverse expansion boundary}). After plugging in (\ref{gamma ansatz}), we get 
\begin{equation} \label{rbd}
\begin{split}
   r_{\text{bd}} (\tau) & = \epsilon  \theta '(\tau )^{\frac{1}{2}} \Biggl\{1+\frac{1}{2} U_1(\theta (\tau ))+\frac{1}{2} \epsilon ^2 \theta'(\tau ) U_2( \theta (\tau ))+\epsilon ^4\left[-\frac{\pi ^4 \theta '(\tau )^2}{4 \beta ^4} +\left(\frac{1}{2} \theta '(\tau )^2 U_3(\theta (\tau ))-\frac{5 \pi ^4 \theta '(\tau )^2 U_1(\theta (\tau ))}{8 \beta ^4}\right.\right.\\
   & \left.\left.+\frac{1}{8} \theta ''(\tau ) U_2'(\theta (\tau ))+\frac{3 \theta ''(\tau )^2 U_2(\theta (\tau ))}{16 \theta '(\tau )^2}\right)\right] \Biggr\}
\end{split}
\end{equation}
which reduces to (\ref{r boundary full}) when $n = 1$ and switched back to Lorentzian signature.
We can then plug (\ref{rbd}) into the metric and compute the tangent vector $t^\mu$ on and normal vector $n^\mu$ to the boundary, which sets the conditions to calculate the extrinsic curvature $K$ at the boundary, 
\begin{equation} \label{K wormhole}
    \begin{split}
        K & = t^\mu t^\nu \nabla_\mu n_\nu = \epsilon l^{-2} \theta '(\tau )^{\frac{1}{2}}\Biggl\{2- U_1(\theta (\tau ))+\epsilon ^2\left[-\frac{\theta ''(\tau )^2}{\theta '(\tau )^3}+\frac{\theta''' (\tau )}{2 \theta '(\tau )^2}+\left(\frac{1}{2} \theta '(\tau ) U_1''(\theta (\tau ))+\frac{\theta ''(\tau )  U_1'(\theta (\tau ))}{2 \theta '(\tau )}\right.\right.\\ 
        & -3 \theta '(\tau )  U_2(\theta (\tau )) \left.\left. - \frac{\theta''' (\tau ) U_1(\theta (\tau ))}{4 \theta '(\tau )^2}-\frac{\theta ''(\tau )^2 U_1(\theta (\tau ))}{2 \theta '(\tau )^3}\right)\right] + \epsilon ^4\left[\frac{\pi ^4 \theta '(\tau )^2}{2 \beta ^4}+\frac{21 \theta ''(\tau )^4}{64 \theta '(\tau )^6}-\frac{3 \theta '''(\tau ) \theta ''(\tau )^2}{16 \theta '(\tau )^5}\right.\\
        & + \left(\frac{3 \pi ^4 \theta '(\tau )^2 U_1(\theta (\tau ))}{4 \beta ^4}+\frac{1}{2} \theta '(\tau )^2 U_2''(\theta (\tau ))+\frac{7}{4} \theta ''(\tau ) U_2'(\theta (\tau ))-\frac{3 \theta ''(\tau )^2 U_2(\theta (\tau ))}{8 \theta '(\tau )^2}-5 \theta '(\tau )^2 U_3 (\theta (\tau ))\right.\\
        & -\frac{3 \theta ''(\tau )^2 U_1''(\theta (\tau ))}{16 \theta '(\tau )^2}+\frac{\theta ''(\tau )^3 U_1'(\theta (\tau ))}{2 \theta '(\tau )^4}-\frac{3 \theta'''(\tau ) \theta ''(\tau ) U_1'(\theta (\tau ))}{8 \theta '(\tau )^3}+\frac{63 \theta ''(\tau )^4 U_1(\theta (\tau ))}{128 \theta '(\tau )^6}+\frac{3 \theta'''(\tau ) U_2(\theta (\tau ))}{4 \theta '(\tau )}\\
        & \left.\left.-\frac{9 \theta'''(\tau ) \theta ''(\tau )^2 U_1(\theta (\tau ))}{32 \theta '(\tau )^5}\right)\right] + o(\epsilon^6)\Biggr\}
    \end{split}
\end{equation}
which reduces to (\ref{K n1  exp}) if we take $n = 1$ --- all $U_i(\theta (\tau ))$'s thus vanish because they are of the order $o(n-1)$ --- and switch from Euclidean back to Lorentzian signature.\\

The dilaton $\phi$ on $\Tilde{\mathcal{M}}_n$ is an integral part of the leading saddle point solution we hypothesized, together with the replica metric (\ref{metric in covering manifold}). Therefore, $\phi$ should take the following form
\begin{equation} \label{phi n}
    \phi = \frac{l^3}{r_n^3}
\end{equation}
where $r_n$ is defined and related to $w_n$ and $w$ through relation (\ref{inverse rn Euclidean}) and (\ref{uniformize}).\\

In order to find the expression of $r_n$ we need to first relate it to $r$, whose boundary expression has already been obtained in (\ref{rbd}). From the equation that defines $r_n$ (\ref{inverse rn Euclidean}), (\ref{uniformize}) and $w = e^{\frac{2\pi}{\beta} (\gamma + i \theta)}$  we have
\begin{multline} \label{inverse rn 2}
    \log\abs{w_n} = \log\abs{\frac{\Tilde{w} + A}{1+ A \Tilde{w}}} = \log \abs{ \frac{{\left( \frac{w-A}{1-Aw} \right)}^{1/n} + A}{1+ A {\left( \frac{w-A}{1-Aw} \right)}^{1/n}}}  = \log \abs{ \frac{{\left( \frac{ e^{\frac{2\pi}{\beta} (\gamma + i \theta)}-A}{1-A e^{\frac{2\pi}{\beta} (\gamma + i \theta)}} \right)}^{1/n} + A}{1+ A {\left( \frac{ e^{\frac{2\pi}{\beta} (\gamma + i \theta)}-A}{1-A e^{\frac{2\pi}{\beta} (\gamma + i \theta)}} \right)}^{1/n}}}  \\ = - \left(\arctan{\frac{\pi r_n}{\beta}} +\arctanh{\frac{\pi r_n}{\beta}}\right).
\end{multline}
Meanwhile from (\ref{inverse r Euclidean}) we also have
\begin{equation} \label{inverse r Euclidean 2}
    \log\abs{w} = \frac{2\pi \gamma}{\beta} = - \left(\arctan{\frac{\pi r}{\beta}} +\arctanh{\frac{\pi r}{\beta}}\right).
\end{equation}
It is convenient to expand $r_n$ as the perturbed ansatz below
\begin{equation} \label{rn ansatz}
    r_n = r + (n-1)  \delta r
\end{equation}
and plug it into (\ref{inverse rn 2}). Next, by subtracting (\ref{inverse r Euclidean 2}) from (\ref{inverse rn 2}) and then expanding both sides in $(n-1)$, we get

\begin{equation} \label{delta r}
\begin{split}
     \delta r & = \frac{\beta ^4-\pi ^4 r^4}{4 \pi \beta ^3 }\left(A^2-1 \right) \left[ \left(-A^2+A e^{\frac{2 \pi  (-\gamma +i \theta )}{\beta }}+A e^{\frac{2 \pi  (\gamma -i \theta )}{\beta }}-1\right) \log \left(\frac{A-e^{\frac{2 \pi  (\gamma -i \theta )}{\beta }}}{-1+A e^{\frac{2 \pi  (\gamma -i \theta )}{\beta }}}\right) \right. \\
     & \left. + \left(-A^2+A e^{\frac{2 \pi  (\gamma +i \theta )}{\beta }}+A e^{-\frac{2 \pi  (\gamma +i \theta )}{\beta }}-1\right) \log \left(\frac{A-e^{\frac{2 \pi  (\gamma +i \theta )}{\beta }}}{-1+A e^{\frac{2 \pi  (\gamma +i \theta )}{\beta }}}\right)\right].
\end{split}
\end{equation}

Plugging (\ref{gamma ansatz}) into (\ref{delta r}), we get a boundary expression of $\delta r$. We then plug it and (\ref{rbd}) into (\ref{rn ansatz}), we finally have the boundary expression for $r_n$,
\begin{equation} \label{rnbd}
\begin{split}
   r_{n,\text{bd}} (\tau) & =  \epsilon  \theta '(\tau )^{\frac{1}{2}} \Biggl\{ 1-\frac{1}{2}  U_1(\theta (\tau ))+\frac{3}{2} \epsilon ^2 \theta '(\tau )  U_2(\theta (\tau ))+\epsilon ^4\left[-\frac{\pi ^4 \theta '(\tau )^2}{4 \beta ^4}+\left(\frac{7 \pi ^4 \theta '(\tau )^2  U_1(\theta (\tau ))}{24\beta^4}\right.\right. +\frac{5}{6} \theta '(\tau )^2  U_3(\theta (\tau )) \\
   & + \left.\left.\frac{1}{8} \theta ''(\tau ) U_2'(\theta (\tau ))+\frac{3 \theta ''(\tau )^2 U_2(\theta (\tau ))}{16 \theta '(\tau )^2}\right)\right] \Biggr\}.
\end{split}
\end{equation}
We are then all set to compute $\phi_{\text{bd}}$ by plugging (\ref{rnbd}) into (\ref{phi n}), it gives
\begin{equation} \label{phi bd}
    \begin{split}
        \phi_{\text{bd}} & = l^3 \Biggl\{ \frac{1}{\epsilon ^3}\left( \frac{1}{\theta '(\tau )^{3/2}}+\frac{ 3 U_1(\theta (\tau ))}{2 \theta '(\tau )^{3/2}} \right)- \frac{1}{\epsilon} \left(\frac{ 9 U_2(\theta (\tau ))}{2 {\theta '(\tau )}^{1/2}}\right)+\epsilon \left[\frac{3 \pi ^4 {\theta '(\tau )}^{1/2}}{4 \beta ^4} + \left(\frac{5 \pi ^4 {\theta '(\tau )}^{1/2} U_1(\theta (\tau ))}{8 \beta ^4}\right.\right.\\
        & \left.\left.-\frac{3 \theta ''(\tau ) U_2'(\theta (\tau ))}{8 \theta '(\tau )^{3/2}}-\frac{9 \theta ''(\tau )^2 U_2(\theta (\tau ))}{16 \theta '(\tau )^{7/2}}-\frac{5}{2} {\theta '(\tau )}^{1/2} U_3(\theta (\tau ))\right)\right]-\epsilon ^3 \left( \frac{9 \pi ^4 \theta '(\tau )^{3/2} U_2(\theta (\tau ))}{2 \beta ^4}\right) \Biggr\}
    \end{split}
\end{equation}
and
\begin{equation} \label{phi bd 23}
    \begin{split}
        \phi_{\text{bd}}^{2/3} & = l^2 \Biggl\{ \frac{1}{\epsilon ^2}\left( \frac{1}{\theta '(\tau )}+\frac{ U_1(\theta (\tau ))}{\theta '(\tau )} \right)- 3 U_2(\theta (\tau )) +\epsilon^2 \left[\frac{ \pi ^4 \theta '(\tau )}{2 \beta ^4} + \left(\frac{ \pi ^4 {\theta '(\tau )} U_1(\theta (\tau ))}{6 \beta ^4} -\frac{5}{3} {\theta '(\tau )} U_3(\theta (\tau ))\right. \right.\\
        & \left.\left.-\frac{ \theta ''(\tau ) U_2'(\theta (\tau ))}{4 \theta '(\tau )}-\frac{3 \theta ''(\tau )^2 U_2(\theta (\tau ))}{8 \theta '(\tau )^{3}}\right)\right] \Biggr\}.
    \end{split}
\end{equation}

The two boundary terms in the replica gravitational action (\ref{action cb in Appendix B}) are thus 
\begin{equation} \label{action 2}
    \begin{split}
     &  I_ {\text{grav, bd, 1}}[\mathcal{M}_n]= \frac{1}{16 \pi G_2} \cdot 2 \int_{\partial \mathcal{M}_n} \phi K  = \frac{1}{8 \pi G_2}  \int_{\partial \mathcal{M}_n} d\tau \cdot \frac{l^2}{\epsilon^2} \cdot \phi K \\
      & = \frac{l^3}{16 \pi  G_2} \int_{\partial \mathcal{M}_n} d\tau \Biggl\{\frac{1}{\epsilon ^4} \left(\frac{4}{\theta '(\tau )}+\frac{4 U_1(\theta (\tau ))}{\theta '(\tau )} \right) + \frac{1}{\epsilon ^2} \left[ \frac{\theta'''(\tau )}{\theta '(\tau )^3}-\frac{2 \theta ''(\tau )^2}{\theta '(\tau )^4} + \left(U_1''(\theta (\tau ))+\frac{\theta ''(\tau ) U_1'(\theta (\tau ))}{\theta '(\tau )^2}\right.\right.\\
      &-\frac{4 \theta ''(\tau )^2 U_1(\theta (\tau ))}{\theta '(\tau )^4} -24 U_2(\theta (\tau )) + \left.\left.\frac{2 \theta'''(\tau ) U_1(\theta (\tau ))}{\theta '(\tau )^3}\right)\right]+\left(\frac{4 \pi ^4 \theta '(\tau ) U_1(\theta (\tau ))}{\beta ^4}+\theta '(\tau ) U_2''(\theta (\tau ))-20 \theta '(\tau ) U_3 (\theta (\tau ))\right.\\
      & -\frac{3 \theta ''(\tau )^2 U_1''(\theta (\tau ))}{8 \theta '(\tau )^3}+\frac{\theta ''(\tau )^3 U_1'(\theta (\tau ))}{\theta '(\tau )^5}+\frac{63 \theta ''(\tau )^4 U_1(\theta (\tau ))}{32 \theta '(\tau )^7}+\frac{2 \theta ''(\tau ) U_2'(\theta (\tau ))}{\theta '(\tau )}+\frac{6 \theta ''(\tau )^2 U_2(\theta (\tau ))}{\theta '(\tau )^3}\\
      & \left.-\frac{3 \theta'''(\tau ) \theta ''(\tau ) U_1'(\theta (\tau ))}{4 \theta '(\tau )^4}-\frac{9 \theta'''(\tau ) \theta ''(\tau )^2 U_1(\theta (\tau ))}{8 \theta '(\tau )^6}-\frac{3 \theta'''(\tau )U_2(\theta (\tau ))}{\theta '(\tau )^2}\right) \Biggr\}
    \end{split}
\end{equation}
and
\begin{equation} \label{action 3}
    \begin{split}
    & I_ {\text{grav, bd, 2}}[\mathcal{M}_n] = - \frac{1}{16 \pi G_2} \cdot  \frac{6}{l} \int_{\partial \mathcal{M}_n}  \phi^{\frac{2}{3}} = - \frac{3}{8 \pi G_2}  \int_{\partial \mathcal{M}_n} d\tau \cdot \frac{l}{\epsilon^2} \cdot \phi^{\frac{2}{3}}\\
    & = \frac{l^3}{16 \pi  G_2} \int_{\partial \mathcal{M}_n} d\tau\Biggl\{ - \frac{1}{\epsilon ^4} \left(\frac{6}{\theta '(\tau )}+\frac{6 U_1(\theta (\tau ))}{\theta '(\tau )} \right)+ \frac{1}{\epsilon^2} \cdot 18 \, U_2(\theta (\tau )) + \left[ - \frac{3 \pi ^4 \theta '(\tau )}{\beta ^4} +\left( -\frac{\pi ^4 \theta '(\tau ) U_1 (\theta (\tau ))}{\beta ^4}\right.\right.\\
    & \left.\left.+ 10 \theta '(\tau ) U_3(\theta (\tau ))+\frac{3 \theta ''(\tau ) U_2'(\theta (\tau ))}{2 \theta '(\tau )}+\frac{9 \theta ''(\tau )^2 U_2(\theta (\tau ))}{4 \theta '(\tau )^3}\right)\right] \Biggr\}.
    \end{split}  
\end{equation}

The remaining challenge that we have never been confronted with in JT gravity --- the problem of evaluating the bulk term in action (\ref{action cb in Appendix B}) which initially involves conical singularities and cosmic brane terms cancelling them --- can be circumvented by converting it to another equivalent evaluation much easier to tackle, namely that we directly evaluate the bulk term on $\Tilde{\mathcal{M}}_n$ and then divide it by $n$, altogether doing away with the singularities from the very beginning of our calculation. One important note before we start off --- though we can cut corners tremendously by skillfully switching between two pictures, the covering manifold $\Tilde{\mathcal{M}}_n$ and the orbifold $\mathcal{M}_n$ of the replica geometry, we must take heed of exactly which boundary time we are integrating over on $\Tilde{\mathcal{M}}_n$.\\

Following the definition $w = e^{\frac{2\pi}{\beta} (\gamma + i \theta)}$, we define "space" coordinate $\gamma_n$ and "time" coordinate $\theta_n$ for the auxiliary coordinate $w_n$ defined on $\Tilde{\mathcal{M}}_n$ by
\begin{equation} \label{gamma n and theta n}
    w_n = e^{\frac{2\pi}{\beta} (\gamma_n  + i \theta_n)}.
\end{equation}
The metric (\ref{metric in covering manifold}) on $\Tilde{\mathcal{M}}_n$ can thus be alternatively written as
\begin{equation} \label{metric in covering manifold 2}
 ds^2_{\text{in}, \Tilde{\mathcal{M}}_n} =   \frac{\beta^2 l^4}{4 \pi^2} \left( \frac{1}{r_n^4} - \frac{\pi^4}{\beta^4}\right)\left( d\gamma_n^2 + d\theta_n^2 \right) = \frac{l^4}{r_n^4} \left[ \left( 1- \frac{\pi^4 r_n^4}{\beta^4} \right) d\theta_n^2 + {\left( 1 - \frac{\pi^4 r_n^4}{\beta^4} \right)}^{-1} dr_n^2 \right]
\end{equation}
where $r_n$ and $\gamma_n$ are related by
\begin{equation}\label{inverse rn Euclidean 2}
 \gamma_n = -\frac{\beta}{2 \pi} \left( \arctan{ \frac{\pi r_n}{\beta} }  + \arctanh{ \frac{\pi r_n}{\beta} }  \right)
\end{equation} in parallel with (\ref{tortoiseDef}) according to (\ref{inverse rn Euclidean}) and (\ref{gamma n and theta n}).\\

Next up, here comes the more intricate part - what is the relationship between the two boundary times --- $\theta_n$ and $\theta$ --- on the boundary? To answer it, we first obtain $\gamma_n$ at the boundary through $r_{n,\text{bd}}(\tau)$ (\ref{rnbd}) and asymptotic expansion of (\ref{inverse rn Euclidean 2}) near/at the boundary, then we extract $\theta_{n,\text{bd}}$ as a function of $\theta$ from (\ref{gamma n and theta n}),(\ref{wn}) and $w = e^{\frac{2\pi}{\beta} (\gamma + i \theta)}$ combined. Since the geometry on $\Tilde{\mathcal{M}}_n$ is the same Narayan black hole of inverse temperature $\beta$, we have from (\ref{inverse rn Euclidean 2}) and (\ref{rnbd})
\begin{equation}\label{gamma n bd}
    \begin{split}
        \gamma_{n,\text{bd}} (\tau) & = - \left( r_{n,\text{bd}} (\tau) + \frac{\pi ^4 {r_{n,\text{bd}}(\tau)}^5}{5 \beta ^4} \right)\\
        & = - \epsilon \theta '(\tau )^{\frac{1}{2}} \Biggl\{ 1 -\frac{1}{2} U_1(\theta (\tau ))+\frac{3}{2} \epsilon ^2    U_2(\theta (\tau )) \theta'(\tau ) + \epsilon ^4\left[-\frac{\pi ^4 \theta '(\tau )^2}{20 \beta ^4} + \left(-\frac{5 \pi ^4 \theta '(\tau )^{5/2} U_1(\theta (\tau ))}{24 \beta ^4}\right.\right.\\
        & \left.\left. + \frac{1}{8} \theta '(\tau )^{\frac{1}{2}} \theta ''(\tau ) U_2'(\theta (\tau ))+\frac{3 \theta ''(\tau )^2 U_2 (\theta (\tau ))}{16 \theta '(\tau )^{3/2}}+\frac{5}{6} \theta '(\tau )^{5/2} U_3(\theta (\tau ))\right)\right] \Biggr\}.
    \end{split}
\end{equation}
Then we have from (\ref{wn}) and $w = e^{\frac{2\pi}{\beta} (\gamma + i \theta)}$ 
\begin{equation} \label{extract theta n}
    \begin{split}
        \theta_{n} & = \frac{1}{i} \left( \frac{\beta}{2 \pi}\log w_n - \gamma_n \right) = - i \left[ \frac{\beta}{2 \pi} \log \left( \frac{{\left( \frac{ e^{\frac{2\pi}{\beta} (\gamma + i \theta)}-A}{1-A e^{\frac{2\pi}{\beta} (\gamma + i \theta)}} \right)}^{1/n} + A}{1+ A {\left( \frac{ e^{\frac{2\pi}{\beta} (\gamma + i \theta)}-A}{1-A e^{\frac{2\pi}{\beta} (\gamma + i \theta)}} \right)}^{1/n}}\right) - \gamma _n\right].
        \end{split}
\end{equation}
We then plug in $\gamma$ at the boundary (\ref{gamma ansatz}) and $\gamma_n$ at the boundary (\ref{gamma n bd}) and expand to the first order of $(n-1)$, it gives $\theta_{n,\text{bd}} (\tau)$ as a function of the boundary mode $\theta(\tau)$
\begin{equation} \label{theta n bd}
    \begin{split}
     \theta_{n,\text{bd}} (\tau) & = \theta (\tau) + (n-1) \cdot \frac{i \beta  e^{-\frac{2 i \pi  \theta (\tau )}{\beta }} \left(A^2 \left(-e^{\frac{2 i \pi  \theta (\tau )}{\beta }}\right)+A e^{\frac{4 i \pi  \theta (\tau )}{\beta }}+A-e^{\frac{2 i \pi  \theta (\tau )}{\beta }}\right) \log \left(\frac{A-e^{\frac{2 i \pi  \theta (\tau )}{\beta }}}{-1+A e^{\frac{2 i \pi  \theta (\tau )}{\beta }}}\right)}{2 \pi  \left(A^2-1\right)}\\
     & +\frac{1}{2}  \epsilon ^2 \theta '(\tau ) U_1'(\theta (\tau )) - \frac{1}{24}  \epsilon ^4 \theta '(\tau )^2 U_1''' (\theta (\tau ))
    \end{split}
\end{equation}
where the $(n-1)$ perturbation to the zeroth order in $\epsilon$ satisfies
 \begin{equation}
    (n-1) \cdot \frac{i \beta  e^{-\frac{2 i \pi  \theta (\tau )}{\beta }} \left(A^2 \left(-e^{\frac{2 i \pi  \theta (\tau )}{\beta }}\right)+A e^{\frac{4 i \pi  \theta (\tau )}{\beta }}+A-e^{\frac{2 i \pi  \theta (\tau )}{\beta }}\right) \log \left(\frac{A-e^{\frac{2 i \pi  \theta (\tau )}{\beta }}}{-1+A e^{\frac{2 i \pi  \theta (\tau )}{\beta }}}\right)}{2 \pi  \left(A^2-1\right)} = - \int d\theta(\tau) U_1(\theta(\tau)).
\end{equation}
Finally, we can calculate the following Jacobian at the $\Tilde{\mathcal{M}}_n$ boundary
\begin{equation} \label{theta n Jacobian}
    \frac{d \theta_{n,\text{bd}}(\tau)}{d \theta (\tau)} = 1 + \left[-U_1(\theta (\tau ))+\epsilon ^2 \left(\frac{1}{2} \theta '(\tau ) U_1''(\theta (\tau ))+\frac{\theta ''(\tau ) U_1'(\theta (\tau ))}{2 \theta '(\tau )}\right) + \epsilon ^4 \left(-\frac{1}{24} \theta '(\tau )^2 U_1^{(4)}(\theta (\tau ))-\frac{1}{12} \theta ''(\tau ) U_1'''(\theta (\tau ))\right)\right]
\end{equation}
This will be exploited below to convert the integral on the boundary of $\Tilde{\mathcal{M}}_n$ to the integrals on n copies of $\mathcal{M}_n$.\\

Putting together all we have derived above, the bulk term in $I_{\text{grav}}$ on $\mathcal{M}_n$ is 
\begin{equation} \label{action 1}
    \begin{split}
          I_ {\text{grav, bulk}}[\mathcal{M}_n]  & = -\left[ (\text{topological on}\; \mathcal{M}_n )  + \frac{1}{16 \pi G_2}  \int_{\mathcal{M}_n} \big(  \phi R + 12 l^{-2} \phi^{\frac{1}{3}} \big)  - \frac{1}{4 G_2} \left( 1 - \frac{1}{n} \right) \sum_{\alpha, \text{Cosmic Branes}} \left[ \phi_0 + \phi(w_\alpha) \right]\right]\\
        & = \frac{1}{n} I_ {\text{grav, bulk}}[\Tilde{\mathcal{M}}_n] = \frac{1}{n} \left[(\text{topological on } \Tilde{\mathcal{M}}_n )   + \frac{1}{16 \pi G_2}  \int_{\Tilde{\mathcal{M}}_n} \big(  \phi R + 12 l^{-2} \phi^{\frac{1}{3}} \big)\right]\\
        & =  \frac{1}{16  \pi n G_2} \int_{\Tilde{\mathcal{M}}_n}  d\theta_n \int_{r_{n,\text{bd}} (\tau)}^{r_{n,\text{horizon}} = \frac{\beta}{\pi}} dr_n \; \left( \frac{l^4}{r_n^4} \right) \cdot \left[ \frac{l^3}{r_n^3}*  \frac{-4 r_n^2}{l^4}  + 12 l^{-2} {\bigg( \frac{l^3}{r_n^3} \bigg)}^{\frac{1}{3}}  \right]\\
        & = \frac{l}{2 \pi n G_2} \int_{\Tilde{\mathcal{M}}_n}  d\theta_n \int_{r_{n,\text{bd}} (\tau)}^{r_{n,\text{horizon}} = \frac{\beta}{\pi}} dr_n \; \frac{1}{r_n^5} = - \frac{l}{8 \pi n G_2}  \int_{\Tilde{\mathcal{M}}_n}  d\theta_n \left(\frac{1}{r_n^4} \right) \Bigg\vert_{r_{n,\text{bd}}(\tau)}^{\frac{\beta}{\pi}} \\
        & =  - \frac{l}{8 \pi n G_2}  \int_0^\beta  d\theta_n \frac{\pi^4}{\beta^4} +  \frac{l}{8 \pi n G_2}  \int_{\partial \Tilde{\mathcal{M}}_n}  d\theta_n \left(\frac{1}{r_n^4} \right) =  - \frac{\pi^3 l}{8 n G_2 \beta^3} +  \frac{l}{8 \pi n G_2}  \int_{\partial \Tilde{\mathcal{M}}_n}  d\theta_n \left(\frac{1}{r_n^4} \right)\\
        & =  \frac{l}{8\pi G_2} \cdot \frac{1}{n} \cdot n \int_{\partial \mathcal{M}_n} d\tau \; \frac{d \theta_{n,\text{bd}}}{d \theta} \theta '(\tau ) \frac{1}{{r_{n,\text{bd}}(\tau)}^4}\\
        & = \frac{l^3}{16 \pi  G_2} \int_{\partial \mathcal{M}_n} d\tau\Biggl\{ \frac{1}{\epsilon^4} \left( \frac{2}{\theta '(\tau )}+\frac{2  U_1(\theta (\tau ))}{\theta '(\tau )} \right) + \frac{1}{\epsilon^2} \left( U_1''(\theta (\tau ))+\frac{\theta ''(\tau ) U_1'(\theta (\tau ))}{\theta '(\tau )^2}-12 U_2(\theta (\tau )) \right)\\
        & + \left[\frac{2 \pi ^4 \theta '(\tau )}{\beta ^4} + \left( \frac{2 \pi ^4 \theta '(\tau ) U_1(\theta (\tau ))}{3 \beta ^4}-\frac{\theta ''(\tau ) U_2'(\theta (\tau ))}{\theta '(\tau )}-\frac{3 \theta ''(\tau )^2 U_2(\theta (\tau ))}{2 \theta '(\tau )^3}-\frac{20}{3} \theta '(\tau ) U_3(\theta (\tau )) \right.\right.\\
        & \left. \left. -\frac{1}{12} \theta '(\tau ) U_1^{(4)}(\theta (\tau ))-\frac{\theta ''(\tau ) U_1'''(\theta (\tau ))}{6 \theta '(\tau )}\right) \right] \Biggr\}
    \end{split}
\end{equation}
where the topological bulk term in the second line and the contribution to the definite integral over $r_n$ from $r_{n,\text{horizon}} = \frac{\beta}{\pi}$ in the fifth line are dropped because these two terms are merely constants and thus totally irrelevant to the boundary modes $\theta (\tau)$. Also it is notable that we made use of the Jacobian (\ref{theta n Jacobian}) on the sixth line to convert the integral over $\theta(\tau)$ on $\Tilde{\mathcal{M}}_n$ boundary to the integral over $\tau$ on n copies of $\mathcal{M}_n$ boundary, hence the factor of $n$ appears in the sixth line, cancelling the factor $\frac{1}{n}$ to give the final answer where $n$ is absent. \\

In the end, we add up the bulk (\ref{action 1}) and boundary parts (\ref{action 2}) (\ref{action 3}) of the on-shell replica gravitational action on $\mathcal{M}_n$ (\ref{action cb in Appendix B}) and get
\begin{equation} \label{replica action in appendix B}
       \begin{split}
      & I_ {\text{grav}}[\mathcal{M}_n] =  \ I_ {\text{grav, bulk}}[\mathcal{M}_n] + I_ {\text{grav, bd, 1}}[\mathcal{M}_n] + I_ {\text{grav, bd, 2}}[\mathcal{M}_n] \\ 
      & = -\frac{l^3}{16 \pi G_2} \int_{\partial \mathcal{M}_n} d\tau \Biggl\{ \frac{1}{\epsilon^2} \left[ \left(\frac{2 \theta ''(\tau )^2}{\theta '(\tau )^4}-\frac{\theta''' (\tau )}{\theta '(\tau )^3}\right)+\left(-2 U_1''(\theta (\tau ))-\frac{2 \theta ''(\tau ) U_1'(\theta (\tau ))}{\theta '(\tau )^2}-\frac{2 \theta '''(\tau ) U_1(\theta (\tau ))}{\theta '(\tau )^3} \right. \right.\\
      & \left. \left. +\frac{4 \theta ''(\tau )^2 U_1(\theta (\tau ))}{\theta '(\tau )^4}+18 U_2 (\theta (\tau ))\right) \right] + \bigg[ \left(\frac{3 \pi ^4 \theta '(\tau )}{\beta ^4}+\frac{21 \theta ''(\tau )^4}{32 \theta '(\tau )^7}-\frac{3 \theta''' (\tau ) \theta ''(\tau )^2}{8 \theta '(\tau )^6} \right)\\
        & +\left(\frac{11 \pi ^4 \theta '(\tau ) U_1 (\theta (\tau ))}{3 \beta ^4}+\theta '(\tau ) U_2''(\theta (\tau ))+\frac{5 \theta ''(\tau ) U_2'(\theta (\tau ))}{2 \theta '(\tau )}+\frac{27 \theta ''(\tau )^2 U_2(\theta (\tau ))}{4 \theta '(\tau )^3}-\frac{50}{3} \theta '(\tau ) U_3 (\theta (\tau )) \right.\\
        & -\frac{\theta ''(\tau ) U_1'''(\theta (\tau ))}{6 \theta '(\tau )}-\frac{3 \theta ''(\tau )^2 U_1''(\theta (\tau ))}{8 \theta '(\tau )^3}+\frac{\theta ''(\tau )^3 U_1'(\theta (\tau ))}{\theta '(\tau )^5}+\frac{63 \theta ''(\tau )^4 U_1(\theta (\tau ))}{32 \theta '(\tau )^7}-\frac{3 \theta''' (\tau ) U_2(\theta (\tau ))}{\theta '(\tau )^2}\\
        & \left.-\frac{1}{12} \theta '(\tau ) U_1^{(4)}(\theta (\tau ))-\frac{3 \theta'''(\tau ) \theta ''(\tau ) U_1'(\theta (\tau ))}{4 \theta '(\tau )^4}-\frac{9 \theta'''(\tau ) \theta ''(\tau )^2 U_1 (\theta (\tau ))}{8 \theta '(\tau )^6} \right) \bigg]  \Biggr\}
  \end{split} 
\end{equation}
where there is no $1/\epsilon^4$ divergence. Although there is still a $1/\epsilon^2$ divergence, it can be cancelled by the Euclidean version of the same counterterm (\ref{epsilon 2 counterterm}), of  course now evaluated in replica geometry, just like in (\ref{on-shell action n=1 0}). So we can safely throw it off at the end of the day.

\newpage
\bibliographystyle{unsrt}
\bibliography{Bib.bib}
\addcontentsline{toc}{section}{References}
\end{document}